%% file: springer_formatted.tex
\documentclass[graybox]{svmult}

\usepackage{type1cm} 
\usepackage{makeidx}         
\usepackage{graphicx}   
\usepackage{multicol}  
\usepackage[bottom]{footmisc}
\usepackage{caption}
\usepackage{newtxtext}   
\usepackage{newtxmath}    

\makeindex 

\usepackage[title]{appendix}
\usepackage{amsmath}

\DeclareMathOperator{\R}{\mathbb{R}}

\begin{document}

\title*{Persistent topology of protein space}
\author{W. Hamilton, J.E. Borgert, T. Hamelryck, J.S. Marron}
\institute{W. Hamilton \at Department of Mathematics, University of Utah, Salt Lake City, UT, USA\\ \email{hamilton@math.utah.edu}
\and J.E. Borgert \at Department of Statistics and OR, University of North Carolina, Chapel Hill, NC, USA\\ \email{elyseb@live.unc.edu} \and T. Hamelryck \at Department of Computer Science \& Department of Biology, University of Copenhagen, DK\\ \email{thamelry@bio.ku.dk} \and J.S. Marron \at Department of Statistics and OR, University of North Carolina, Chapel Hill, NC, USA\\ \email{marron@unc.edu}}
%
%
\maketitle

\abstract*{
	Protein fold classification is a classic problem in structural biology and bioinformatics. We approach this problem using persistent homology. In particular, we use alpha shape filtrations to compare a topological representation of the data with a different representation that makes use of knot-theoretic ideas. We use the statistical method of Angle-based Joint and Individual Variation Explained (AJIVE) to understand similarities and differences between these representations.
}

\abstract{
	Protein fold classification is a classic problem in structural biology and bioinformatics. We approach this problem using persistent homology. In particular, we use alpha shape filtrations to compare a topological representation of the data with a different representation that makes use of knot-theoretic ideas. We use the statistical method of Angle-based Joint and Individual Variation Explained (AJIVE) to understand similarities and differences between these representations.
}

\section{Introduction}

%
%


In this study we use persistent homology (PH) and Wasserstein distances between topological representations to quantify the geometry and topology of protein folds. We present our analysis of the resulting geometry of \textit{protein space}, and compare our results to existing protein structure quantification approaches, notably the Gaussian Integral Tuned (GIT) \cite{roegen.2003.PNAS} vector representation. The GIT vector approach been shown to be very effective at automated protein classification \cite{gronbaek.2020.PJ,roegen.2003.PNAS}.

\begin{figure}[ht]
	\centering
	
	\includegraphics[width = 0.45\textwidth]{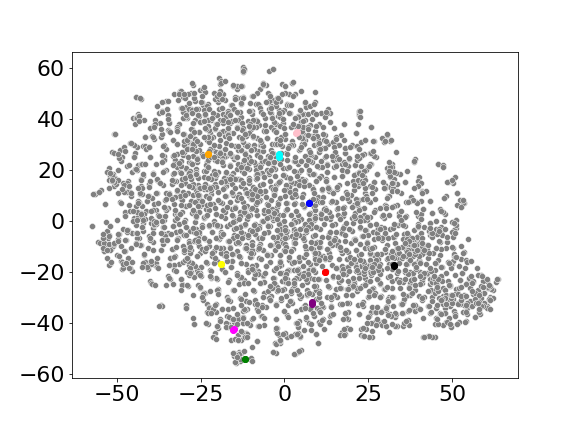}
	\includegraphics[width = 0.45\textwidth]{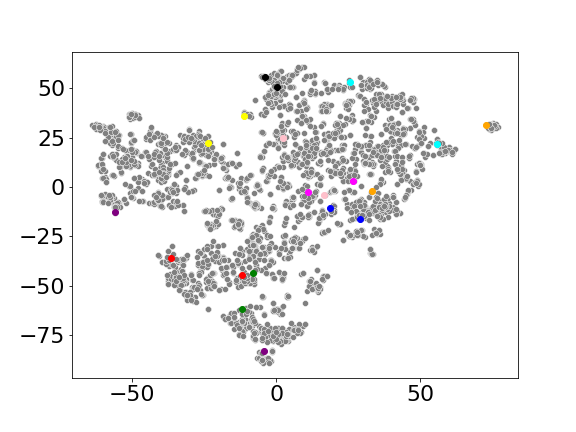}
	\caption{Ten pairs of proteins with common CAT label that are closely related by PH but not by GIT. The left plot shows the t-SNE coordinates of the individual PH matrix. The right plot shows the t-SNE coordinates of the original GIT matrix. Each pair of proteins is colored uniquely. The pairs are close together (overlapping) in PH but widely separated in GIT.}
	\label{fig:CAT_pairs}
\end{figure}

A major goal is to understand what insights the PH approach reveals that GIT vectors do not. The GIT representation, which effectively quantifies local curvature, corresponds well to the conventional structural classification labels called CATH \cite{burley.2020.NAR}. The PH representation, on the other hand, focuses more on global structure. To investigate the different structural aspects captured by these two methods, we apply statistical methodology developed for multi-block data. Such data refers to the setting where disparate sets of features on a common set of samples are represented by multiple data blocks. The method, Angle-based Joint and Individual Variation Explained (AJIVE) \cite{feng.2018.JMvA}, decomposes the data blocks into joint and individual modes of variation. Applying AJIVE to the data blocks generated by each representation (GIT and PH) allows us to compare what is jointly captured by both methods and, more interestingly, to contrast the individual features. In particular, we find topologically similar pairs of proteins in the PH individual matrix that are far apart in the GIT representation, as shown in Figure \ref{fig:CAT_pairs}. 

We compute the \textit{t-distributed stochastic neighbor embedding} (t-SNE) \cite{vanderMaaten.2008.JMLR} coordinates of each data matrix for visual comparison. The t-SNE method is a nonlinear dimensionality reduction technique that is useful for embedding high-dimensional data in a lower dimensional space for visualization; moreover, t-SNE tends to separate data into clusters and is very popular for visualizing clustered data. To investigate proteins that are considered similar from the PH view but not by GIT, we first calculate the AJIVE PH individual data matrix. Next, we identify pairs of proteins with the same CATH labels whose distance from each other is relatively close in the t-SNE coordinates of the PH individual matrix compared to the t-SNE coordinates of the original GIT matrix, as shown by Figure \ref{fig:CAT_pairs}. On the left is the t-SNE plot of the PH individual matrix with proteins colored by pair and on the right is the t-SNE plot of the original GIT vectors matrix with the same coloring. Recall that the PH individual matrix gives us components of the variation that are captured by PH but not by the GIT vectors. Since the pairs overlap in the left plot of Figure  \ref{fig:CAT_pairs} and spread out in the right plot, we see that the topological similarities of these proteins identified by PH are not detected by the GIT approach.

There is already a growing literature on using persistent homology for protein structure analysis, including protein compressibility \cite{gamerio.2015.JIAM,ichinomiya.2020.BJ,xia.2014.IJNMBE} and protein classification \cite{cang.2015.CMB}. As mentioned above, the goal of this study was to compare global topological descriptors of protein structure, via persistent homology, to a popular local curvature-based descriptor, via GIT vectors. 

\subsection{Acknowledgements}

Peter R{\o}gen provided helpful comments and suggestions regarding the GIT vector discussion. We also thank the anonymous reviewers for substantial comments and suggestions. The research of J. E. Borgert and J. S. Marron were partially supported by the grants NIH/NIAMS, P30AR072580 and NIH/NIH, R21AR074685.

\section{Methods}

\subsection{Alpha complexes and filtrations}

\begin{figure}[h]
    \centering
    \includegraphics[width = 0.20\textwidth]{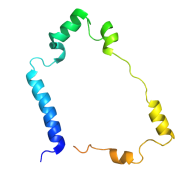}
    \includegraphics[width = 0.24\textwidth]{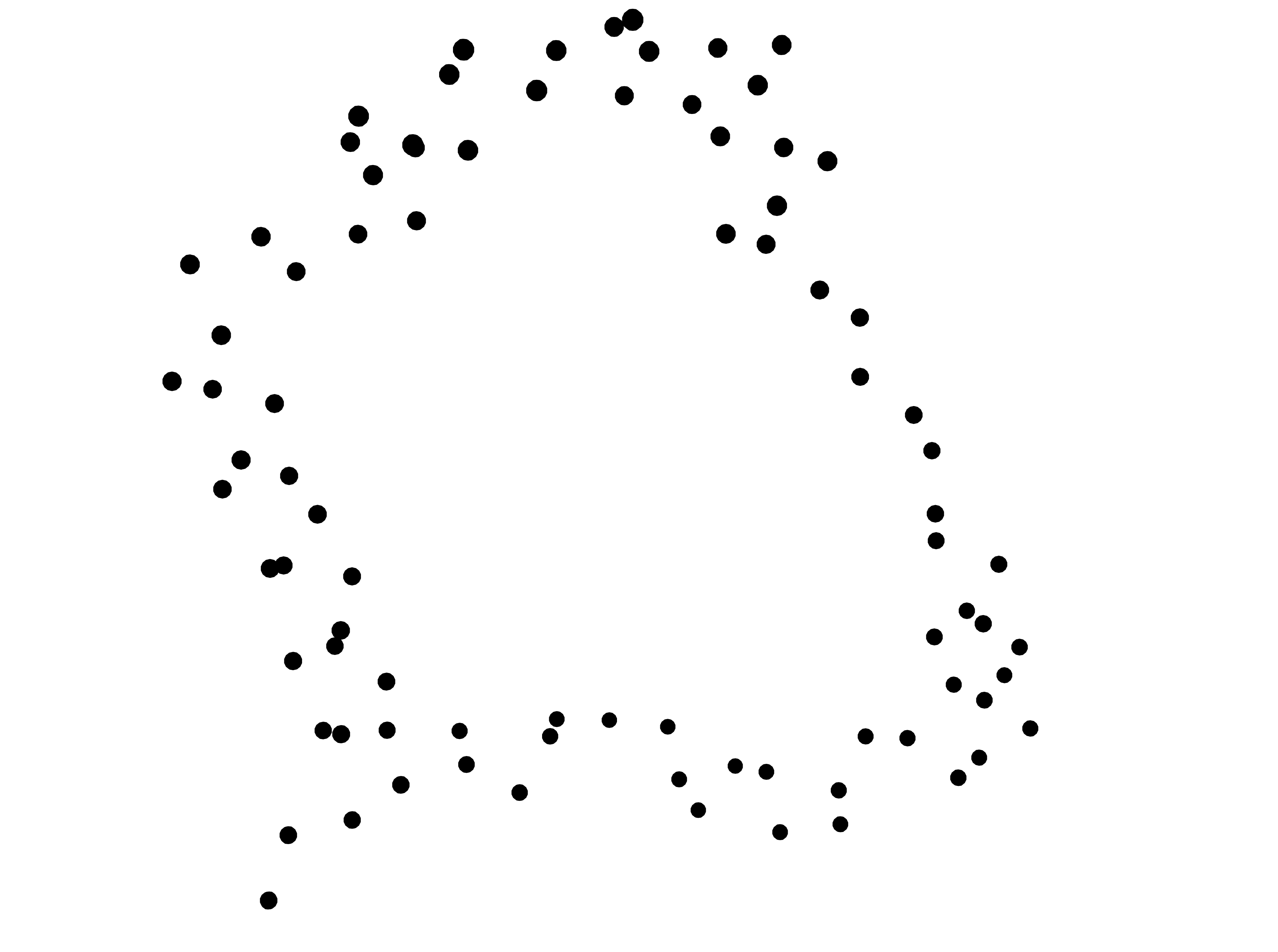}
    \includegraphics[width = 0.24\textwidth]{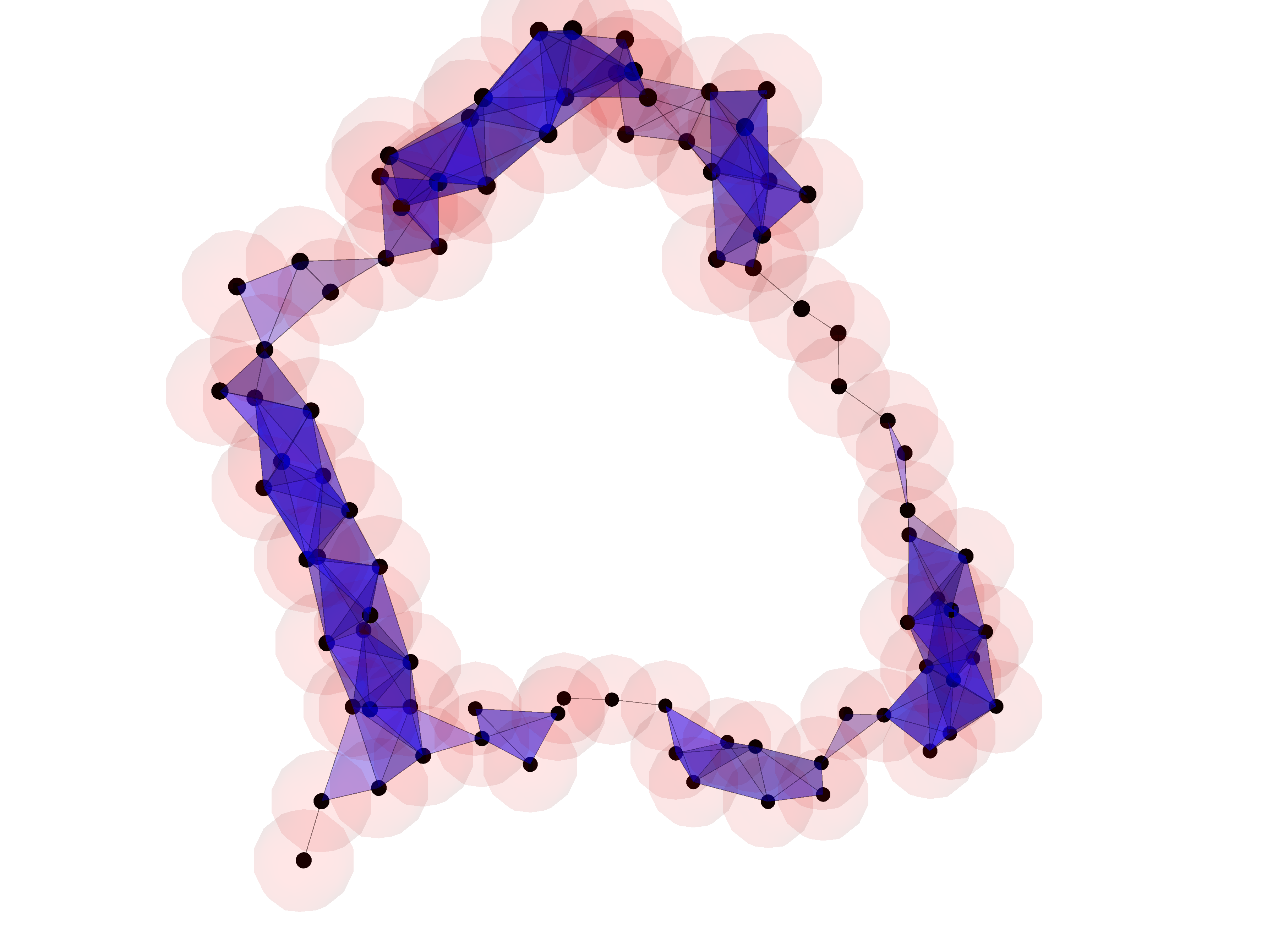}
    \includegraphics[width = 0.24\textwidth]{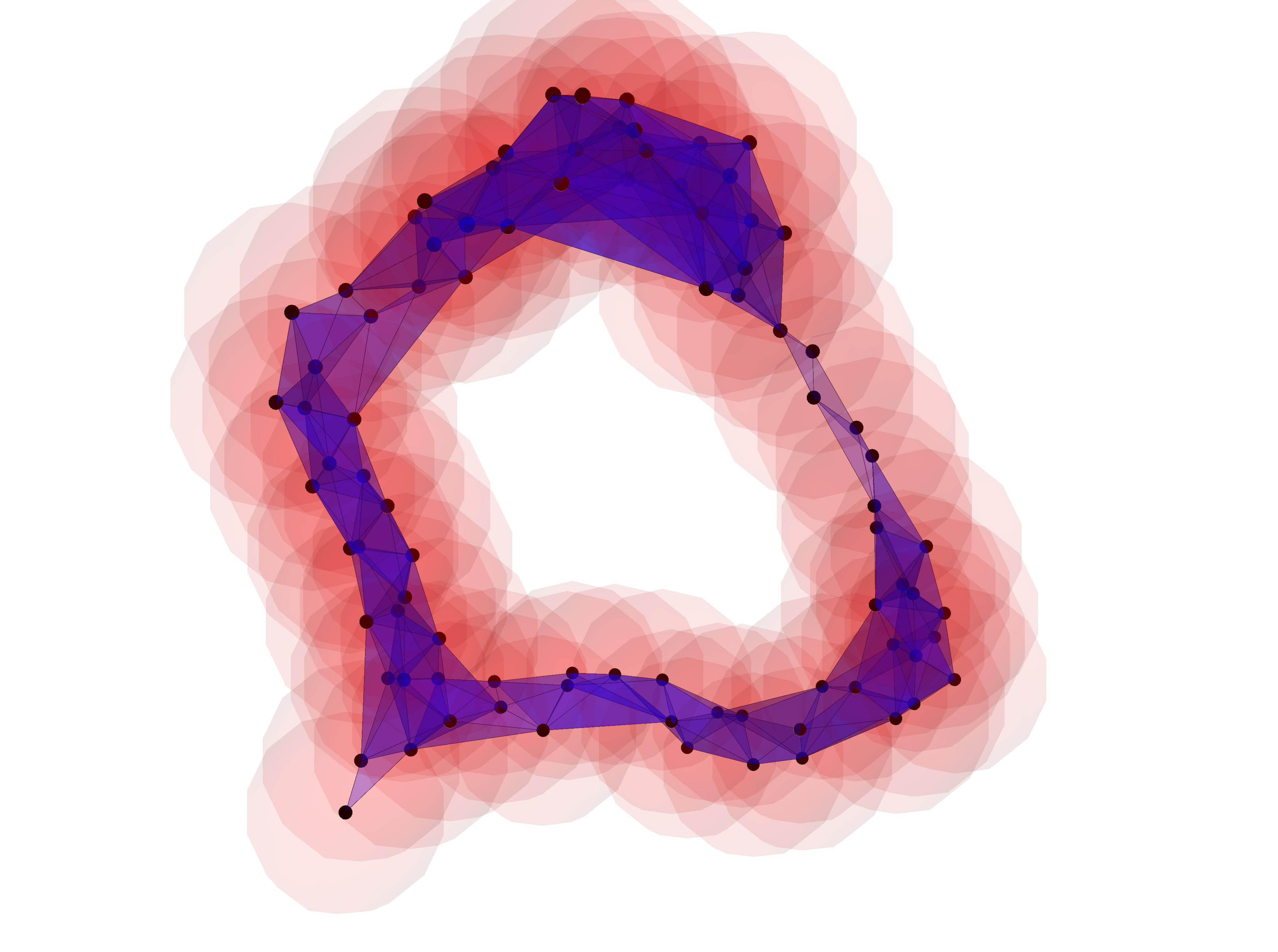}
    
    \caption{Left: a visualization of the 3D structure of the protein chain 1HFE-T. Center-left: the point cloud of $C_\alpha$ atoms used to build the alpha filtration. Center-right: the alpha shape for 1HFE-T with alpha $=1.7$\AA. At this level, the 5 $\alpha$-helices in 1HFE-T can be inferred from the locations of triangles, and the hole they surround has as boundary the union of red balls. Right: the alpha shape for 1HFE-T for alpha$=2.6$ \AA.}
    \label{fig:alpha_example}
\end{figure}

Alpha complexes and alpha filtrations are families of triangulated surfaces associated to a point cloud in $\R^n$ \cite{edelsbrunner.1995.DCG,edelsbrunner.1983.TIT}, and have been proposed as effective geometrizations of protein tertiary structure \cite{winter.2009.IEEE}. In this section we describe the construction of alpha shapes and alpha filtrations and illustrate these objects on a protein chain in Figure \ref{fig:alpha_example}.

Alpha complexes are one approach to defining a combinatorial structure on a point cloud in Euclidean space. The idea is to place a ball of some fixed radius around each point, and connect points with an edge if the corresponding balls intersect, connect three points with a triangle if all three balls intersect, etc. Since we want each edge, triangle, etc., to be geometrically meaningful, we also ask that the balls are intersected with the Voronoi cells of their centers. Recall that the Voronoi cell
$V_{x_i}$ of a point $x_i$ in a point cloud $x_1, x_2, ..., x_N$ is the set $$V_{x_i} = \{x \colon |x_i - x| \leq |x_j - x| \text{ for } j\neq i\}.$$ This intersection requirement ensures that edges cannot ``jump over'' points to connect non-``adjacent'' points in the point cloud, and that no combinatorial objects of dimension higher than the ambient space can appear. For an explicit definition of an alpha complex $X_\alpha$, let $B_\alpha(x_i) = \{x \colon |x - x_i| \leq \alpha \text{ and } x \in V_{x_i} \}$ be the ball of radius $\alpha$ centered at $x_i$ and define the $k$-skeleton of $X_\alpha$, denoted $X_\alpha^{(k)}$, to be the collection of all $k$-simplices in the complex: $$X^{(k)}_\alpha = \{ (x_{i_1},..., x_{i_k}) \colon \cap_{j = 1}^{k} B_{\alpha}(x_{i_j}) \neq \emptyset \}.$$ The alpha complex for radius $\alpha$ is then the union of $k$-skeletons for $k=1, 2, ..., d$ for a point cloud in $\R^d$. In the rest of this paper we focus on alpha shapes in $\mathbb{R}^3$, since the chemically relevant protein structure we study is 3-D.

One natural question is: what radius of ball should we use to construct the alpha complex? Note that if $\alpha_1 < \alpha_2$, we have an inclusion of complexes $X_{\alpha_1}\subset X_{\alpha_2}$. This enables us to keep track of the alpha complexes for various radii and to utilize a multiscale view of the data. Mathematically, a {\em filtration} is a family of complexes $X_1\subset X_2 \subset \cdots X_n \subset \cdots$ such that $X_i \subset X_{i+1}$, and an alpha filtration is a family of alpha complexes parametrized by the radius of balls used in the construction. Even though we can vary the radius continuously, new simplices in an alpha filtration are added at discrete stages. Thus, in practice, this family of alpha complexes for a continuously varying radius can be completely described by a discrete family of alpha complexes $X_{\alpha_1} \subset X_{\alpha_2} \subset \cdots \subset X_{\alpha_n}$ for some largest radius $\alpha_n$. Filtrations are computationally useful when we consider persistent homology and intrinsic topological features across scales.

In Figure \ref{fig:alpha_example} we show a 3-D rendering of the protein chain 1HFE-T (left) \cite{burley.2020.NAR}, along with the collection of $C_\alpha$ atoms (center-left) and two alpha complexes in the filtration (center-right, right). The growing balls in the alpha shape construction are indicated in red, edges between points are indicated in black, and triangles between triples of points are shown in blue. Note that the average distance between adjacent $C_\alpha$ atoms along a protein is approximately 3.5 \AA. Using balls of radius 1.7 \AA (center-right), about half the average intermolecular distance, the alpha shape has a single connected component. Hence, in our context, connectivity alone as a topological feature contains no useful information. The center-right plot shows the 5 $\alpha$-helices of 1HFE-T as the regions with many triangles present. Using balls of radius 2.6 \AA (right) the individual $\alpha$-helices are no longer visible and the ``hole'' in the center of the protein is more clearly defined. This example illustrates how filtrations can reveal multiscale structure in data, of which persistent homology is able to quantify.

\subsection{Persistent homology}

Persistent homology (PH) is a mathematical approach to quantify multiscale topological features, such as connected components ($0$-dimensional), non-contractible loops ($1$-dimensional), and higher-dimensional analogues such as cavities. At each stage of a filtration topological features are computed, and then matched with the corresponding features at other stages (if the features still exist). The collection of all persistent features across the filtration for dimension $d$ is referred to as $H_k$, or $H_k(X_\alpha)$ to emphasize the filtration used. For an introductory article to PH see \cite{ghrist.2008.BAMS,carlsson.2009.BAMS}, and for a textbook introduction see \cite{edelsbrunner.2010.AMS,ghrist.2014.createspace}. We do not consider $H_0$ further here because, as noted above, connectivity alone contains no useful chemical information. It would be chemically interesting to consider $H_2$, which focuses on cavities in the data, as well. We do not use $H_2$ here since our focus is on comparing to GIT, which is about 1-dimensional structure in the data.



\begin{figure}[ht]
	\centering
	\includegraphics[width = 0.32\textwidth]{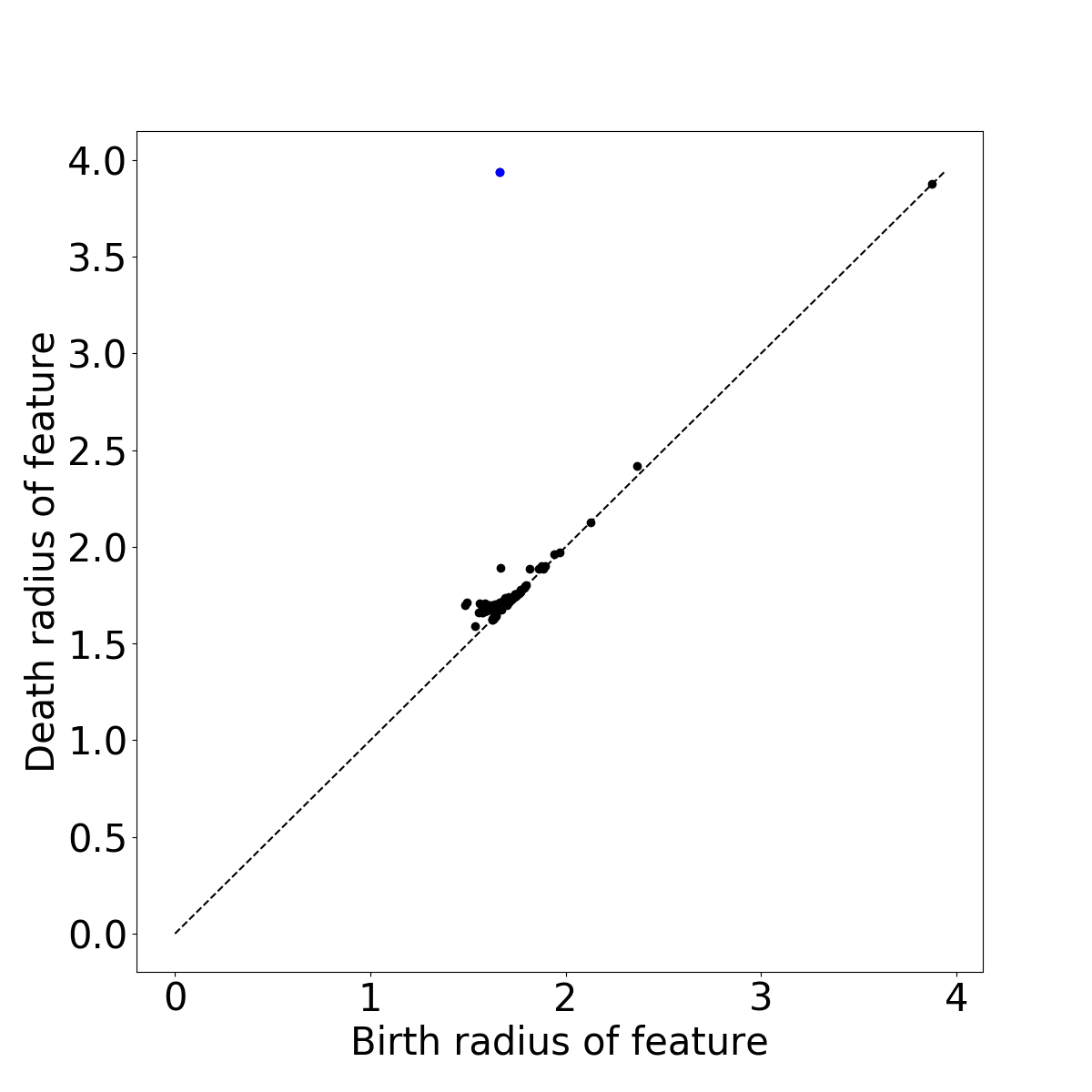}
	\includegraphics[width = 0.32\textwidth]{plots/protein_alpha_example_bt_2p6.png}
	\includegraphics[width = 0.32\textwidth]{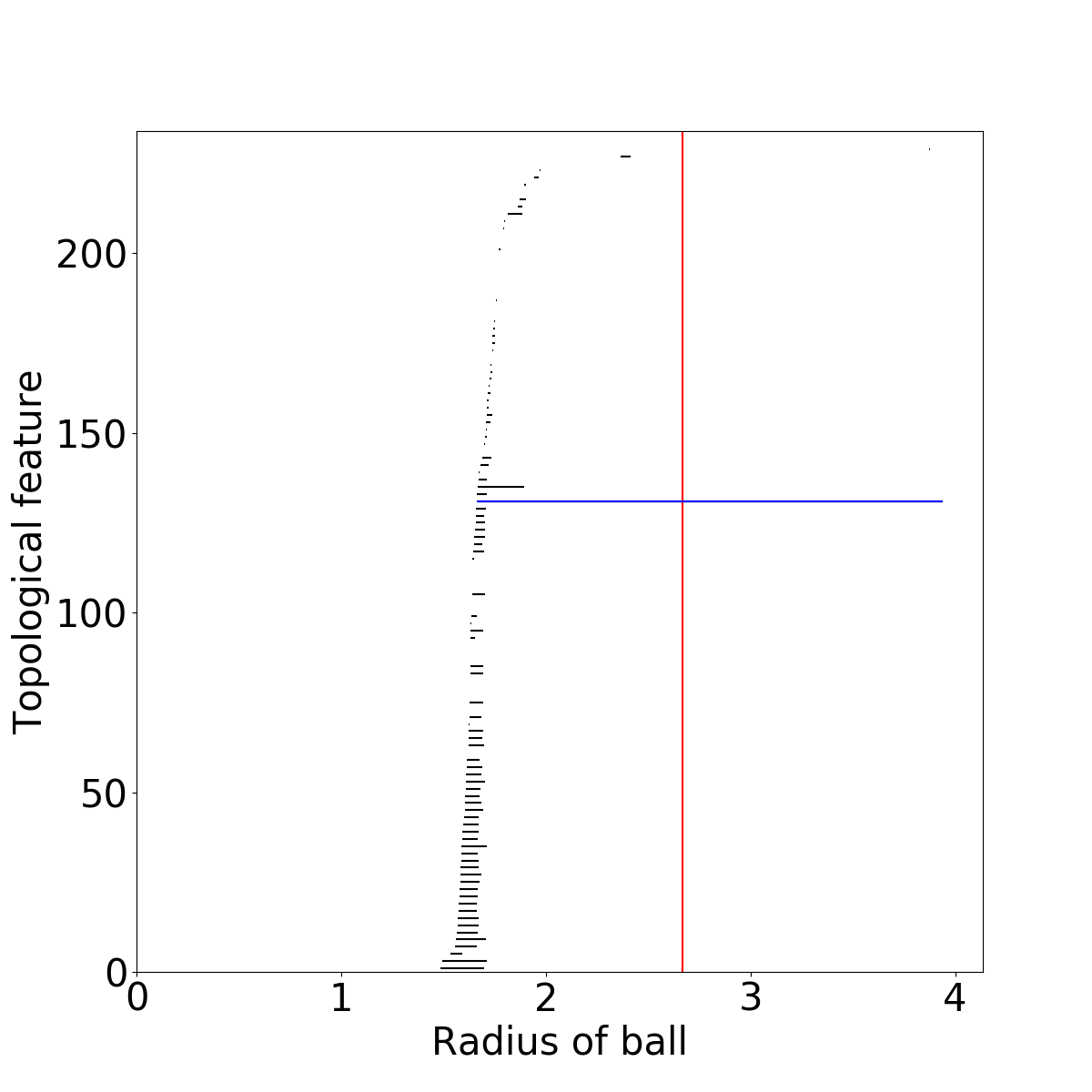}
	
	\caption{Left: the persistence diagram (PD) of the protein chain 1HFE-T. Each dot corresponds to a 1-dimensional feature that appear anywhere in the filtration, and the blue dot corresponds to the persistent ``hole'' seen in the alpha shapes of 1HFE-T. Center: the alpha shape of 1HFE-T for alpha $= 2.6$ \AA. Right: the barcode representation of the PD for 1HFE-T. The vertical red line is $x = 2.6$ \AA, the radius used for the alpha shape in the center plot.}
	\label{fig:barcode_example}
\end{figure}

The PH of a filtration is summarized as a persistence diagram (PD), which is a multiset of points $\{(b_i,d_i)\}_{i}$ that encode the filtration index where a feature (indexed by $i$) appears ($b_i$) and where that feature disappears ($d_i$). In the literature $b_i$ is called the birth time, and $d_i$ the death time. Points $(b_i, d_i)$ farther from the diagonal have a larger persistence, which is the difference between the birth and death time, and are interpreted as intrinsic topological features. Since a PD consists of points in $\R^2$, we can visualize the topological features of a point cloud by plotting its PD, as shown in the left of Figure \ref{fig:barcode_example}; the plotted points are features, whose $x$-coordinates are usually the birth time and whose $y$-coordinates are usually the death time. An alternative visualization, shown in the right plot of Figure \ref{fig:barcode_example}, uses a horizontal line for each feature. The initial $x$-coordinate is its birth time $b_i$ and terminal $x$-coordinate its death time $d_i$; this kind of representation is referred to as a barcode. The $y$-axis corresponds to persistent feature index $i$, and the user has freedom to choose how the intervals are sorted. Popular choices are to sort the intervals by their birth time, or by their death time. The right plot in Figure \ref{fig:barcode_example} shows the barcode representation for the same PD displayed in the left plot, with the features sorted by birth time.

As an example of PDs and their visualization, we revisit the alpha filtration built on 1HFE-T, shown in Figure \ref{fig:alpha_example}, and consider its $1$-dimensional persistent features $H_1$ in Figure \ref{fig:barcode_example}. In the center the alpha complex for alpha $=2.6$ {\AA} is displayed; note the large ``hole'' visible in the center of the protein. In the left plot the PD is plotted, and there is one feature plotted far above the diagonal colored blue. This blue dot corresponds to the large hole in the center of 1HFE-T. In the right plot the barcode representation is displayed. The long blue, horizontal bar corresponds to the blue dot in the PD plot, which corresponds to the large intrinsic hole in 1HFE-T. The red, vertical bar in the barcode plot indicates the filtration value for the alpha complex shown in the center; the red bar is a line segment showing $x=2.6$ \AA. The vertical line only intersects the horizontal blue line, and none of the black lines, meaning that the alpha shape only has one 1-dimensional feature present.

\subsection{Wasserstein distances}

The Wasserstein distance is a metric on the space of PDs with the important property of continuity with respect to small perturbations of the input point cloud \cite{edelsbrunner.2010.AMS}. Features from two different barcodes are paired, say by drawing line segments between them, and the sum of lengths of these line segments is computed. Since two PDs will generically have a different number of features, this pairing process needs to be modified to allow features to be paired with zero-persistence (on the diagonal) features. 

\begin{figure}[ht]
	\centering
	\includegraphics[width = 0.49\textwidth]{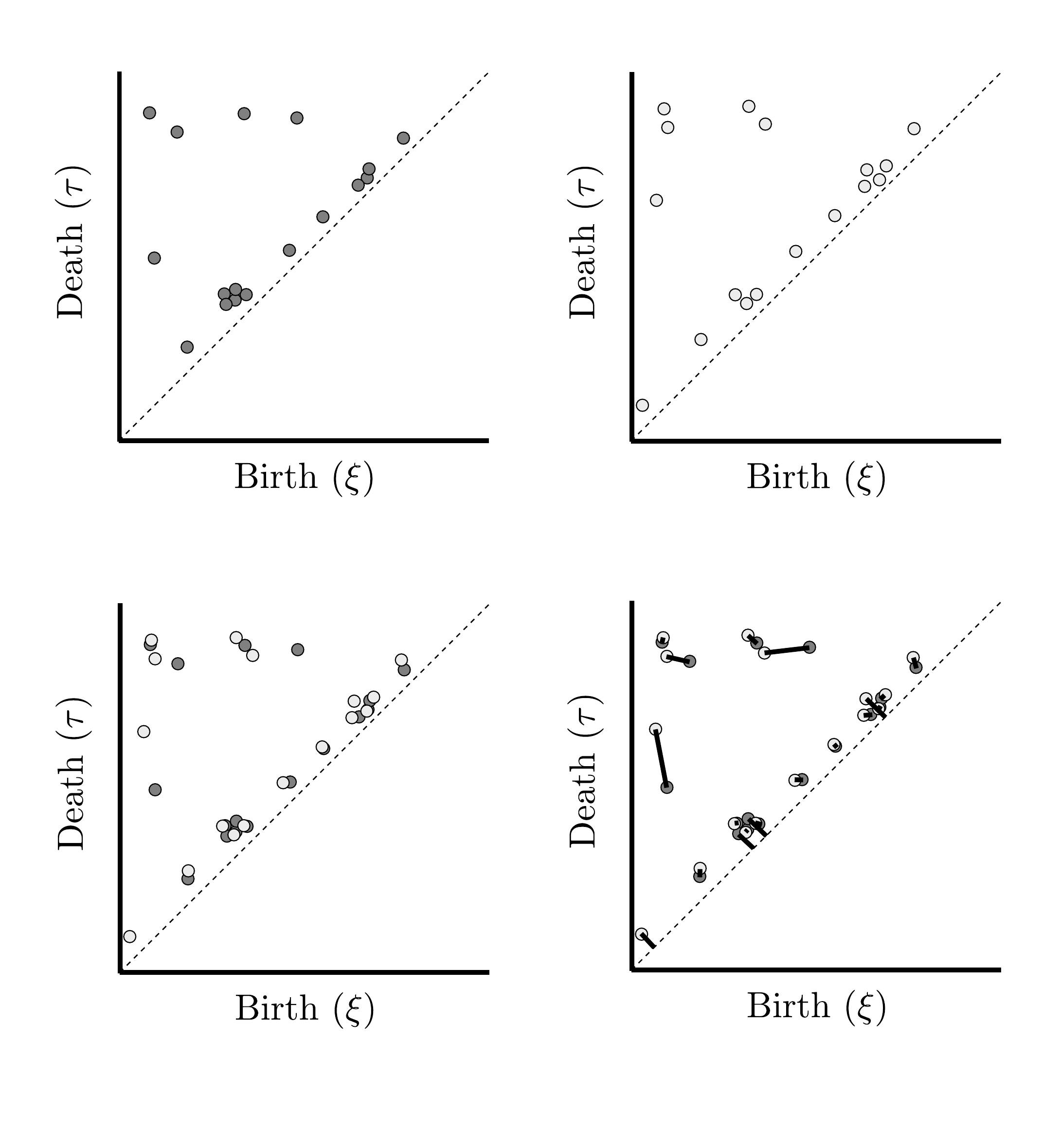}
	\caption{An illustration of the Wasserstein distance between two persistence diagrams. Top row: two persistence diagrams. Bottom left: the persistence diagrams overlaid. Bottom right: the optimal matching between the diagrams indicated by solid black bars.}
	\label{fig:wass_dist_example}
\end{figure}

For two PDs $D_1 = \{(b_1^i, d_1^i)\}_i$ and $D_2 = \{(b_2^j,d_2^j)\}_j$, the $q$-Wasserstein distance is defined as the cost of the optimal matching between the two point sets: $$W_q(D_1, D_2) := \inf_{\gamma\colon D_1^* \to D_2^*} \left( \sum_{x \in D_1^*} \| x - \gamma(x) \|_{\infty}^q \right)^{1/q},$$ where $D_i^* := D_i \cup \{(x,x) \colon x\in [0,\infty)\}$ and the infimum ranges over all bijective maps $\gamma\colon D_1^* \to D_2^*$ \cite{edelsbrunner.2010.AMS,panaretos.2020.springer}. Points along the diagonal are included in the matching process since PDs often contain a different number of points. Since diagonal points have the same birth and death time, they do not correspond to any topological features appearing in the filtration. Figure \ref{fig:wass_dist_example} illustrates an optimal matching between two persistence diagrams: the top-left and top-right plots contain PDs, and the bottom-left plot has the two PDs overlaid. In the bottom-right plot, the optimal matching is shown as black lines representing minimal length matchings of features to features, or features to the diagonal. The (squared) 2-Wasserstein distance between these two PDs is the sum of the squared $L^\infty$ norms (lengths) of the black line segments.

%
%

Note that $W_q$ is a metric on the space of PDs for $1\leq q \leq \infty$. In our study we used the 2-Wasserstein distance $W_2$ to compute distances between PDs, since we are interested in comparing PDs based on every feature, in contrast to the $\infty$-Wasserstein distance which only accounts for the matched pair of PD points that are furthest apart. We used Dionysus \cite{morozov} for our Wasserstein distance computations, which relies on novel geometric approaches to efficiently compute optimal matchings \cite{kerber.2017.ACM}.


\subsection{GIT vectors}

One approach to quantifying protein structure is through representations called Gaussian Integral Tuned (GIT) vectors. This approach interprets the structure of a protein chain as a curve in 3-D Euclidean space that is characterized using a vector in $\R^{30}$ consisting of geometric invariants \cite{roegen.2005.JPCM,roegen.2003.PNAS} originating in knot theory. The inspiration for this technique is through a quantity called the writhe $\text{Wr}$ of a closed curve $\gamma\colon [0,1]\to \mathbb{R}^3$, defined as $$\text{Wr}(\gamma) = \frac{1}{2\pi} \int \int_{0<t_1 < t_2<L(\gamma)} \omega(t_1, t_2) dt_1 dt_2,$$ where $L(\gamma)$ is the arc length of $\gamma$ and $$\omega(t_1, t_2) := \text{Det} (\gamma'(t_1), \gamma(t_1) - \gamma(t_2), \gamma'(t_2)) / |\gamma(t_1) - \gamma(t_2)|^3.$$ The writhe can be thought of as the average number of discrete, signed crossings $\omega(t_1, t_2)$ seen in planar projections of the curve. For a piecewise linear curve $\gamma$, consisting of $N$ linear segments, the signed writhe is defined as $$I_{(1,2)}(\gamma) = \text{Wr}(\gamma) = \sum_{0< i_1 < i_2 < N } W(i_1, i_2),$$ with $$W(i_1, i_2) = \frac{1}{2\pi} \int_{i_1 = t_1}^{i_1+1} \int_{i_2=t_2}^{i_2 + 1} \omega(t_1, t_2)dt_1 dt_2.$$ The quantity $W(i_1, i_2)$ can be interpreted as the probability of seeing the $i_1$th and $i_2$th line segments crossing, when viewed from an arbitrary direction. The unsigned writhe is $\text{Wr}_{|1,2|}(\gamma) := \sum_{0<i_1<i_2<N}|W(i_1, i_2)|,$ and a family of similar descriptors is constructed, including $$I_{|1,3|(2,4)}(\gamma) := \sum_{0<i_1 < i_2 < i_3 < i_4 < N} |W(i_1,i_3)| W(i_2, i_4) $$ and $$ I_{(1,5)(2,4)(3,6)}(\gamma) := \sum_{0<i_1 < i_2 < i_3 < i_4 < i_5 < i_6 < N} W(i_1,i_5) W(i_2, i_4) W(i_3, i_6).$$ 

Given a protein, we consider the $C_\alpha$ atoms as forming a piecewise linear curve in $\R^3$ and define the corresponding GIT vector as the vector containing the measures: the number $N$ of $C_\alpha$ atoms; $I_{|1,2|}$ and $I_{(1,2)};$ the three quantities $I_{(i_1,i_2)(i_3,i_4)}$ where $(i_1, i_2, i_3, i_4)$ is one of three permutations of $(1,2,3,4)$; all nine possible signed/unsigned versions of $I_{(i_1,i_2)(i_3,i_4)}$; and all fifteen signed quantities $I_{(i_1,i_2)(i_3,i_4)(i_5,i_6)}$ where $(i_1, ..., i_6)$ is a permutation of $(1, ..., 6)$. One can also consider the quantities of the form $I_{(i_1,i_2)...(i_7,i_8)}$, and the unsigned combinations. The GIT approach uses 30 such quantities, resulting in 30-dimensional vectors as descriptors of protein structure. Each geometric invariant is divided by $N^a$ where $a$ is chosen to make the measures independent of N and normalized to correlate with the root mean squared distance for small pertubations of the structures.

GIT vectors are fast to compute, and have proven effective at classifying proteins based on their fold simply by clustering their GIT vectors. The appeal of this approach is that a computationally expensive pairwise comparison of protein structures can be avoided. Recently, GIT vectors were also used to identify unusual conformations in protein structures that could indicate the presence of errors or biologically interesting structural features \cite{gronbaek.2020.PJ}.

\subsection{Dimensionality reduction}

The object-oriented data analysis approach we take in this paper converts data objects to barcodes, which we compare using the 2-Wasserstein distance. The collection of all barcodes is an infinite-dimensional space, and we only have the distances between barcodes at our disposal, so we employ a number of dimensionality reduction techniques throughout this study to embed the data objects into a low-dimensional space and visualize the resulting structure.

Two related techniques we use are Principal Component Analysis (PCA) and Multidimensional Scaling (MDS). PCA recovers directions of maximal variation from a collection of vectors, which are computed to minimize the vector norm from the data to each direction \cite{jolliffe.2002.springer}. MDS is a variation of PCA which is especially useful for non-Euclidean data objects. This method seeks coordinates for each data object whose Euclidean distances are close to given distances \cite{torgerson.1952.psych}; in our case, the distances provided are 2-Wasserstein distances between protein barcodes. When the given distances are Euclidean, MDS and PCA return essentially the same scores. Formally, given a collection of distances $\{d_{ij}\}$ between data objects $x_i, x_j$, MDS works to minimize the stress functional $\sum_{i\neq j} \left(d_{ij} - \|X_i - X_j\|_2 \right)^2,$ where $\|X_i - X_j\|_2$ is the vector (Euclidean) norm between vectors $X_i, X_j$; variants of this method exist which minimize different, but related, stress functionals. Since the stress is non-linear in the unknown coordinates, optimization techniques beyond SVD and other linear algebraic methods need to be implemented.

The third technique we employ is the t-distributed stochastic neighbor embedding (t-SNE) \cite{vanderMaaten.2008.JMLR}, which finds embedding coordinates using a probabilistic approach. In particular, a probability measure $\pi_1$ is defined on pairs of data objects via precomputed distances, and Euclidean coordinates $X_1,...,X_n$ are computed by minimizing the Kullback-Leibler divergence between $\pi_1$ and a kernel density estimate constructed with $X_1, ..., X_n$ as centers; in practice the estimate is built using Gaussians. The visualisation t-SNE provides an effective display of local aspects such as clusters and community structures, versus PCA and MDS which more accurately reflect global positions.

\subsection{AJIVE}
Angle-based Joint and Individual Variation Explained (AJIVE) \cite{feng.2018.JMvA} is a statistical method for decomposing the joint and individual variation in a multi-block data setting, where each data block contains a disparate set of features taken on a common sample set. AJIVE works in three steps to implement an angle-based solution that is efficient and guarantees identifiability. Similar to PCA, AJIVE uses SVD to decompose data matrices and relates singular values to principal angles to determine segmentation of joint and individual components. First, a low-rank approximation of each data block is found by separately applying SVD. Next, the joint components are determined by performing SVD of the concatenated bases of the row spaces from the first step and thresholding based on principal angles and bounds from perturbation theory. Finally, joint and individual approximation matrices are given by projecting the joint row space and its orthogonal components onto the original data blocks. AJIVE does not require normalization and is not sensitive to potential issues arising from heterogeneity in scale and dimension between data blocks. 

As detailed in the paper \cite{feng.2018.JMvA}, given data matrices $X_k$, $k \in \{1,...,K\}$, which have the same number of samples (columns) but possibly different numbers of features (rows), we define the following decomposition,
\begin{equation}
X_k = A_k + E_k = J_k + I_k + E_k.
\end{equation} 
In this model, the $A_k$ are deterministic low-dimensional signal matrices, $J_k$ are joint matrices, $I_k$ are individual matrices, and $E_k$ are random error matrices. The matrices $J_k$ representing the shared variation need not be identical but must have a shared row space, while the matrices $I_k$ representing individual variation are constrained to have row spaces orthogonal to the shared row space of the $J_k$.



Segmentation of the joint spaces is based on studying the relationship between these score subspaces using theoretical bounds from Principal Angle Analysis and assessing noise effects using perturbation theory. 
An important point is that by focusing on segmentation of the row spaces, the methodology captures variation patterns across the data objects which give interpretable meaning of the components. Moreover, AJIVE is distinct from other methods that simultaneously analyze multi-block data in that it identifies \textit{individual} variation patterns. Segmentation of the individual components is a particularly important property for the analysis done in Section 3.

\section{Results}

\subsection{Data processing}

For our analysis, we use the top8000 protein data set \cite{chen.2010.AC}, consisting of 7957 high quality protein chains from the RCSB Protein Data Bank (PDB) \cite{burley.2020.NAR}.  Proteins were chosen based on the resolution of their PDB files, and then a single chain (a linear polymer of amino acids) is selected; more details can be found on the top8000 repository in \cite{top8000}. Some of the top8000 proteins were chopped chains, meaning they were a smaller portion of a longer contiguous chain. Other proteins in the data set had chain breaks, meaning small portions of the polypeptides were missing due to experimental error or other reasons.

Since we were interested in structural features highlighted by our topological approach, we utilize the CATH database \cite{CATH.2019}. CATH labels characterize a protein’s fold using a four-fold hierarchical descriptor, consisting of Class, Architecture, Topology and Homology. For our study we restrict our attention to 2949 protein chains that (1) did not have any chopping (these had suffix 00 in the top8000 directory), (2) had no chain breaks, (3) had CATH labels available, and (4) allowed the calculation of GIT vectors. These choices were made both for convenience and quality of data. We focus on individual chains, rather than proteins with possibly multiple chains, because fold classification is done at the chain level. In what follows, we will refer to a protein and its chain (in the top8000 data set) interchangeably. Also we refer to our subset of the top8000 data set as the CATH00 data set.

\subsection{Visualization of protein space}

We compute the protein $H_1$ PDs using the persistent homology of alpha-filtrations built on the chains' $C_\alpha$ atoms. Using $C_\alpha$ atoms is a standard approach in protein fold classification \cite{gronbaek.2020.PJ}: each amino acid in a protein has a $C_\alpha$ atom, the different groups attached to the $C_\alpha$ atom determine different amino acids, and $C_\alpha$ atoms are generally equidistributed along the protein backbone. For the statistical analysis, we compute all pairwise $2$-Wasserstein distances between PDs and then apply MDS and t-SNE to visualize the geometry and topology of this space of proteins.

\begin{figure}[ht]
	\centering

	\includegraphics[width = 0.5\textwidth]{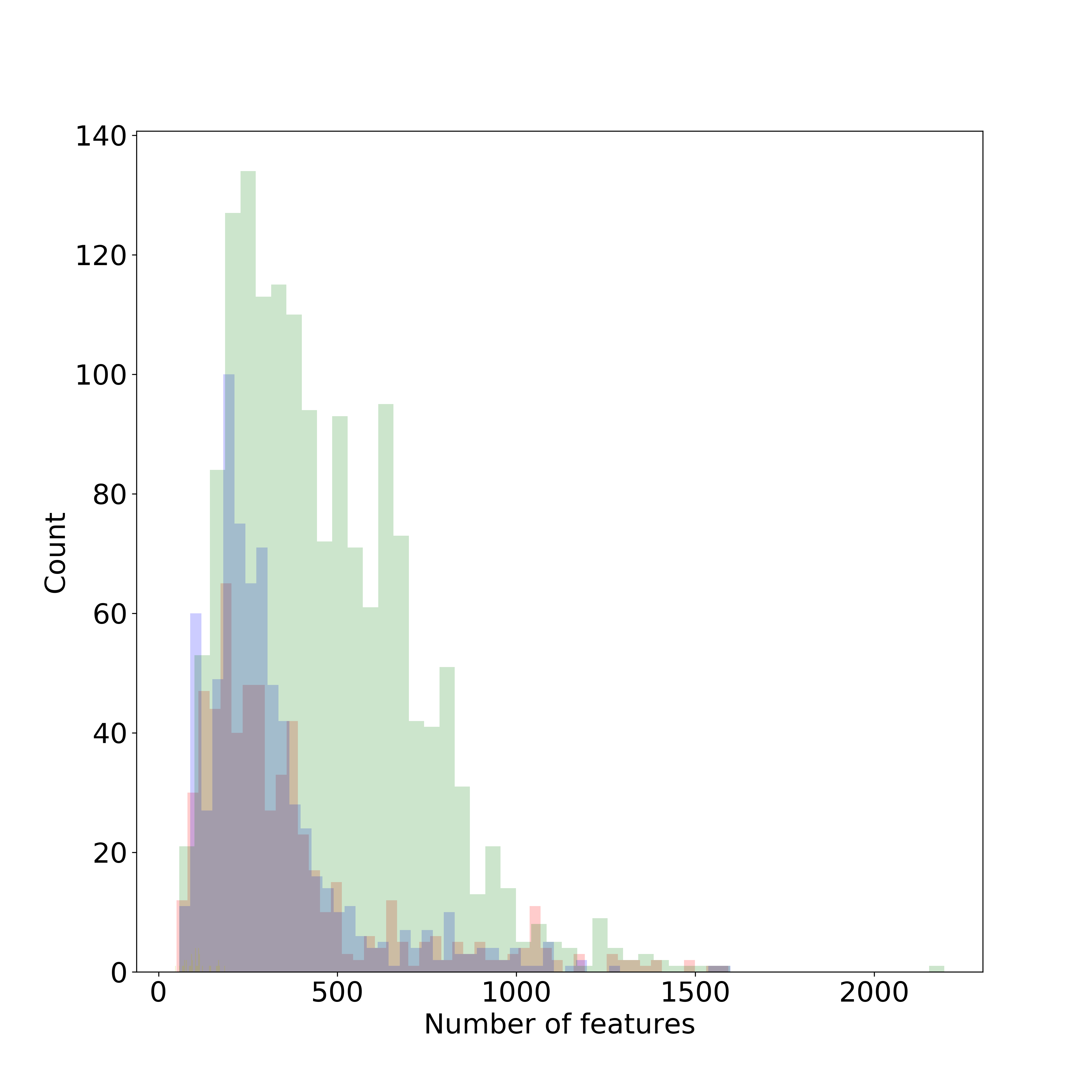}
	
	\caption{ Histograms of PD sizes for proteins overlaid on top of each other. For each of the plots, the x-axis is number of features in a PD, and the y-axis is the corresponding count of proteins; 50 bins were used in each. Note that Class 4 contains relatively few proteins, and so is not clearly visible in the overlaid histogram.}
	\label{fig:CATH00_BC_histograms}
\end{figure}

Figure \ref{fig:CATH00_BC_histograms} shows histograms of the number of features in each protein's PD, with all classes overlaid. Classes 1 and 2 (red and blue, respectively) qualitatively have the same distribution, whereas Class 3 (green) generally consists of proteins with more features in their PDs. Finally, Class 4 (orange) generally consists of the smallest proteins and has relatively few data objects, and hence is not clearly visible when overlaid with the other classes. Table \ref{table:protein_counts} shows the number of proteins in each Class.

\begin{table}[h!]
\centering
\captionsetup{justification=centering}
 \begin{tabular}{||c c||}
 \hline
 CATH Class label & Number of Proteins \\ [0.5ex] 
 \hline\hline
 1 & 605 \\ 
 \hline
 2 & 730 \\
 \hline
 3 & 1577 \\
 \hline
 4 & 37 \\ [1ex]
 \hline
\end{tabular}
\caption{Number of proteins in the CATH00 data set per Class label.}
\label{table:protein_counts}
\end{table}

\begin{figure}[ht]
	\centering
	
	\includegraphics[width = 0.5\textwidth]{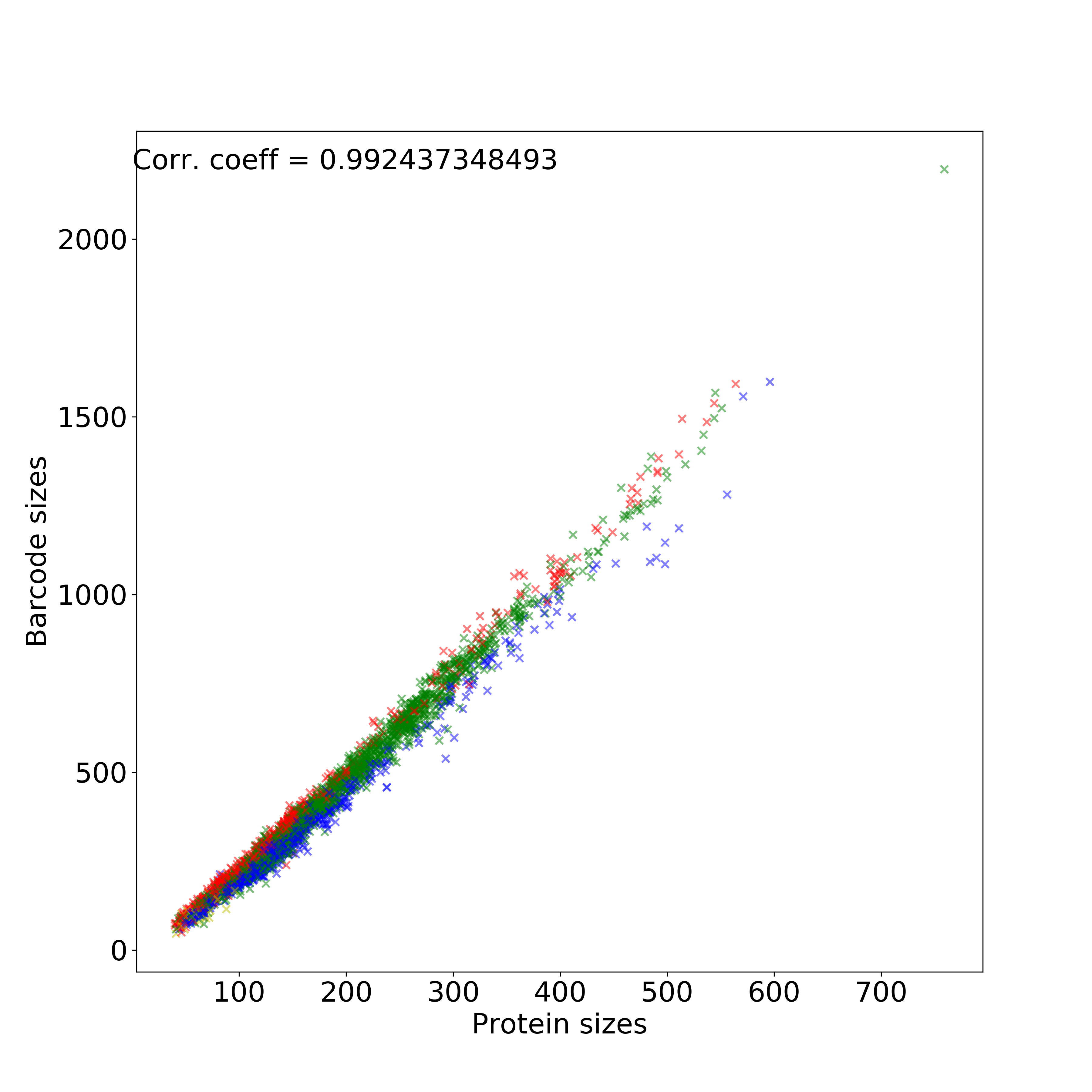}
	
	\caption{ Scatter plot of protein size against PD size. Each marker corresponds to a protein, the x-axis indicates protein size (number of $C_\alpha$ atoms in the backbone), and the y-axis indicates PD size. Protein size and PD size have a correlation coefficient of $0.992$, indicating these are essentially the same.}
	\label{fig:CATH00_BC_size_correlation}
\end{figure}

In these figures and below, we adopted the following color scheme when indicating CATH Class: 
\begin{itemize}
    \item Red - Class 1, primarily $\alpha$-helices
    \item  Blue - Class 2, primarily $\beta$-sheets
    \item Green - Class 3, primarily a mix of $\alpha$-helices and $\beta$-sheets
    \item Orange - Class 4, no significant content of $\alpha$-helices or $\beta$-sheets
\end{itemize}
We also use proteins' sizes, measured by the number of $C_\alpha$-atoms in the backbone, as a color scheme, with blue corresponding to smaller proteins and red corresponding to larger proteins. For our data, proteins' sizes correlates strongly with their PD sizes as seen in Figure \ref{fig:CATH00_BC_size_correlation}, where the correlation coefficient is $0.992.$

\begin{figure}[ht]
	\centering
	\includegraphics[width = 0.7\textwidth]{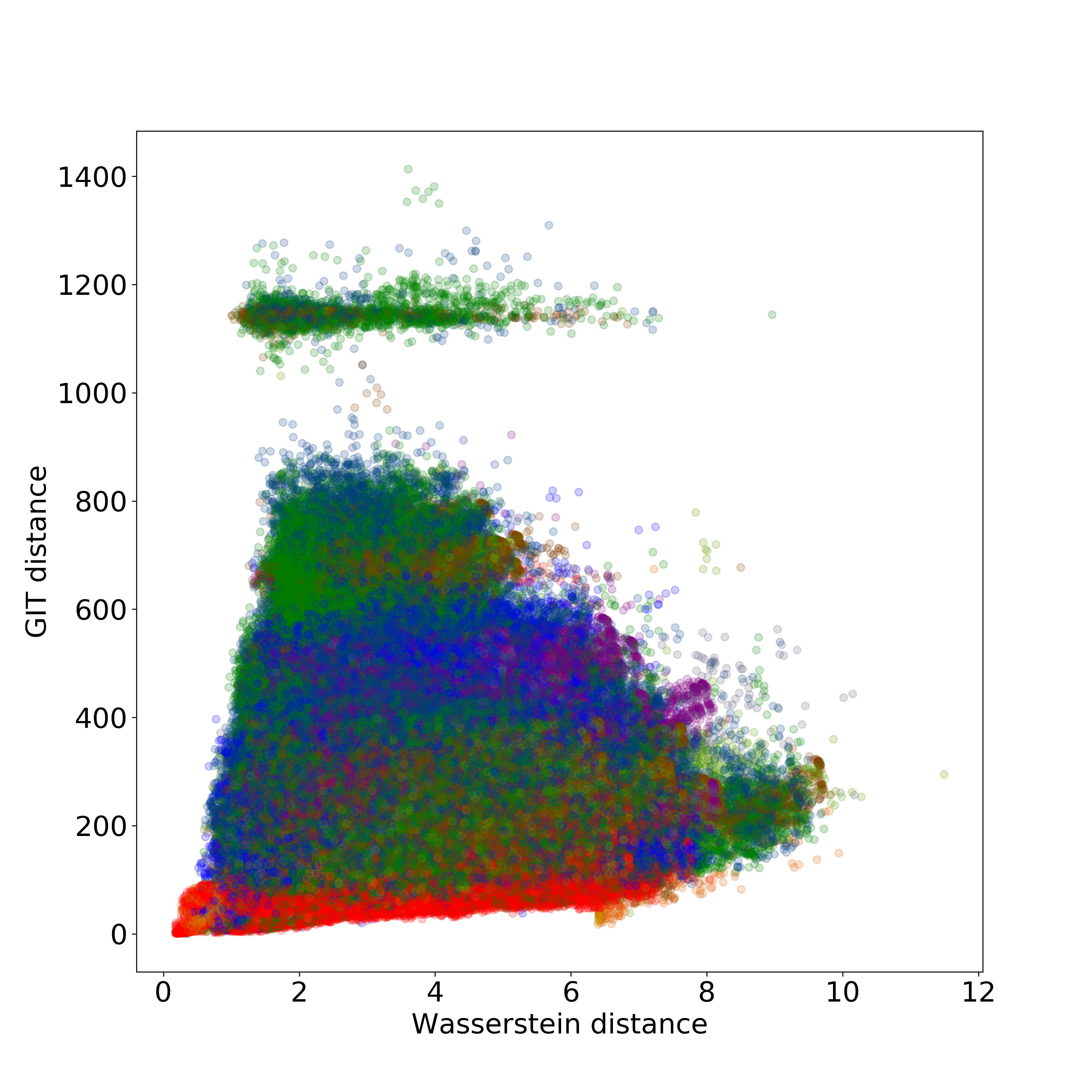}
	
	\caption{All pairwise distances between proteins, with x-coordinate the Wasserstein distance and y-coordinate the GIT distance. There is no apparent correlation between Wasserstein and GIT distances, indicating that these two metrics are driven by far different structural aspects.}
	\label{fig:CATH00_Wass_GIT_dists}
\end{figure}

A natural first question is whether the Wasserstein and GIT metrics quantify very different structural aspects of this space of proteins. This is answered in Figure \ref{fig:CATH00_Wass_GIT_dists} using a scatter plot of all pairwise distances: each marker corresponds to a pair of proteins $P_1, P_2$, with x-coordinate Wasserstein distance $d_{\text{Wass}}(P_1,P_2)$ between the corresponding 1-homology PDs, and y-coordinate the distance between corresponding GIT vectors $d_{\text{GIT}}(P_1,P_2)$. If the proteins belong to the same CATH-C class then the marker has the corresponding CATH-C color, and if the proteins belong to different CATH-C classes then the marker color is an average of RGB values for the corresponding CATH-C colors. If the Wasserstein and GIT distances were detecting the same kind of geometry in protein space, we would expect the two distances to correlate strongly. As seen in Figure \ref{fig:CATH00_Wass_GIT_dists}, the two distances are not at all correlated, indicating that Wasserstein and GIT distances do in fact detect different structural aspects of protein space. The strong cluster at the top of Figure \ref{fig:CATH00_Wass_GIT_dists} is a consequence of the large pairwise distances between the small set of outliers seen in the upper right of Figure \ref{fig:CATH00_GIT} with the rest of the data. That small cluster seems to consist of small proteins with little secondary structure, typically proteins with metals and/or disulphide bridges.

\begin{figure}[ht]
	\centering
	\includegraphics[width = 0.32\textwidth]{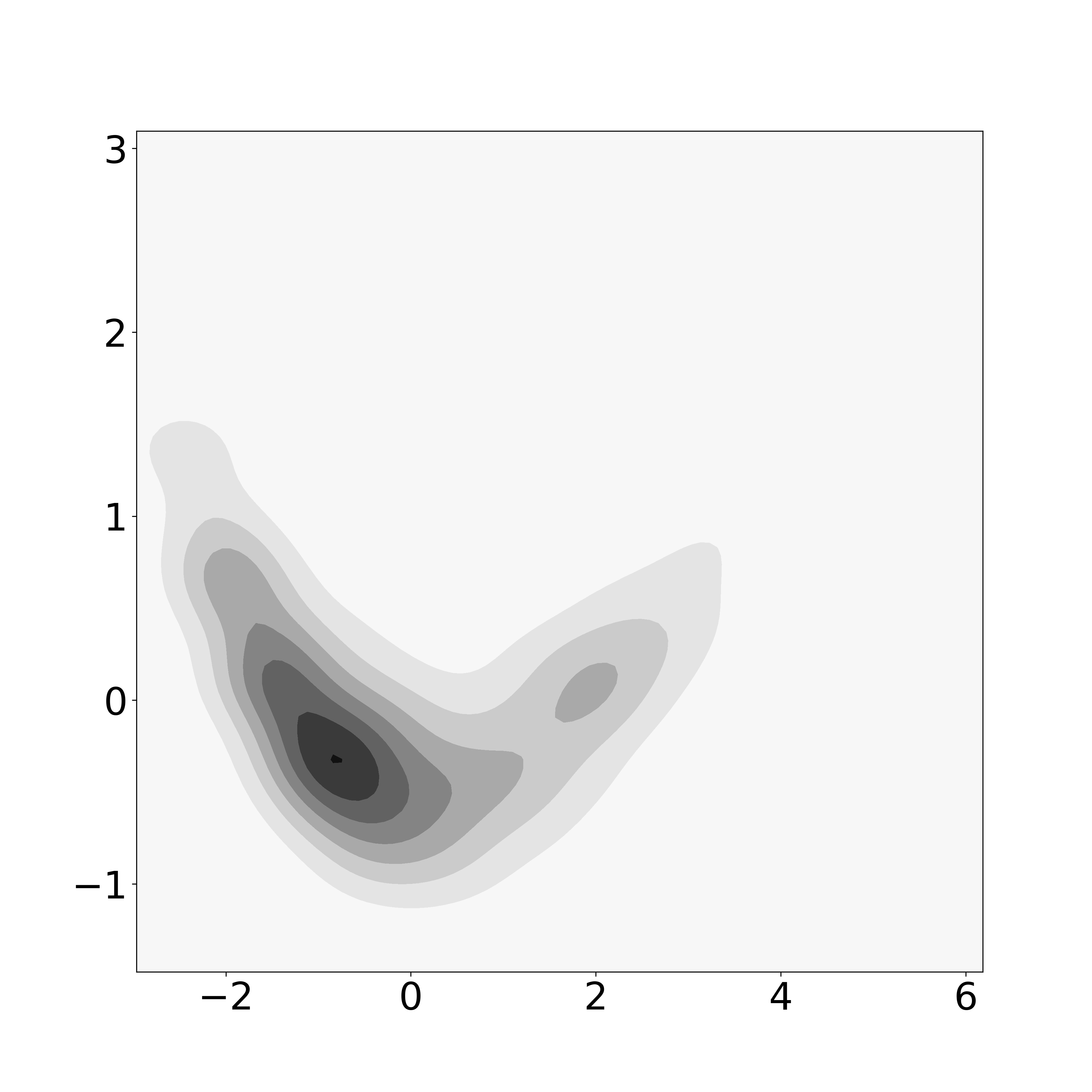}
	\includegraphics[width = 0.32\textwidth]{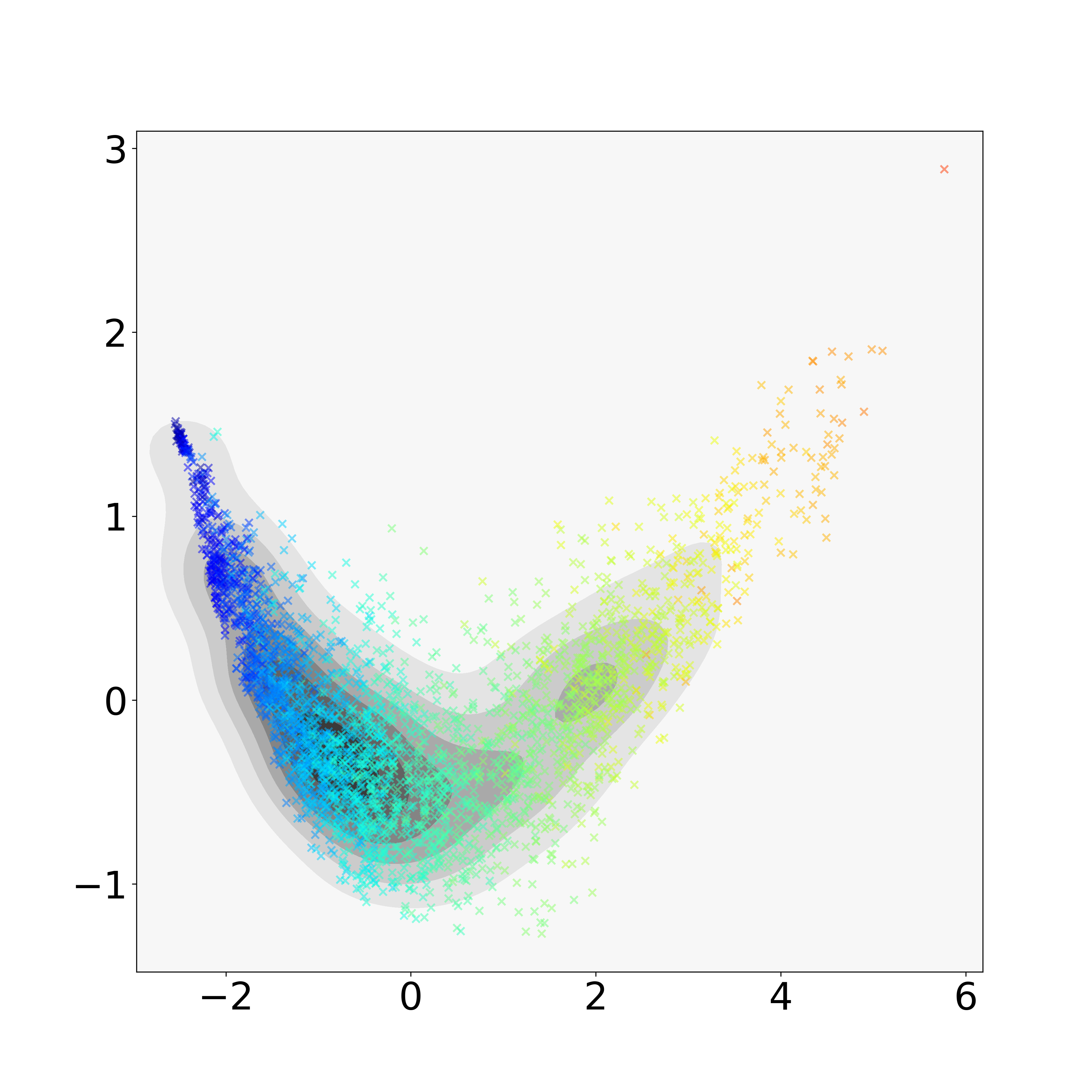}
	\includegraphics[width = 0.32\textwidth]{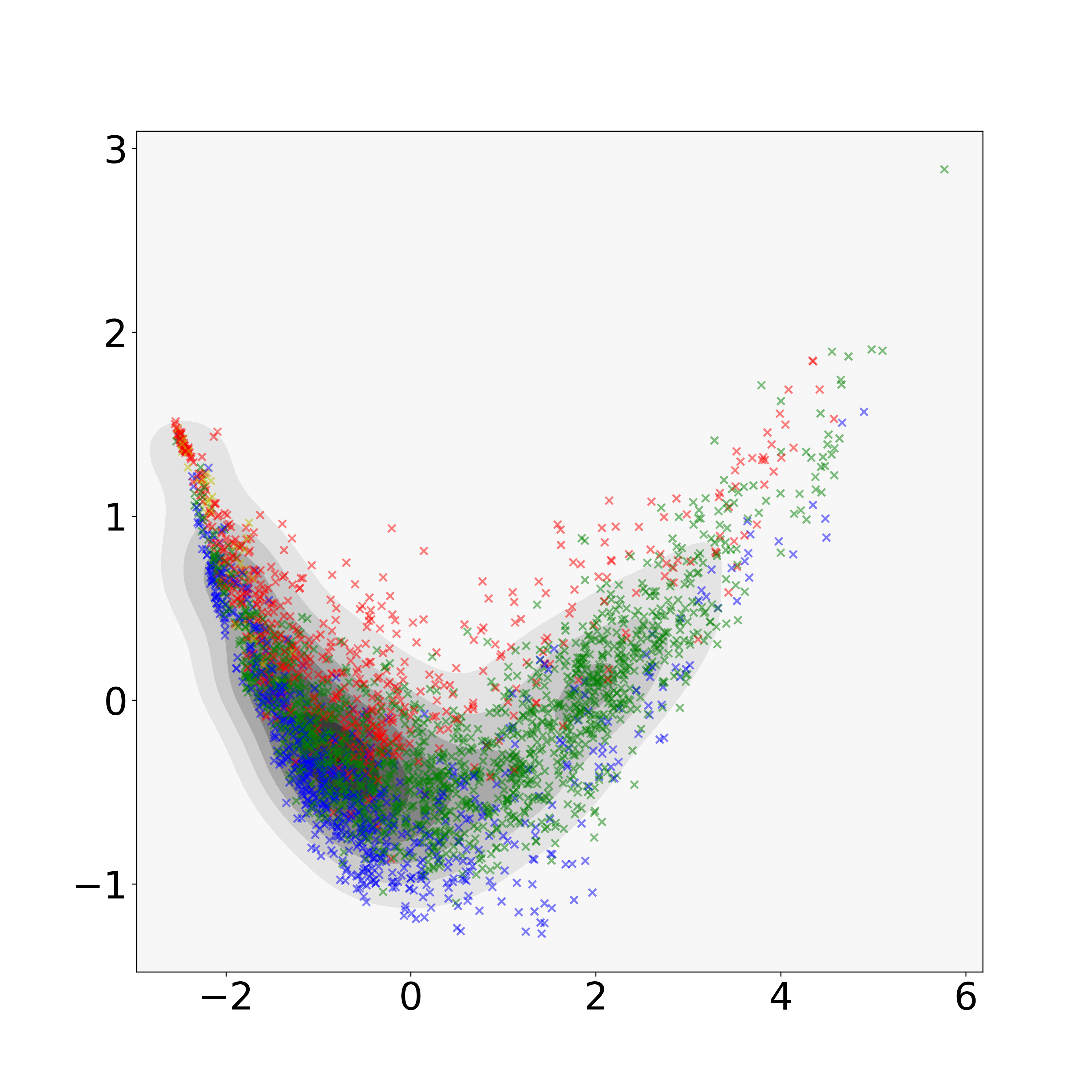}
	\caption{The first two PCs of the MDS coordinates using the $2$-Wasserstein metric on the full barcodes for the CATH00 data set. The x-axis corresponds to PC 1, and the y-axis PC 2, in each plot. On the left, the kernel density estimate for the MDS embedding; in the middle, proteins are colored by protein size; on the right, proteins are colored by CATH-C label. Here we have clear separation for both barcode size and CATH-C labels, and we see that protein size (CATH-C) corresponds to PC 1 (PC 2, respectively).}
	\label{fig:CATH00_full_wass_MDS_PC12}
\end{figure}

In Figure \ref{fig:CATH00_full_wass_MDS_PC12} we display the results of using MDS to embed the proteins into $\mathbb{R}^{25}$ and then plot the projections onto the first two principal directions. On the left is the kernel density estimate of the projected points, and in the middle and on the right are the PC 1 and PC 2 scatter plots using two different color schemes. The colors in the middle plot are based on the sizes of each proteins' PD (blue for fewer features in the PDs, red for more features), while the colors in the right-most plot correspond to Class labels as in Figures \ref{fig:CATH00_BC_histograms} and \ref{fig:CATH00_BC_size_correlation}. We see separation of the proteins based on protein size along PC 1, and separation by Class label along PC 2.

\begin{figure}[ht]
	\centering
	\includegraphics[width = 0.49\textwidth]{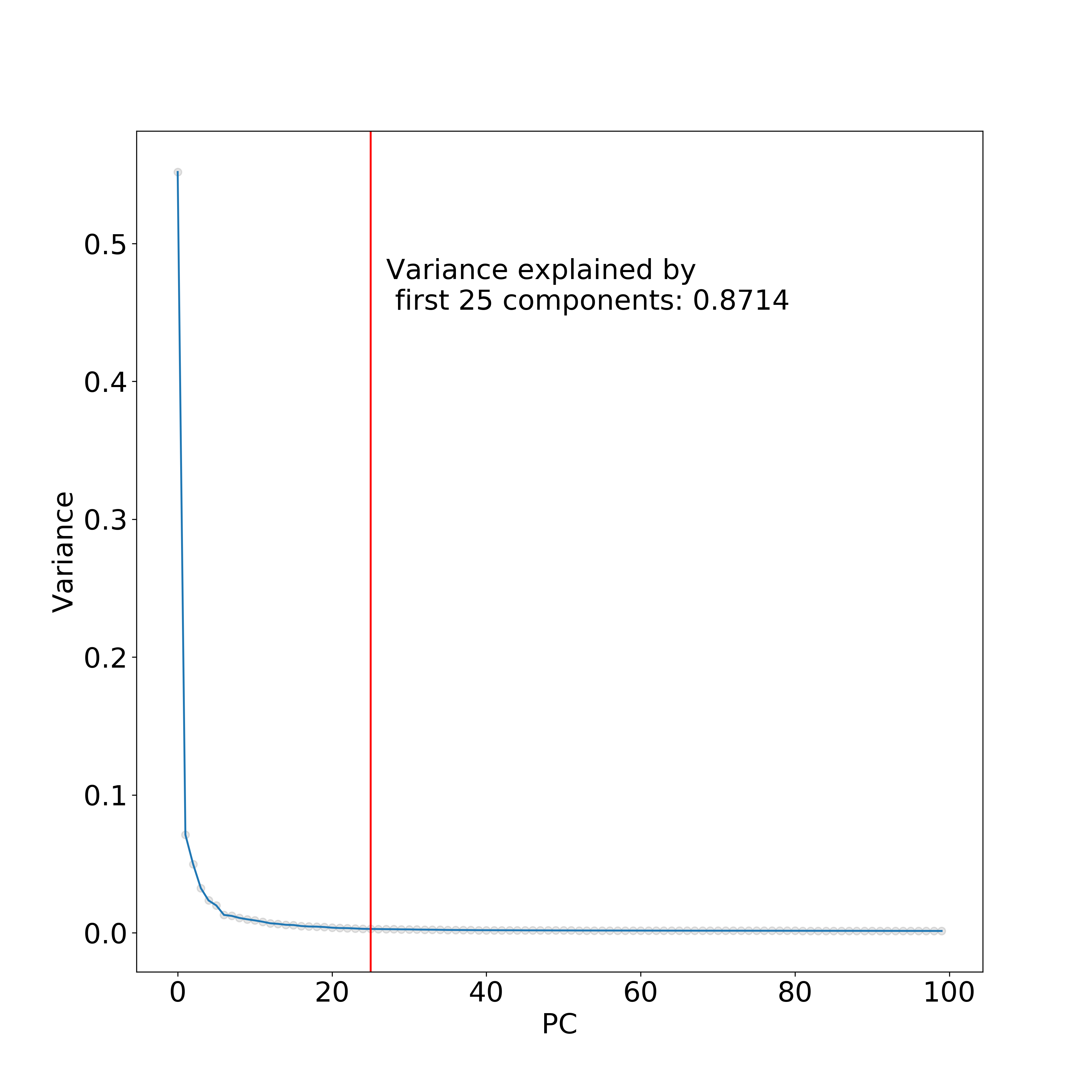}
	\includegraphics[width = 0.49\textwidth]{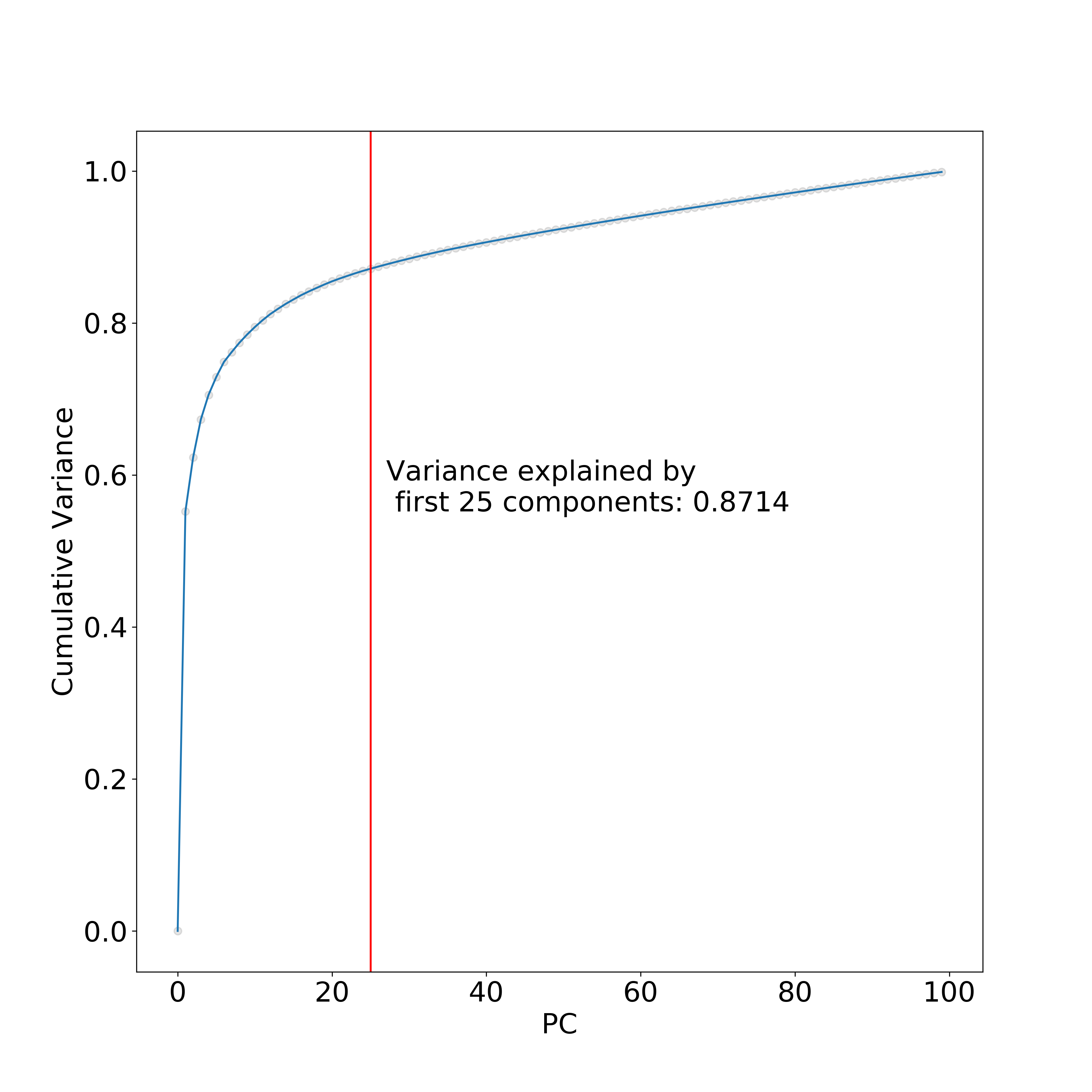}
	
	\caption{The plots of variance (left) and cumulative variance (right) captured in the $\mathbb{R}^{100}$ MDS embedding; the vertical red line indicates PC 25. The first 25 PCs capture 87.14\% of the total variation.}
	\label{fig:CATH00_MDS_quality}
\end{figure}

The quality of the MDS embedding into $\mathbb{R}^{25}$ is measured by further embedding the same Wasserstein distance matrix into $\mathbb{R}^{100}$ and computing the amount of variation captured in the first 25 PCs. As shown in Figure \ref{fig:CATH00_MDS_quality}, the first 25 PCs captured 87.14\% of the total variation. Hence, in the rest of this paper, we use 25 dimensional MDS representations since that is approximately the number of features provided by GIT. Moreover, for this data set the stress scores were essentially monotonically decreasing as a function of dimension. Finally, we visually compared the first 2 PC scatter plot for the 25-dimensional embedding with the corresponding scatter plot for the 100-dimensional embedding, and observed they were visually very similar.

\begin{figure}[ht]
	\centering
	\includegraphics[width = 0.32\textwidth]{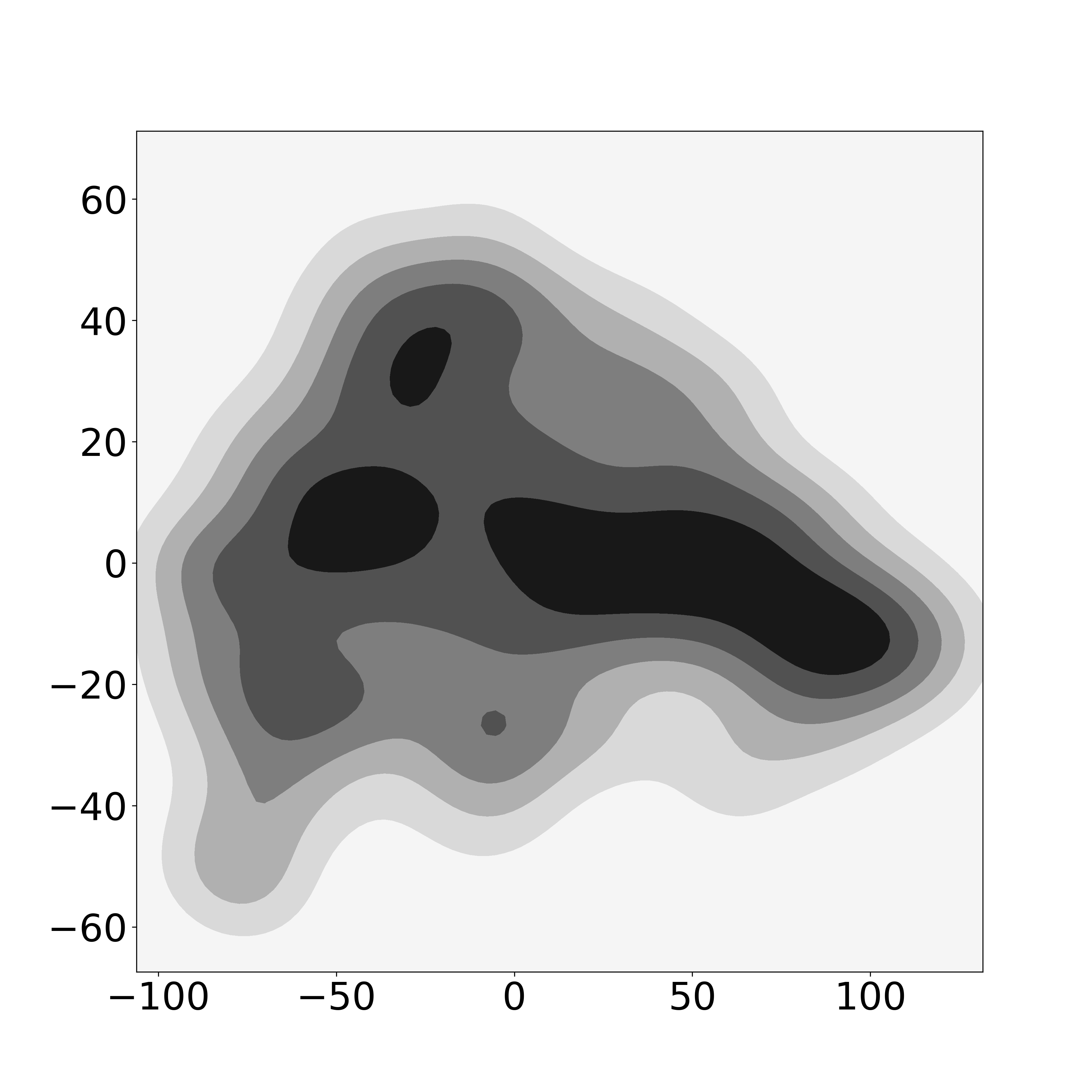}
	\includegraphics[width = 0.32\textwidth]{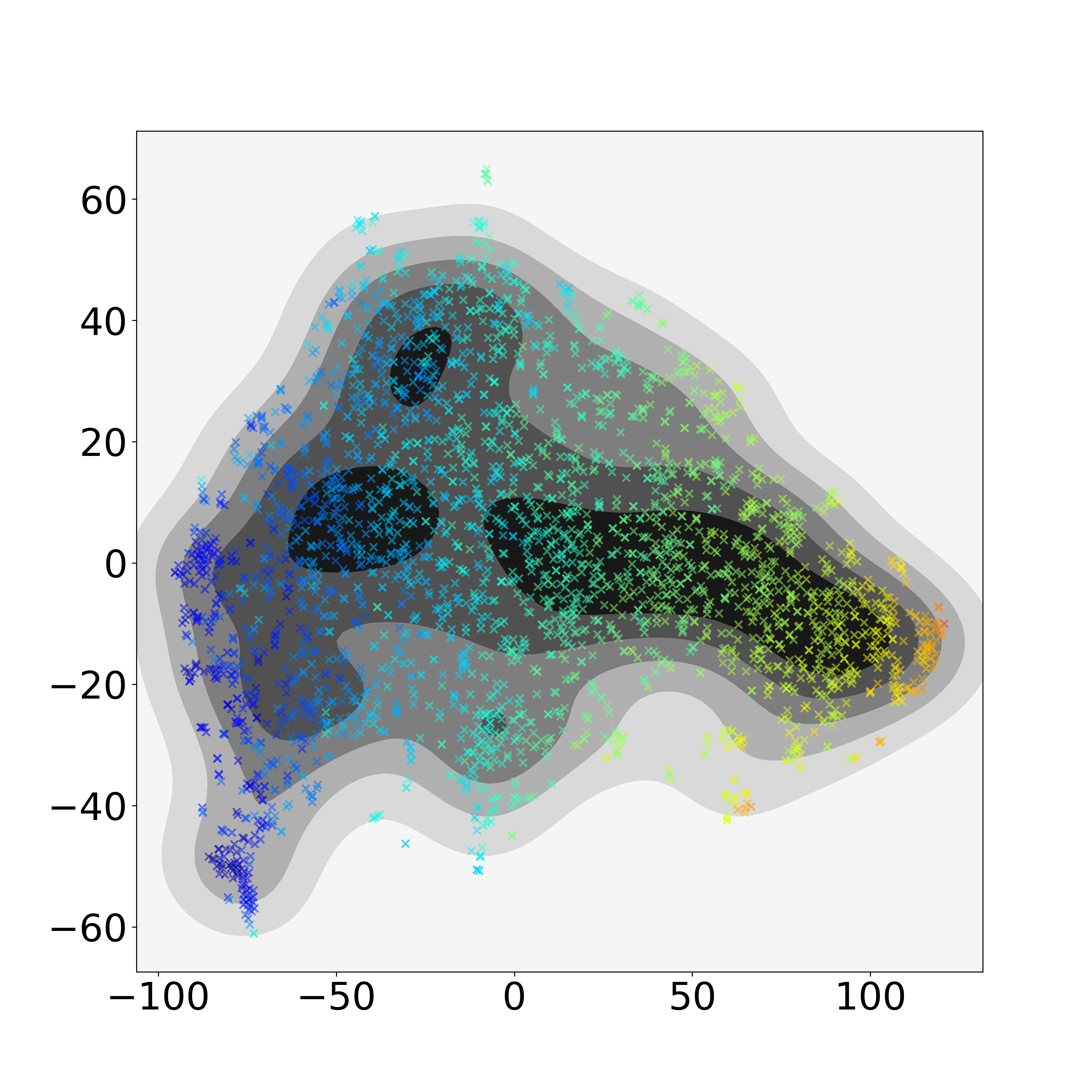}
	\includegraphics[width = 0.32\textwidth]{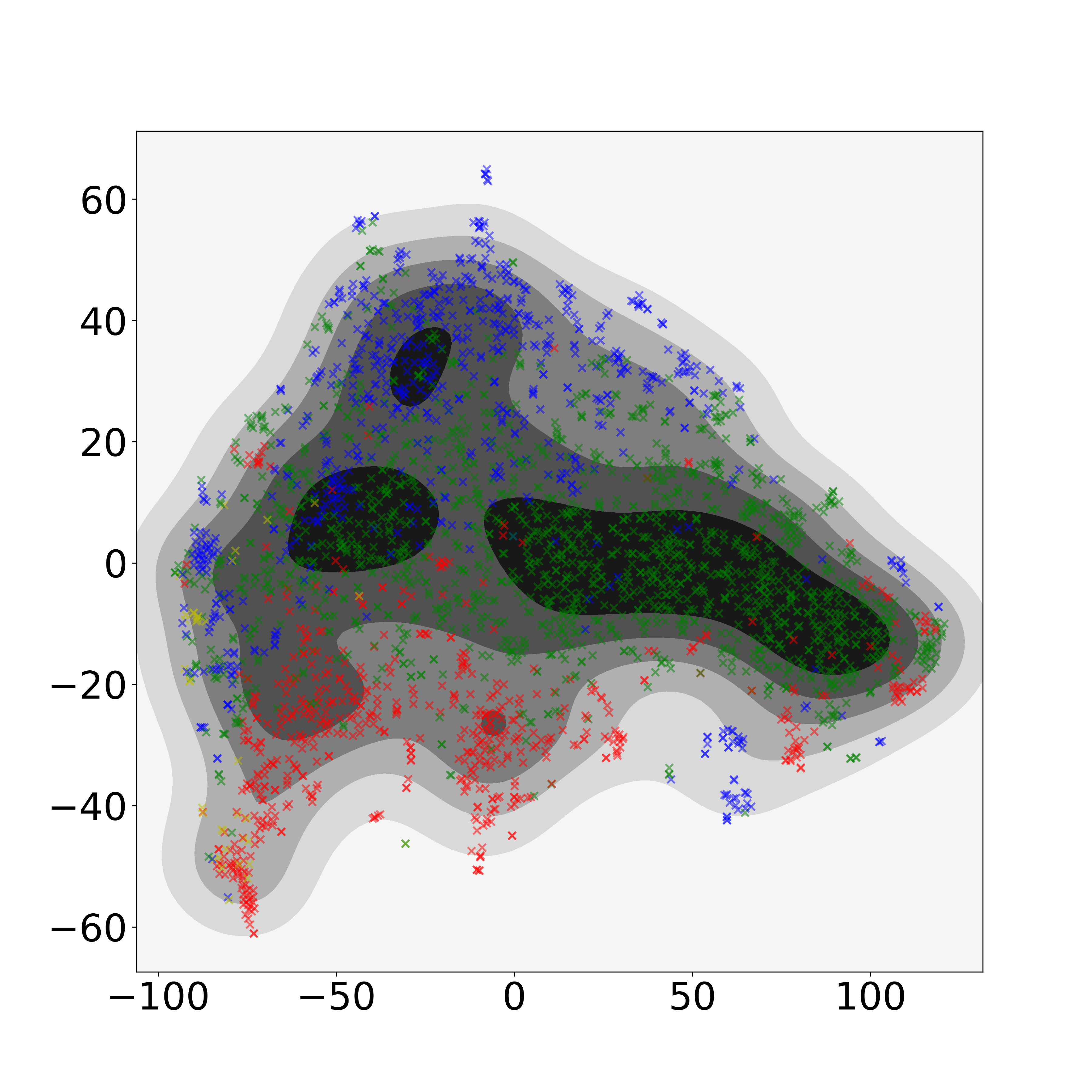}
	\caption{For the CATH00 data set, t-SNE coordinates using the $2$-Wasserstein metric on the full barcodes, perplexity parameter $30$. The x-axis corresponds to PC 1, and the y-axis PC 2, in each plot. On the left, the kernel density estimate for the t-SNE embedding; in the middle, proteins are colored by protein size; on the right, proteins are colored by CATH-C label. Barcode size (CATH-C) correspond to PC 1 (PC 2, respectively).}
	\label{fig:CATH00_full_wass_tsne}
\end{figure}

We also use t-SNE to visualize the relationships between these proteins, and the results are shown in Figure \ref{fig:CATH00_full_wass_tsne}. As in Figure \ref{fig:CATH00_full_wass_MDS_PC12}, we show the kernel density estimate (left) together with the 2D scatter plots of the t-SNE embedding with protein size colors (center) and Class label colors (right). Similar to the MDS embedding, the t-SNE embedding shows separation based on protein size and Class labels.

%
%


%
%

\subsection{Statistical Analysis of PDs and GIT vectors}

\begin{figure}[ht]
	\centering
	\includegraphics[width = 0.32\textwidth]{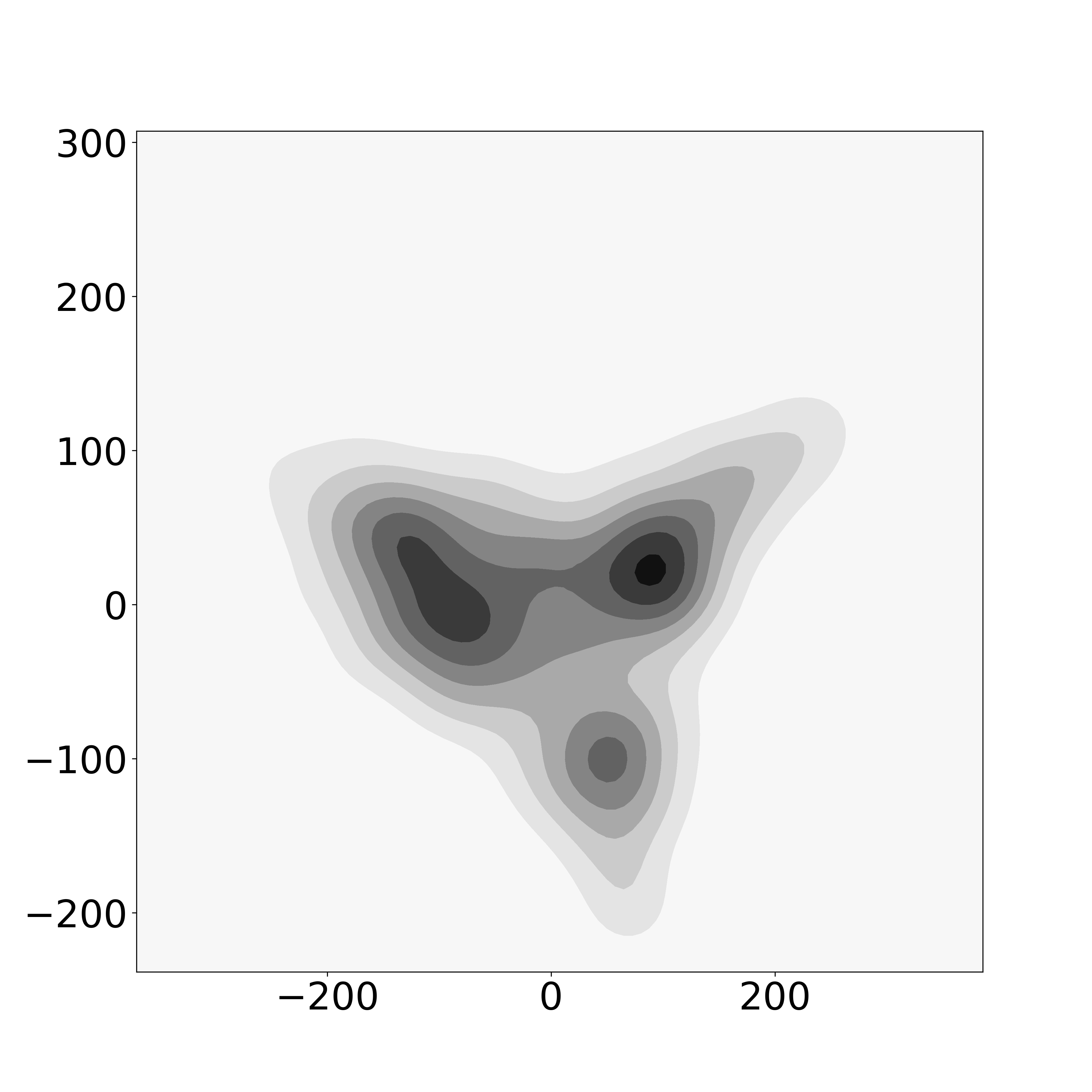}
	\includegraphics[width = 0.32\textwidth]{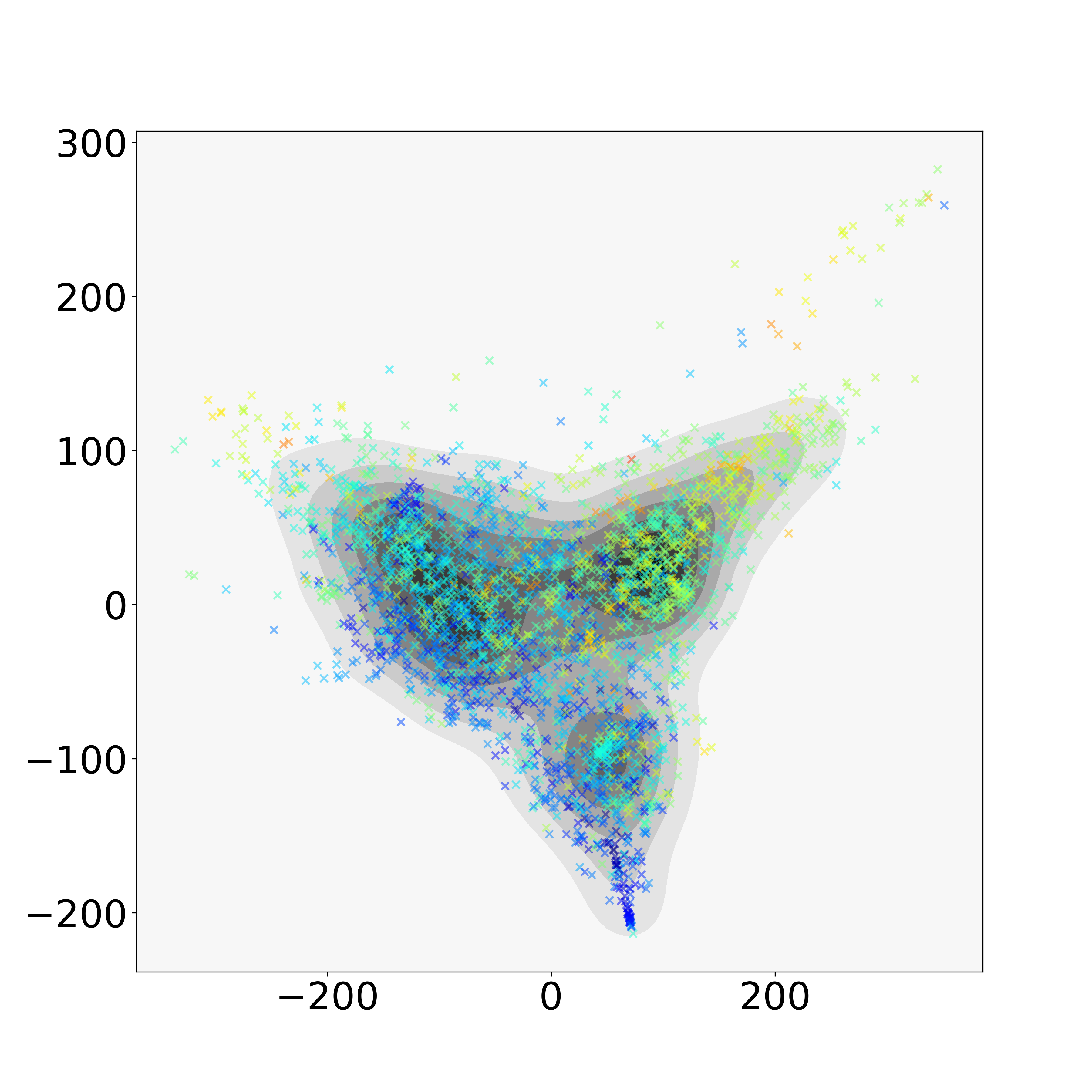}
	\includegraphics[width = 0.32\textwidth]{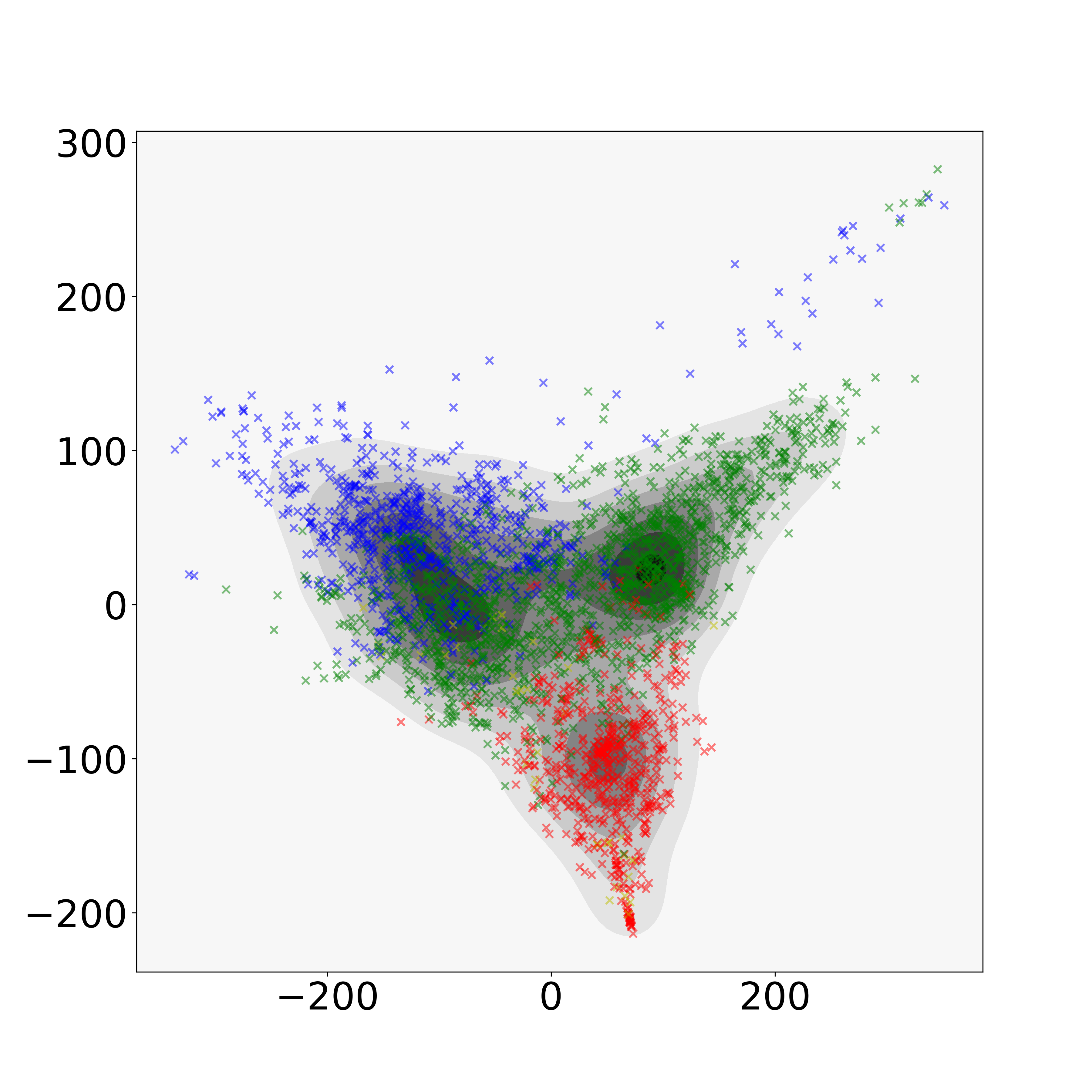}
	\caption{For the CATH00 data set, GIT vectors are computed and then projected onto the first two principal directions. The x-axis corresponds to PC 1, and the y-axis PC 2, in each plot. On the left, the kernel density estimate for the GIT embedding; in the middle, proteins are colored by protein size; on the right, proteins are colored by CATH-C label. We see separation based on CATH-C label, like the MDS and t-SNE embeddings, but not by protein size.}
	\label{fig:CATH00_GIT}
\end{figure}

For each protein we compute its GIT vector in $\R^{30}$, perform PCA and project the points onto the first two principal directions; the results are displayed in Figure \ref{fig:CATH00_GIT}. On the left is the kernel density estimate for the projected coordinates. In the center and on the right are scatter plots of the coordinates with protein size and Class label colors, respectively. As in the MDS and t-SNE figures (Figures \ref{fig:CATH00_full_wass_MDS_PC12} and \ref{fig:CATH00_full_wass_tsne}), GIT vectors do a good job of sorting proteins based on their Class labels. Unlike the previous two plots, GIT vectors do not give a strong separation via protein size.

Since we are most interested in aspects of the data found by PH that are not contained in the GIT vectors, we compare the two approaches using AJIVE. The representations from each method are taken as data blocks, $X_1$ (PH) and $X_2$ (GIT), in the AJIVE algorithm. The estimated joint components in the AJIVE decomposition, $\hat{J_1}$ and $\hat{J_2}$, reveal features of protein structure captured well by both methods.

\begin{figure}[ht]
	\centering
	
	\includegraphics[width = 0.32\textwidth]{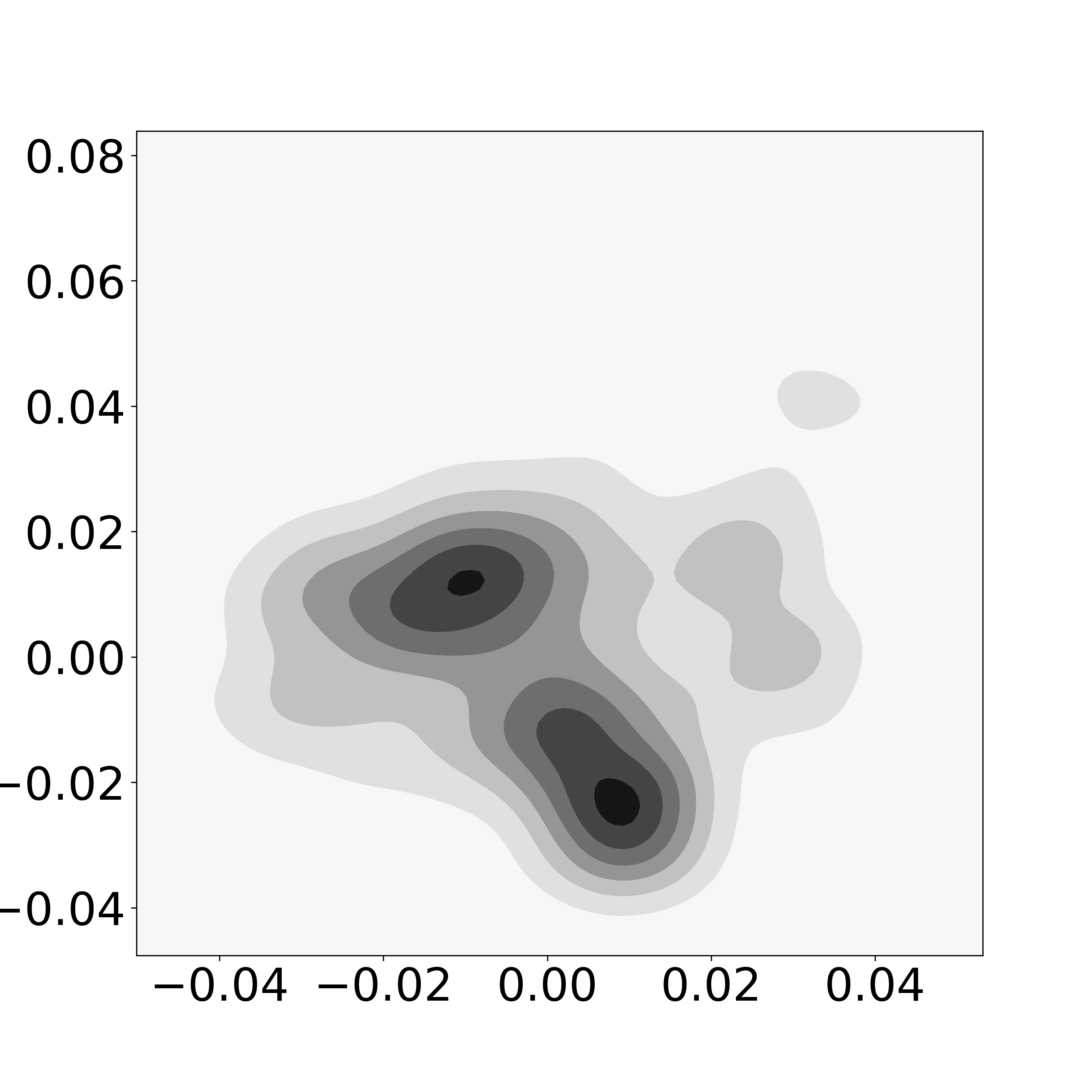}
	\includegraphics[width = 0.32\textwidth]{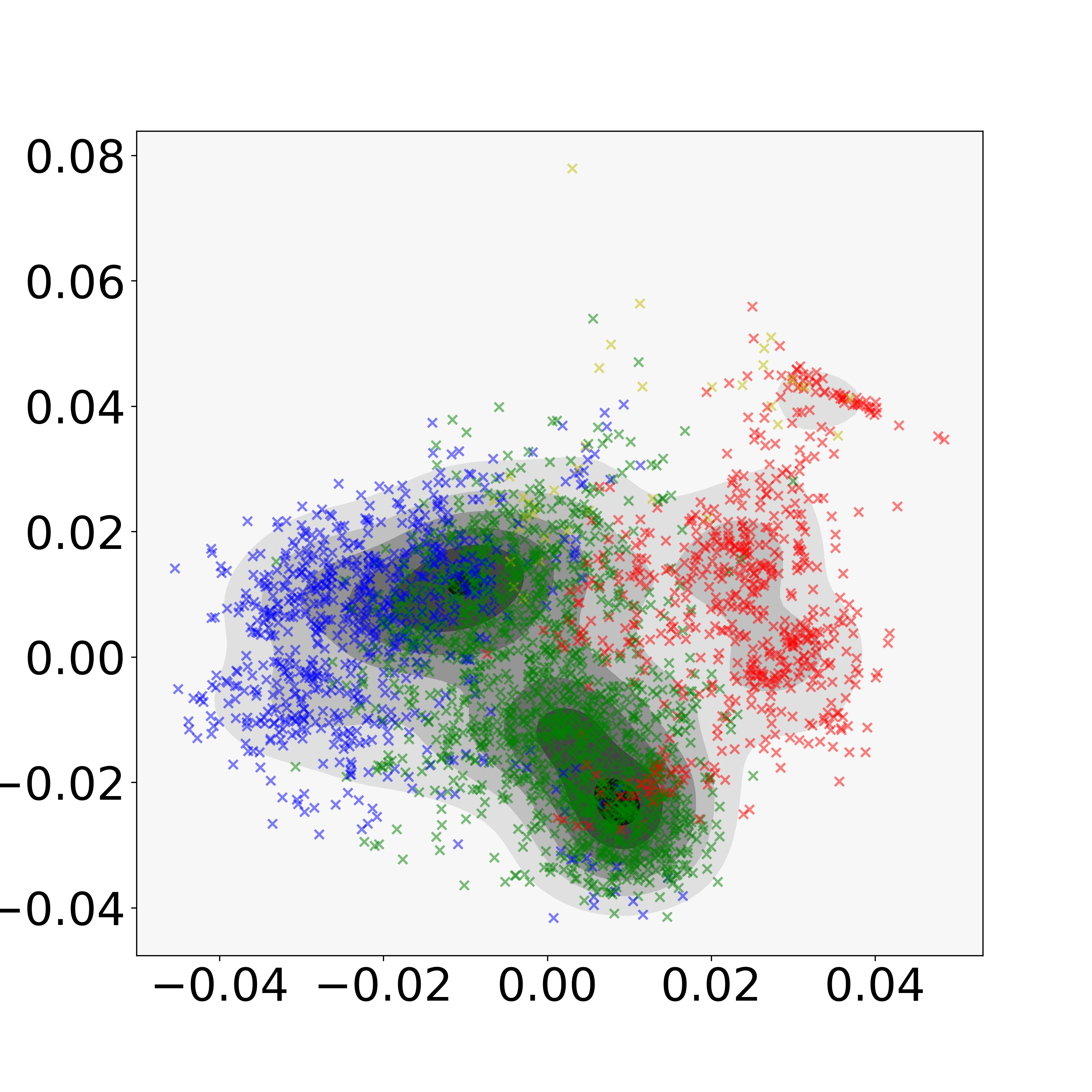}	
	\includegraphics[width = 0.32\textwidth]{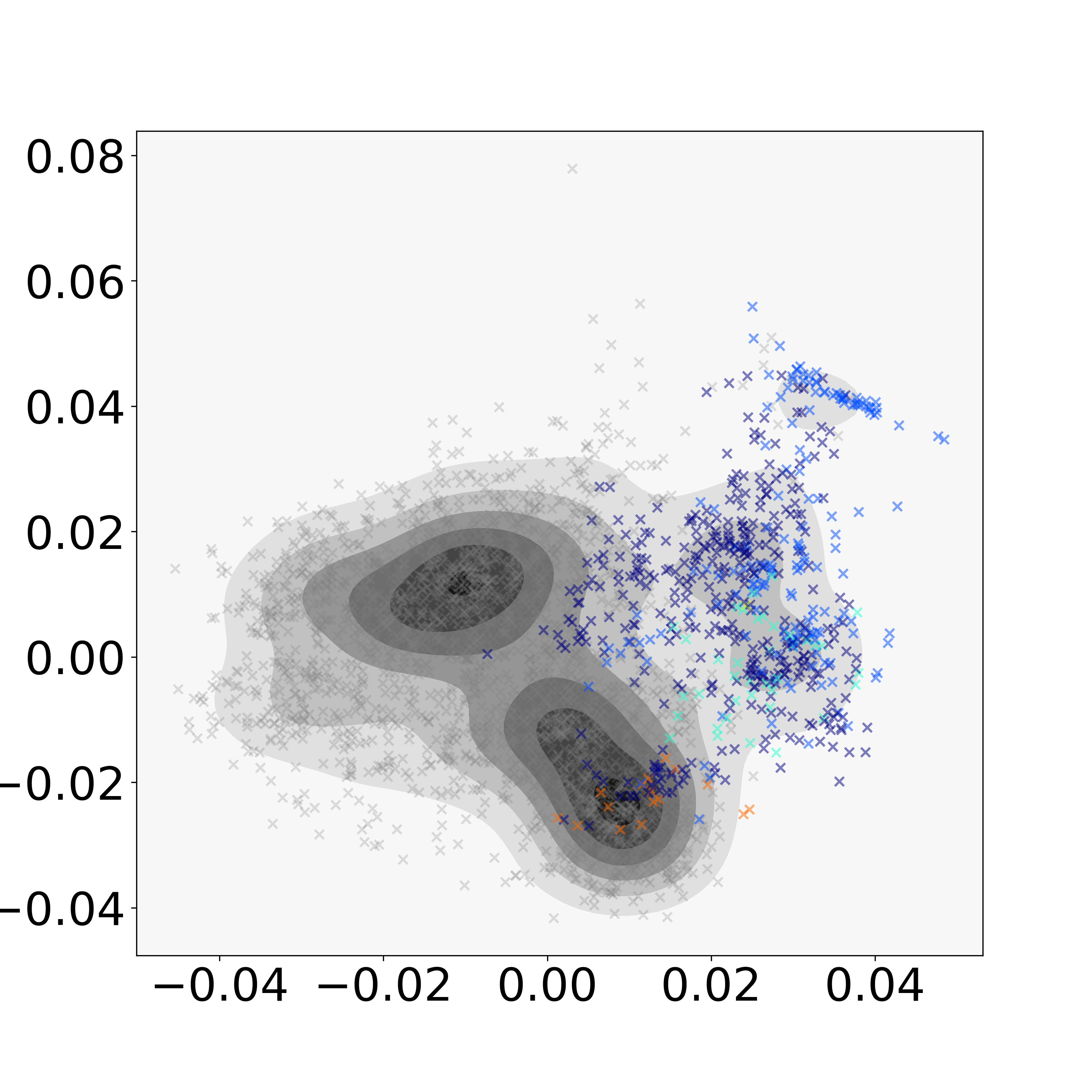}
	\caption{Plots of the common normalized scores (CNS) from AJIVE. The x-axis corresponds to PC 1, and the y-axis PC 2, in each plot. The left plot shows the kernel density estimate of the first two components. The middle plot shows the first two components with proteins colored by Class label. Separation of Class label is a mode of variation common to both methods. On the right, $\alpha$-helices colored by their Architecture label.}
	\label{fig:CNS_12}
\end{figure}

The first two components of the common normalized scores (CNS) from AJIVE are shown in Figure \ref{fig:CNS_12}. The CNS represent a common basis of $row(\hat{J})$, where $row(\hat{J})$ denotes the estimated common score subspace of the joint matrices $J_1$ and $J_2$. On the left, the kernel density estimate shows two denser regions in the center of the plot. The middle panel shows separation of the Class labels. The green points, representing the class of mixed $\alpha$-helices and $\beta$-sheets (Class label 3), are concentrated in the dense center regions of the density estimate. Note that the red points, which represent the $\alpha$-helix class of proteins, appear to lie in a distinct region of the density estimate and have noticeable clusters. The right panel investigates the next CATH level of the red points ($\alpha$-helix) with proteins colored by the second level of CATH classification, Architecture. Separation by Architecture label is not clear and clusters do not appear homogeneous. This indicates that separation by Class label is a joint feature of PH and GIT vectors, however separation by Architecture label is not a driver of the joint components. 

\begin{figure}[ht]
	\centering
	
	\includegraphics[width = 0.24\textwidth]{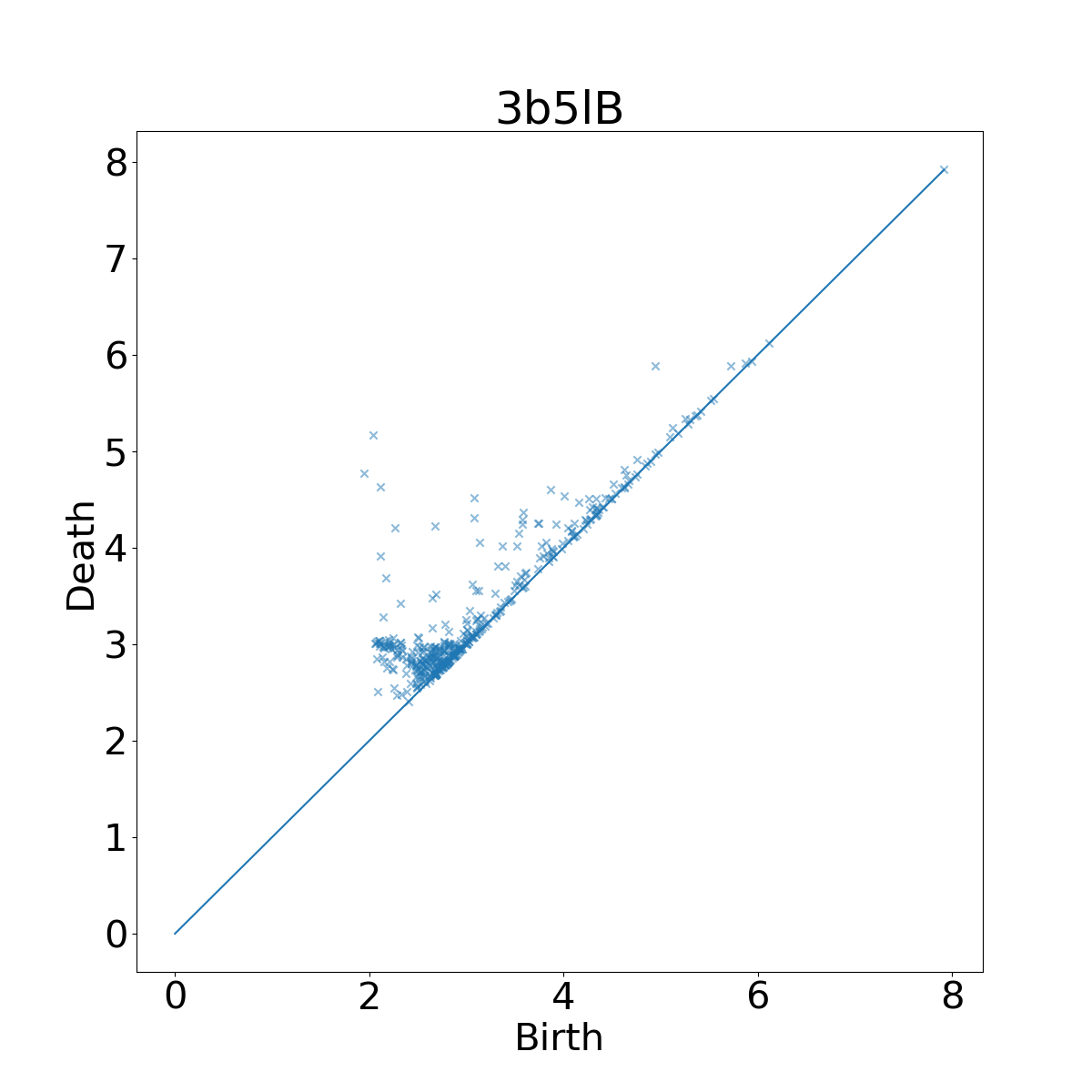}
	\includegraphics[width = 0.24\textwidth]{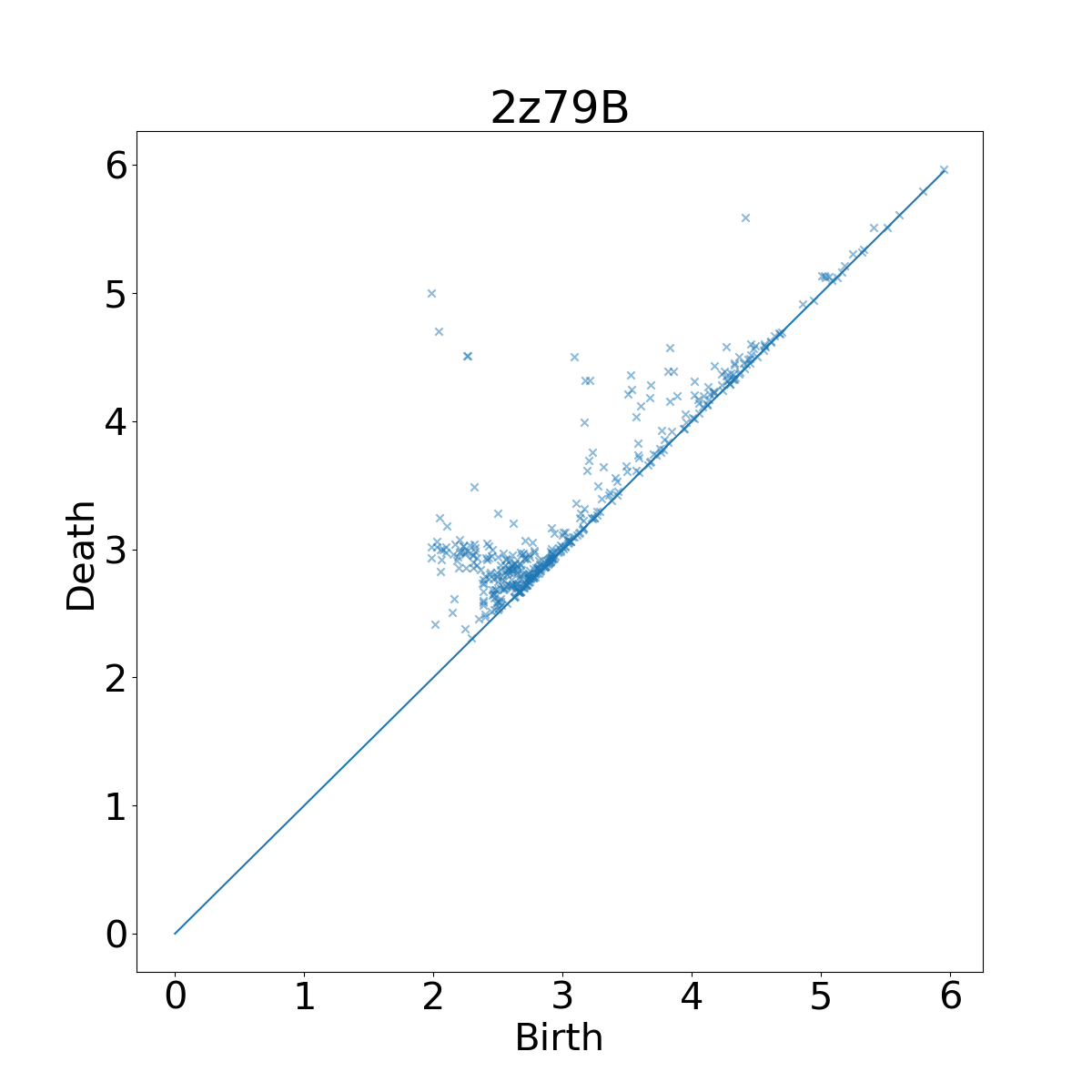}
	\includegraphics[width = 0.24\textwidth]{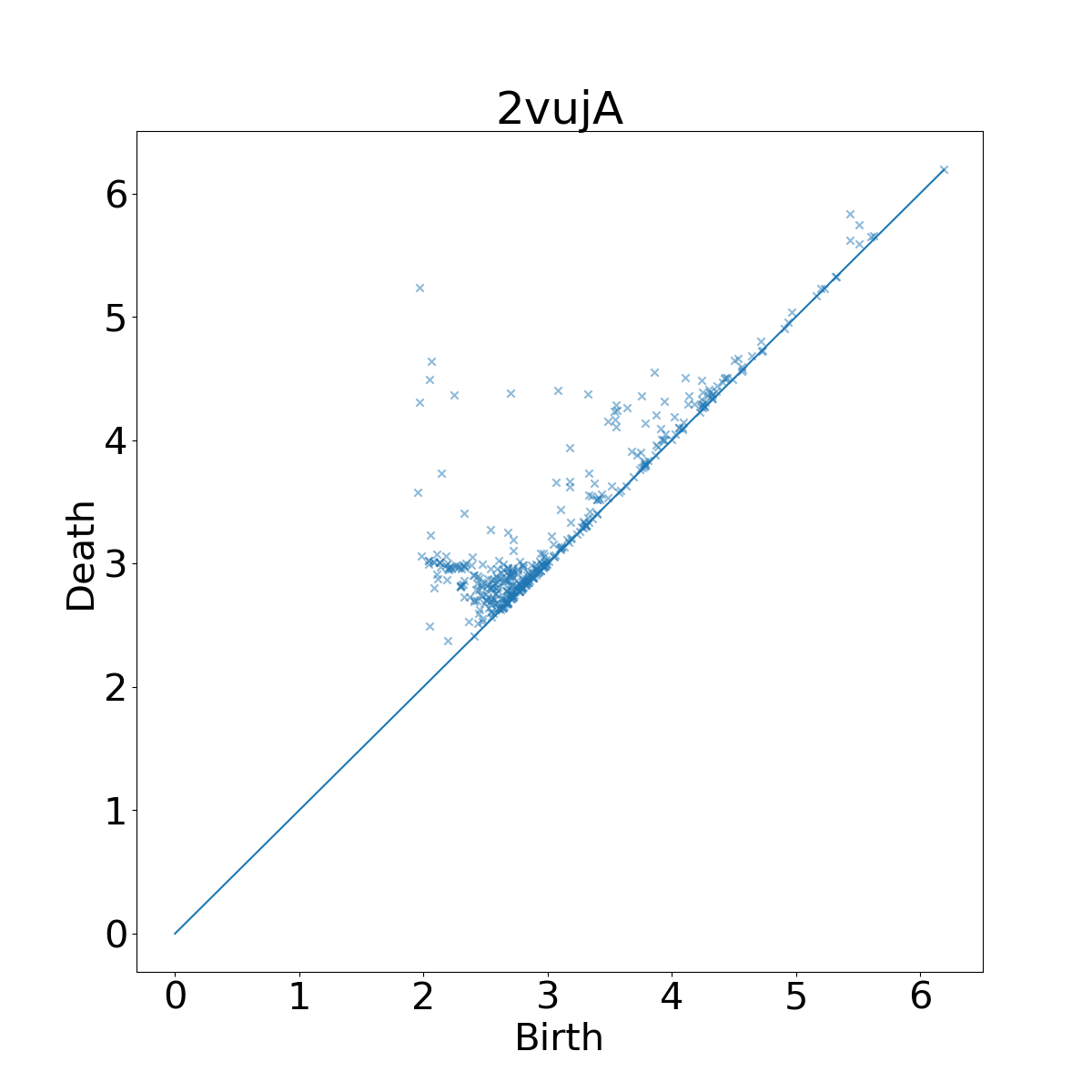}
	\includegraphics[width = 0.24\textwidth]{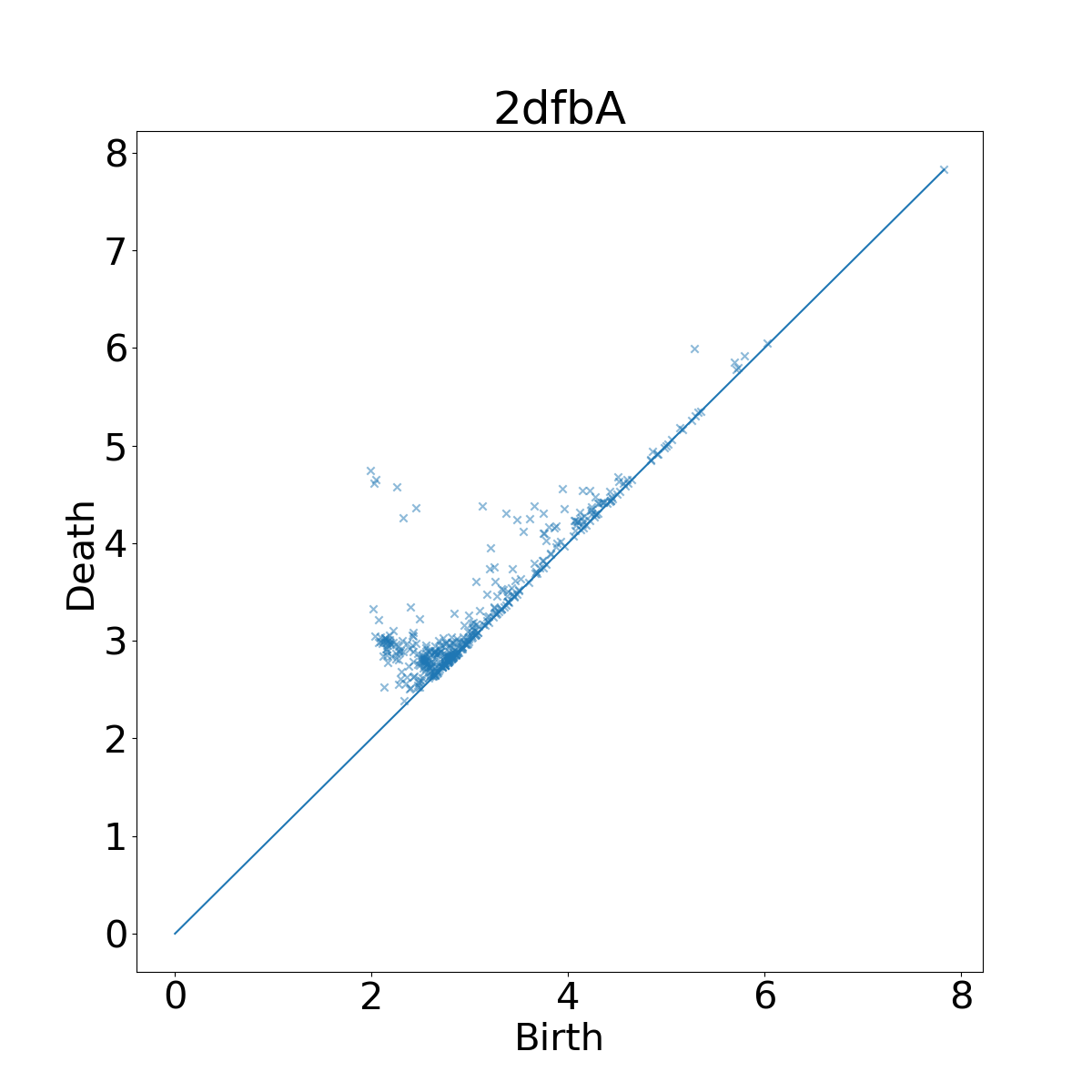}
	
	\includegraphics[width = 0.24\textwidth]{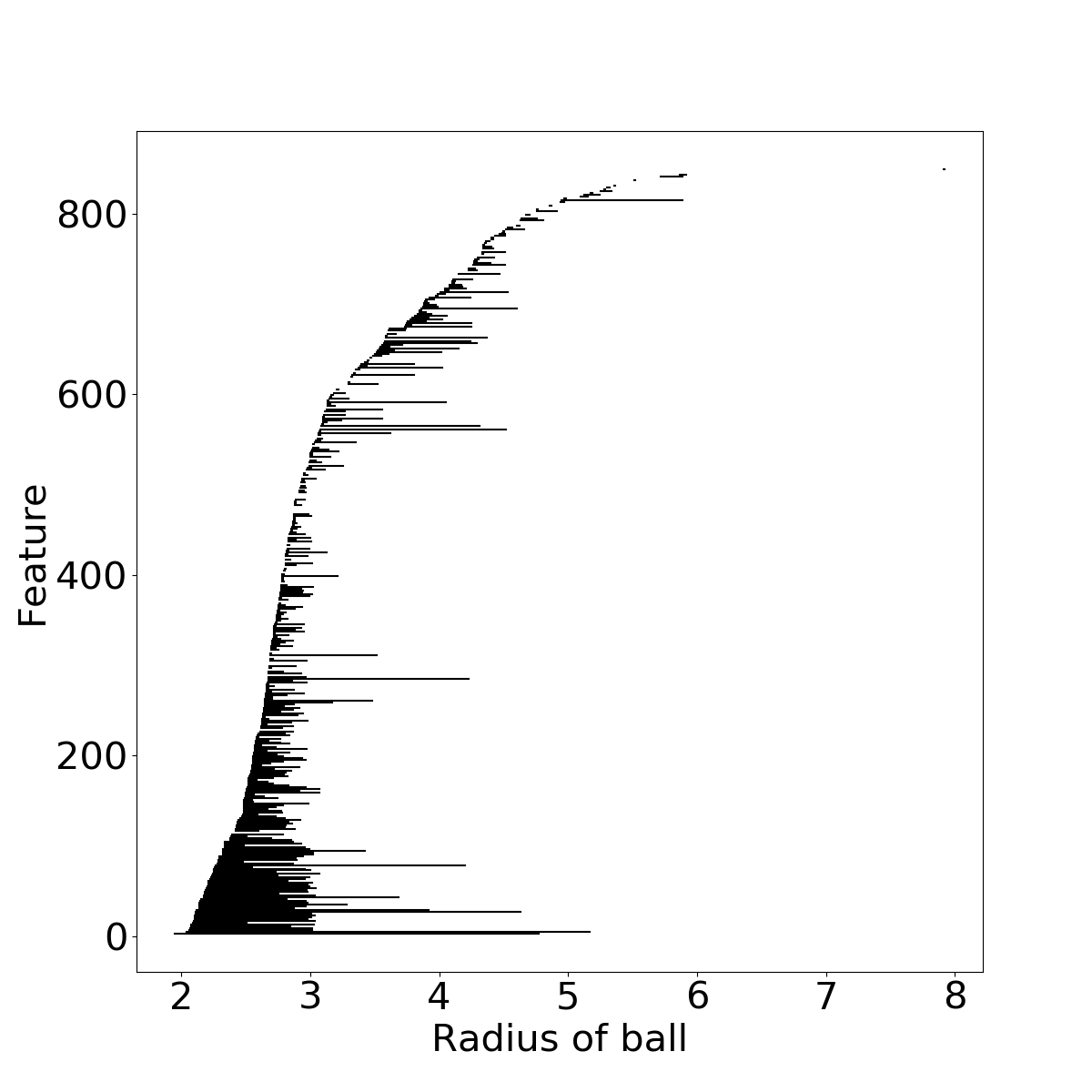}
	\includegraphics[width = 0.24\textwidth]{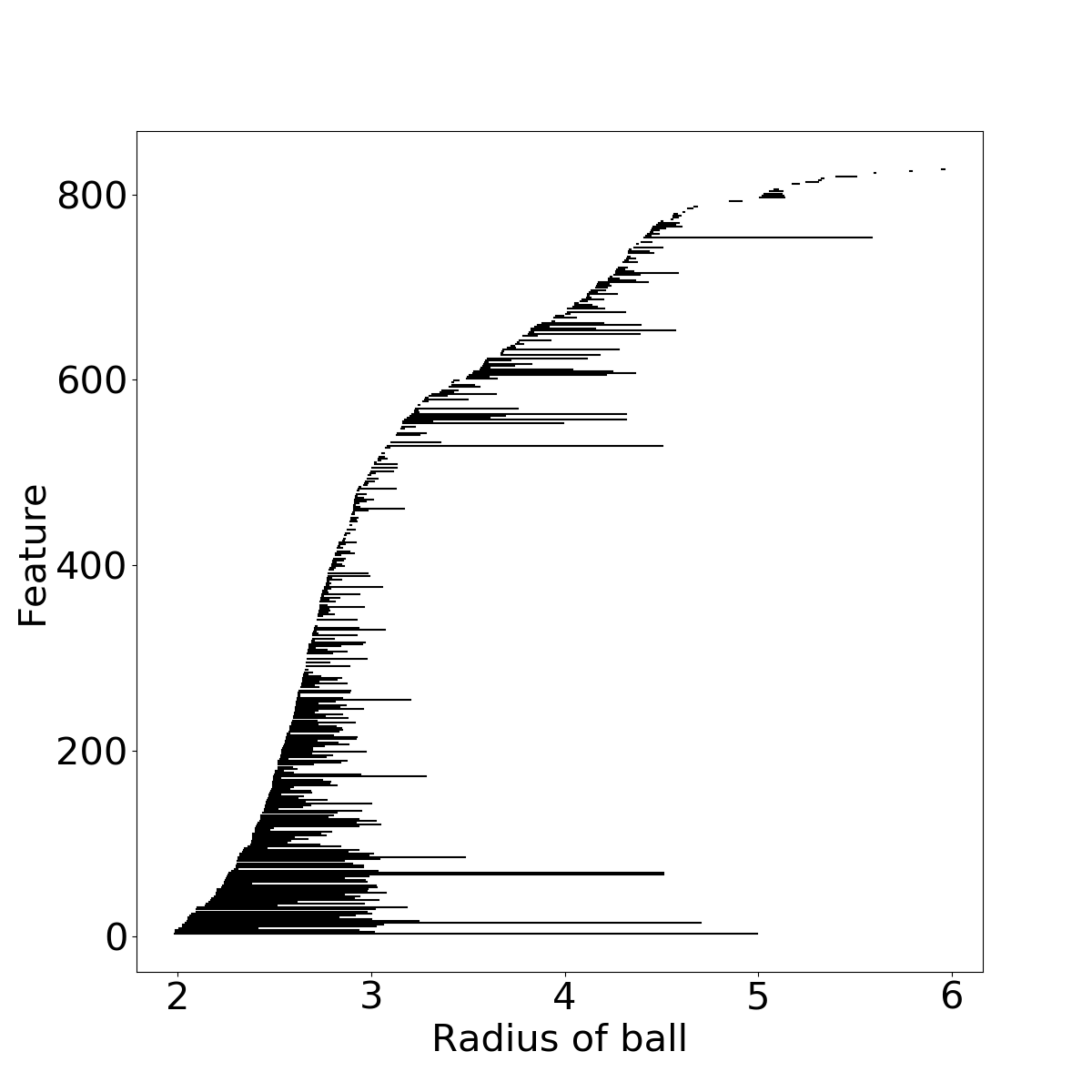}
	\includegraphics[width = 0.24\textwidth]{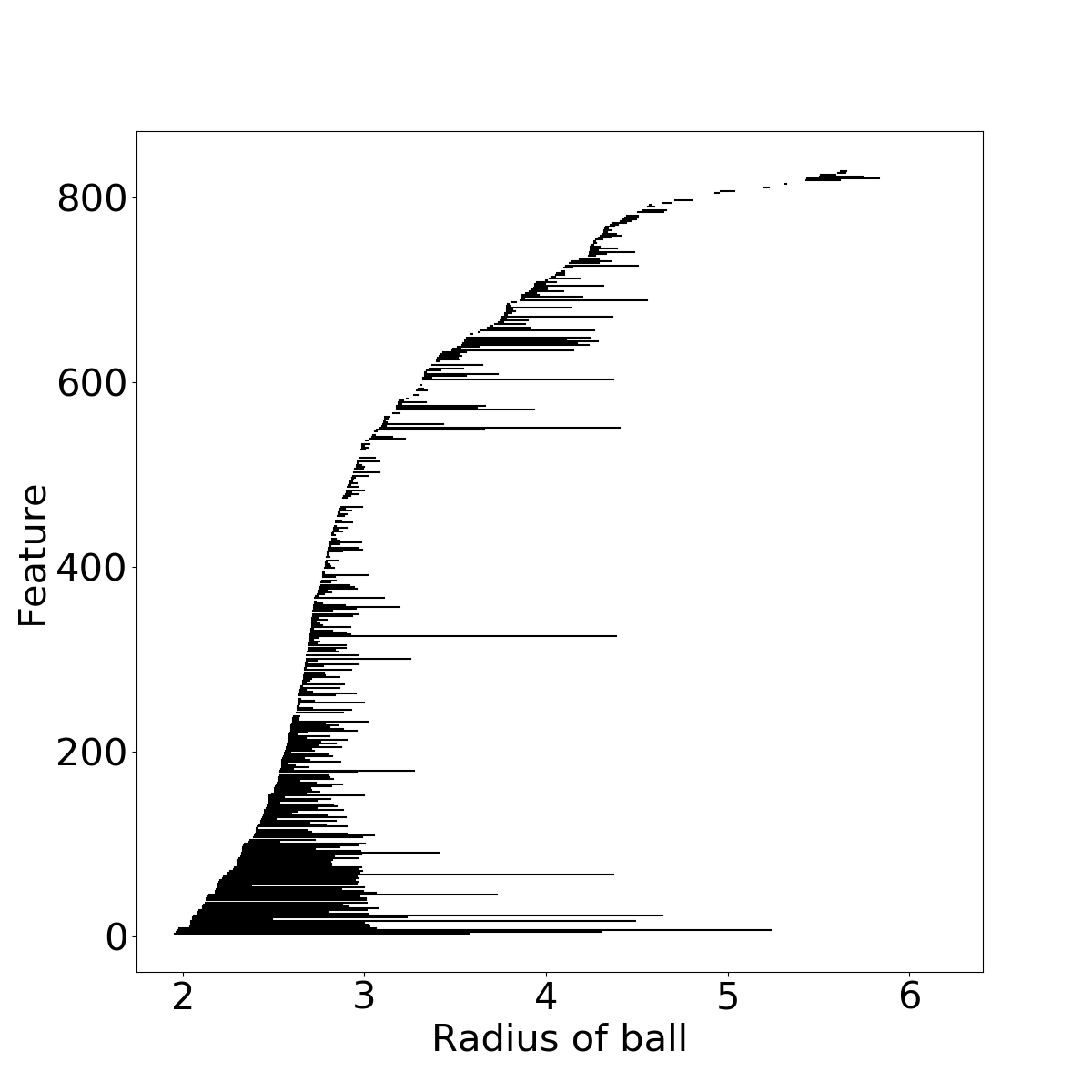}
	\includegraphics[width = 0.24\textwidth]{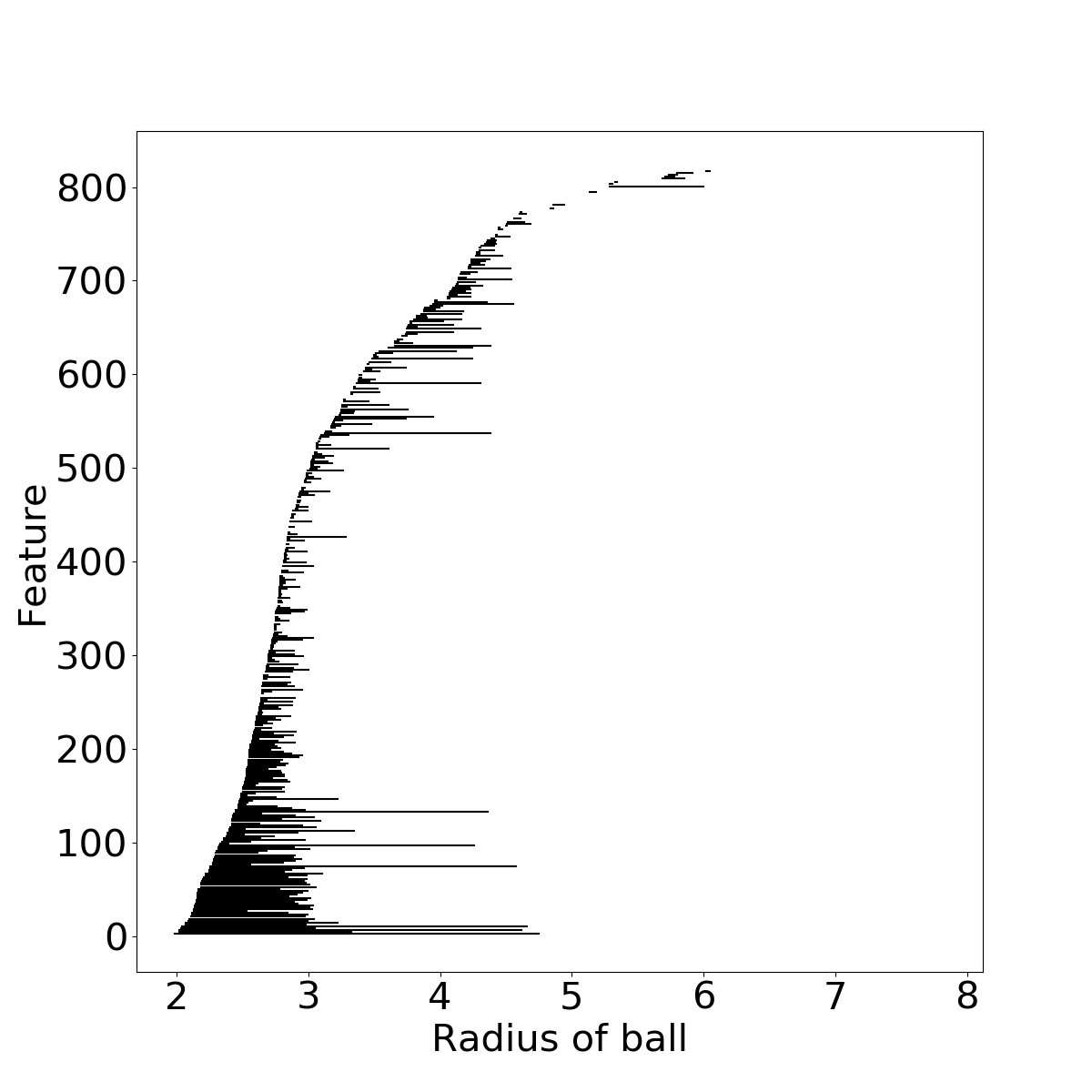}
	
	\includegraphics[width = 0.24\textwidth]{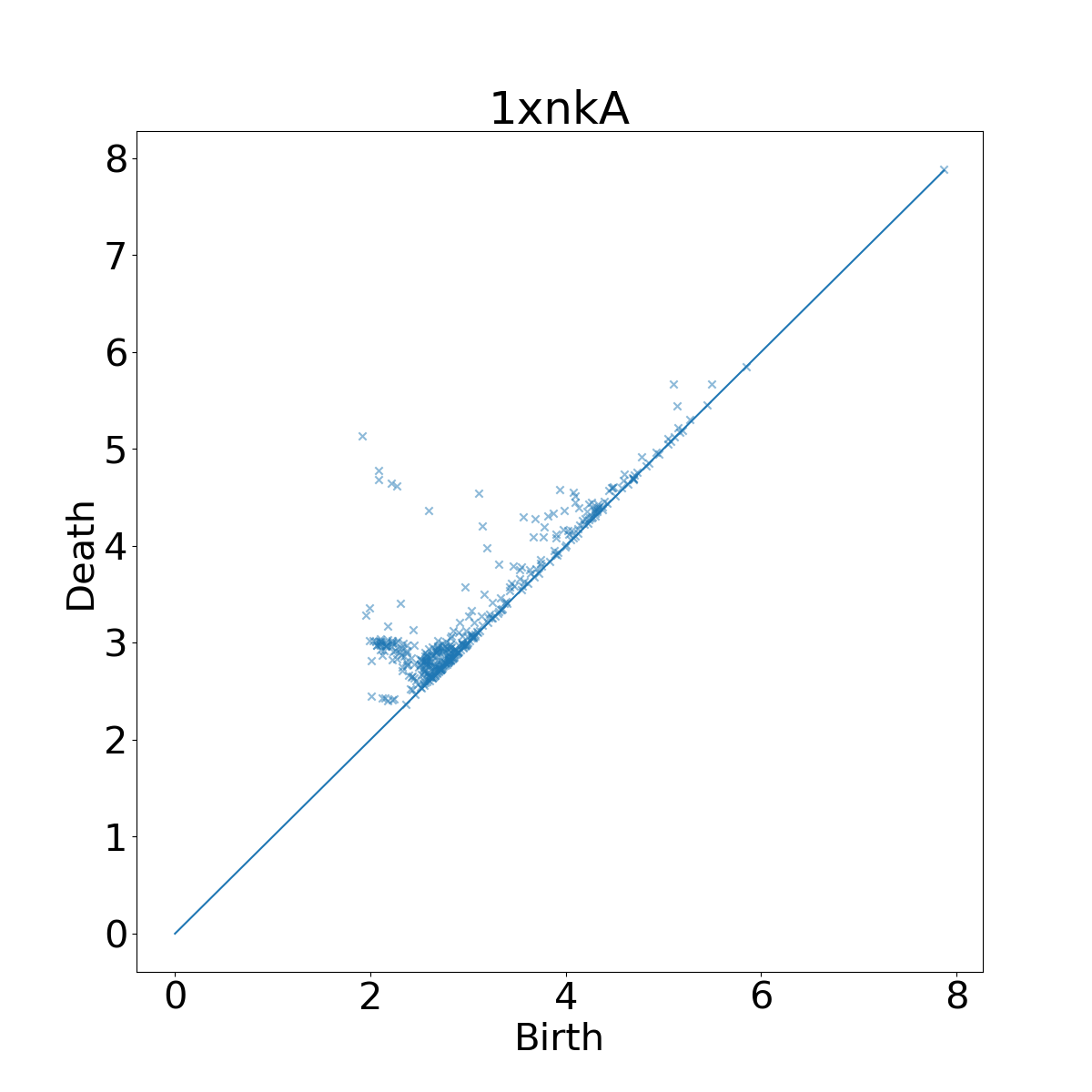}
	\includegraphics[width = 0.24\textwidth]{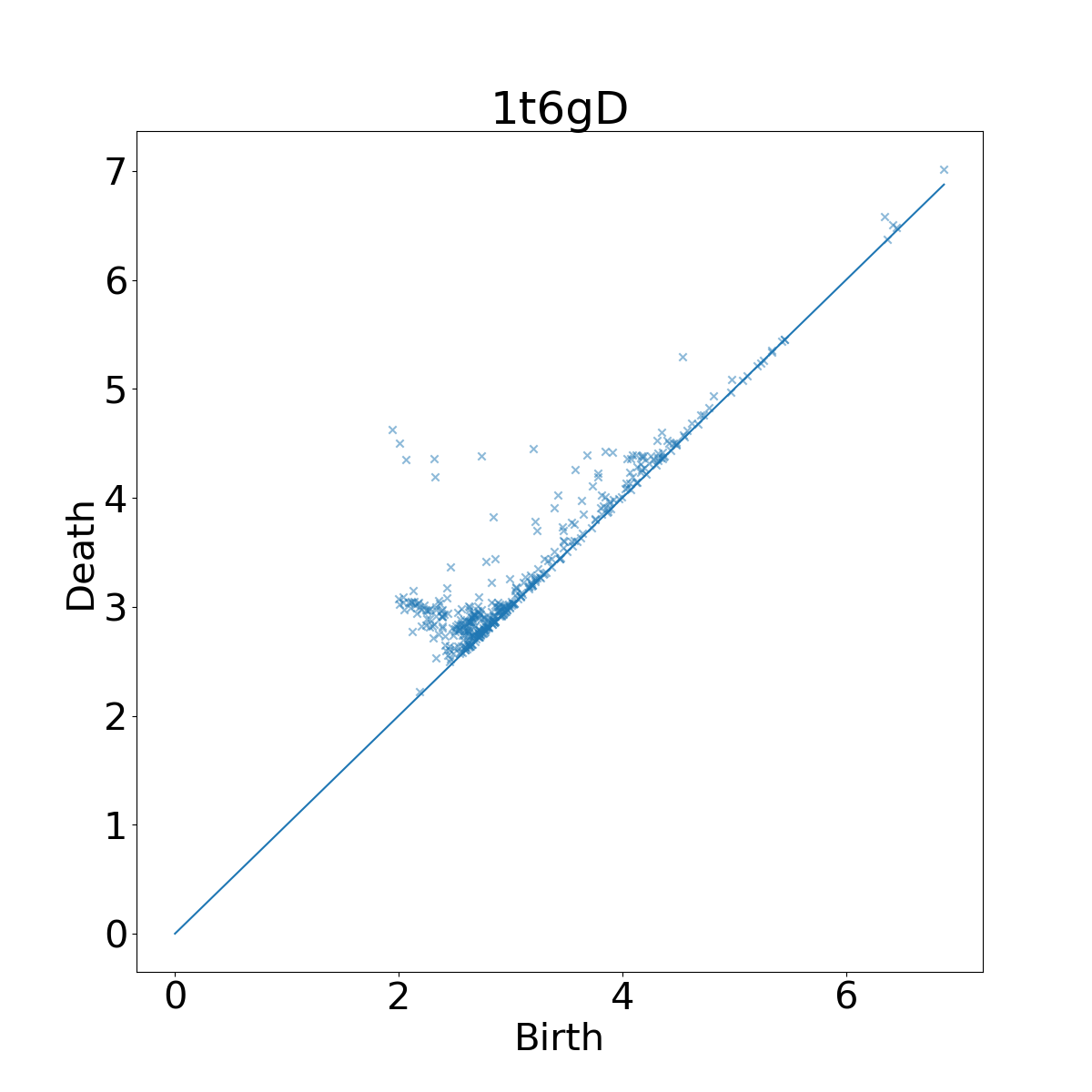}
	\includegraphics[width = 0.24\textwidth]{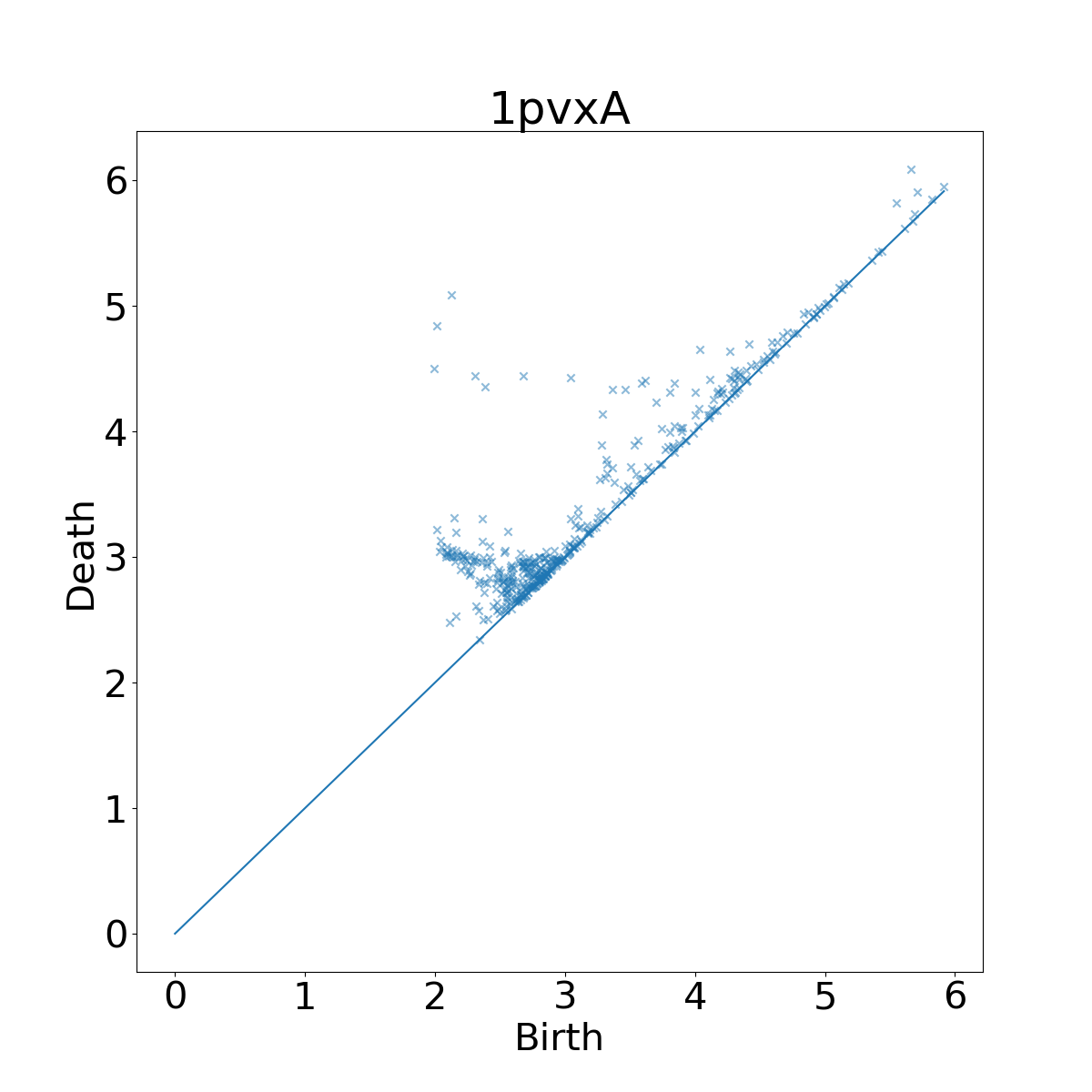}
	
	\includegraphics[width = 0.24\textwidth]{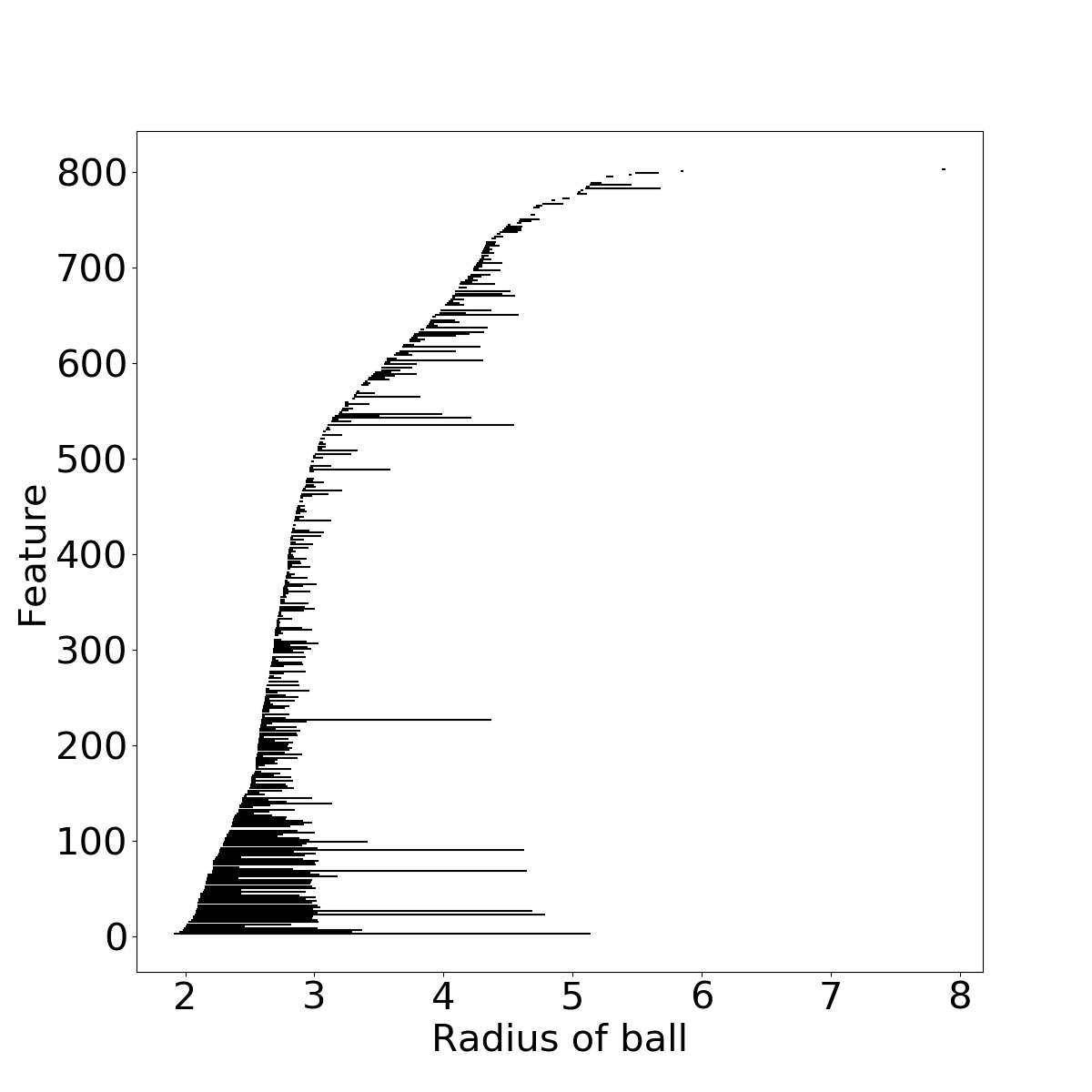}
	\includegraphics[width = 0.24\textwidth]{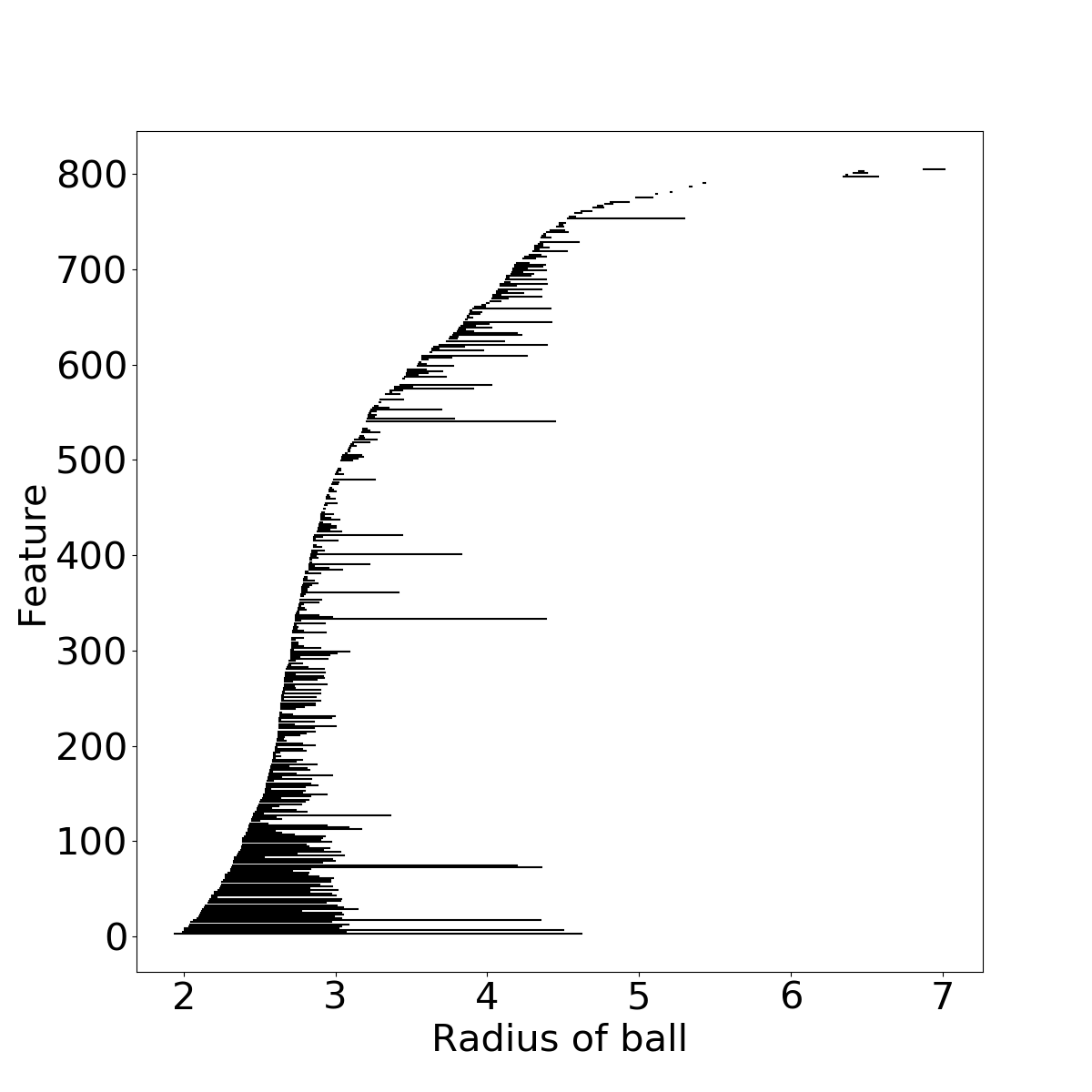}
	\includegraphics[width = 0.24\textwidth]{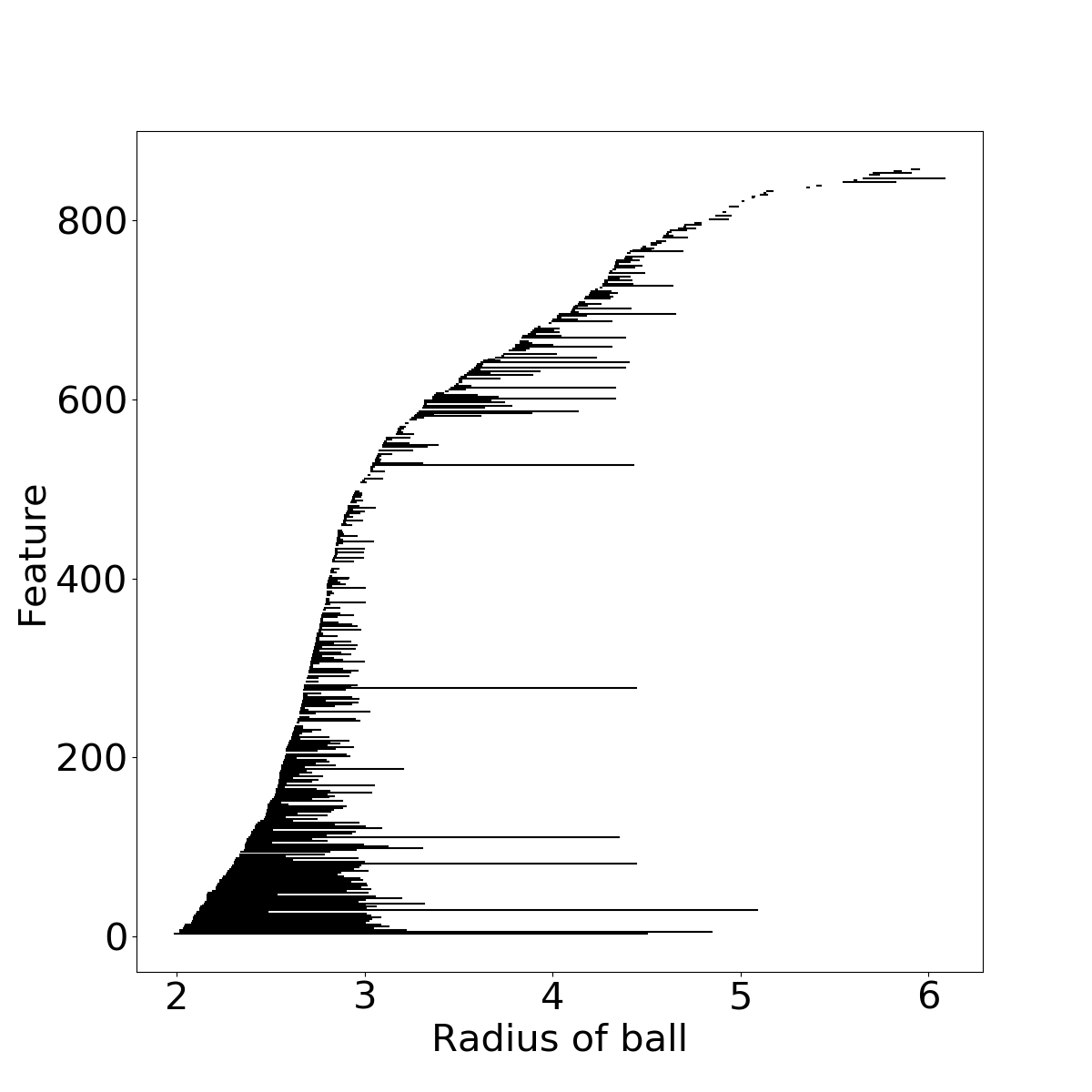}
	
	\caption{The PDs (top rows) and barcodes (bottom rows) of the proteins in an interesting cluster from the PH individual data representation; each vertical pair of plots corresponds to a single protein. The common letter-C shaped feature of the proteins in this cluster corresponds to the early born persistent feature that persists to approximately 6 {\AA}.}
	\label{fig:ind_cluster_21}
\end{figure}

Next we study structural aspects identified by PH that were not identified using the GIT approach. This is based on the individual block specific scores (BSS), which give the estimated matrices $\hat{I}_1$ and $\hat{I}_2$ in the AJIVE decomposition. In particular, we are interested in proteins that are closely related by PH but not by GIT. We compute the t-SNE coordinates of the PH individual BSS matrix and perform hierarchical clustering on these coordinates using Ward's linkage and number of clusters $n=300$. The clustering is performed on the t-SNE coordinates rather than the original individual matrices since t-SNE gives a useful representation for identifying and visualizing clusters. We then select the 10 most homogeneous clusters in terms of their Class and Architectures labels, where homogeneity was measured by calculating the entropy in each label of each of the 300 clusters. These clusters demonstrate aspects of protein structure that are revealed by PH but not by GIT. One particular cluster is carefully investigated in Figure \ref{fig:ind_cluster_21}.

\begin{figure}[ht]
	\centering
	
	\includegraphics[width = 0.95\textwidth]{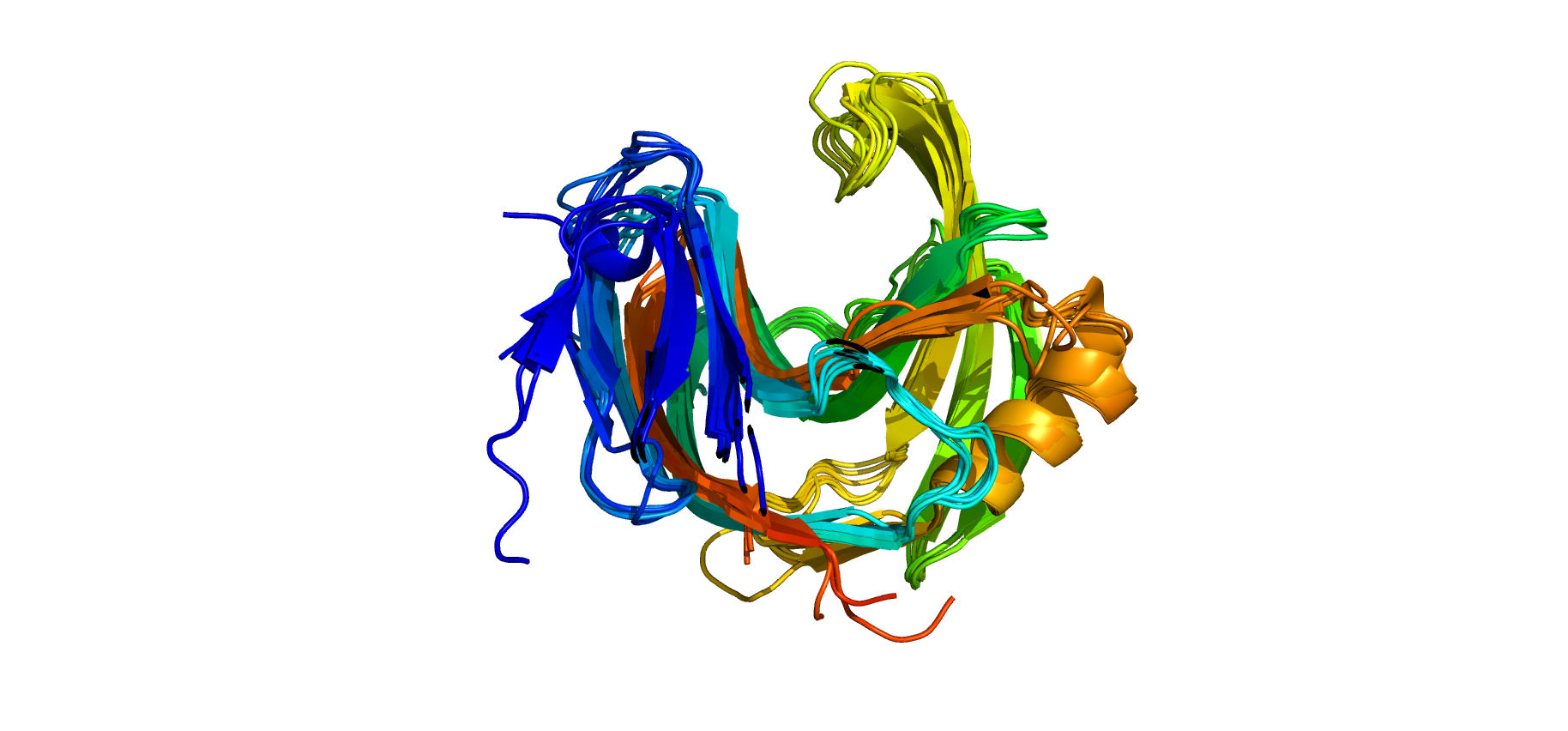}
	\caption{Proteins from the cluster in Figure \ref{fig:ind_cluster_21} overlaid on top of each other. Note the C-shaped cavity (top) common to these proteins corresponding to the long bars that were born late in each barcode on the bottom of Figure \ref{fig:ind_cluster_21}. This C-shaped cavity is biologically significant and gives an example of the aspects of protein structure that drive PH.}
	\label{fig:ind_cluster_21_proteins}
\end{figure}

The barcodes in the bottom row all have a persistent (long) bar that is born late. Such topologies can be generated by a letter C-shaped loop where the difference between the endpoints corresponds to the late birth time. Figure \ref{fig:ind_cluster_21_proteins} shows the 3-D protein structure of the seven members of this cluster overlaid on top of each other. The structures have been aligned using the standard visualization software PyMol to highlight the common C-shape. The top of this overlay shows a clear C-shaped cavity common to these proteins. This has biological significance as it serves as a catalytic site for the binding and hydrolysis of oligosaccharides \cite{DAVIES1995853}. Hence, this is chemical structure readily apparent from the PH point of view, but not by GIT. The protein names and CATH labels of the proteins in this cluster are given in Table \ref{table:cluster_table} of Appendix \ref{tables}.

	

\subsection{Comparison of PDs to GIT vectors}

To analyze the aspects of protein structure that are captured specifically by PH, we focus on pairs of proteins closely related by PH but not by GIT. Figure \ref{fig:CAT_pairs} shows the t-SNE plots colored by pairs of proteins identical in CAT label whose distance from each other is relatively close in the t-SNE coordinates of the PH individual matrix compared to the t-SNE coordinates of the original GIT matrix. This was identified by calculating the Euclidean distance between proteins in both coordinates, normalized by the total sum of squared distances from the mean of the corresponding coordinates, and finding the ratio of distance in (individual) PH to distance in GIT. Figure \ref{fig:CAT_pairs} shows the protein pairs colored in each view. Note that proteins in each pair are close together in PH and overlap each other in the left plot of Figure \ref{fig:CAT_pairs}, but are widely separated in GIT and can be individually distinguished in the right plot of Figure \ref{fig:CAT_pairs}. These pairs exhibit cases where PH accurately groups related proteins while GIT does not. See Table \ref{table:pairs_table} in Appendix \ref{tables} for the identities and CAT labels of proteins in each pair as well as the ratio of their distances in PH versus GIT.

\begin{figure}[ht]
	\centering
	
	\includegraphics[width = 0.45\textwidth]{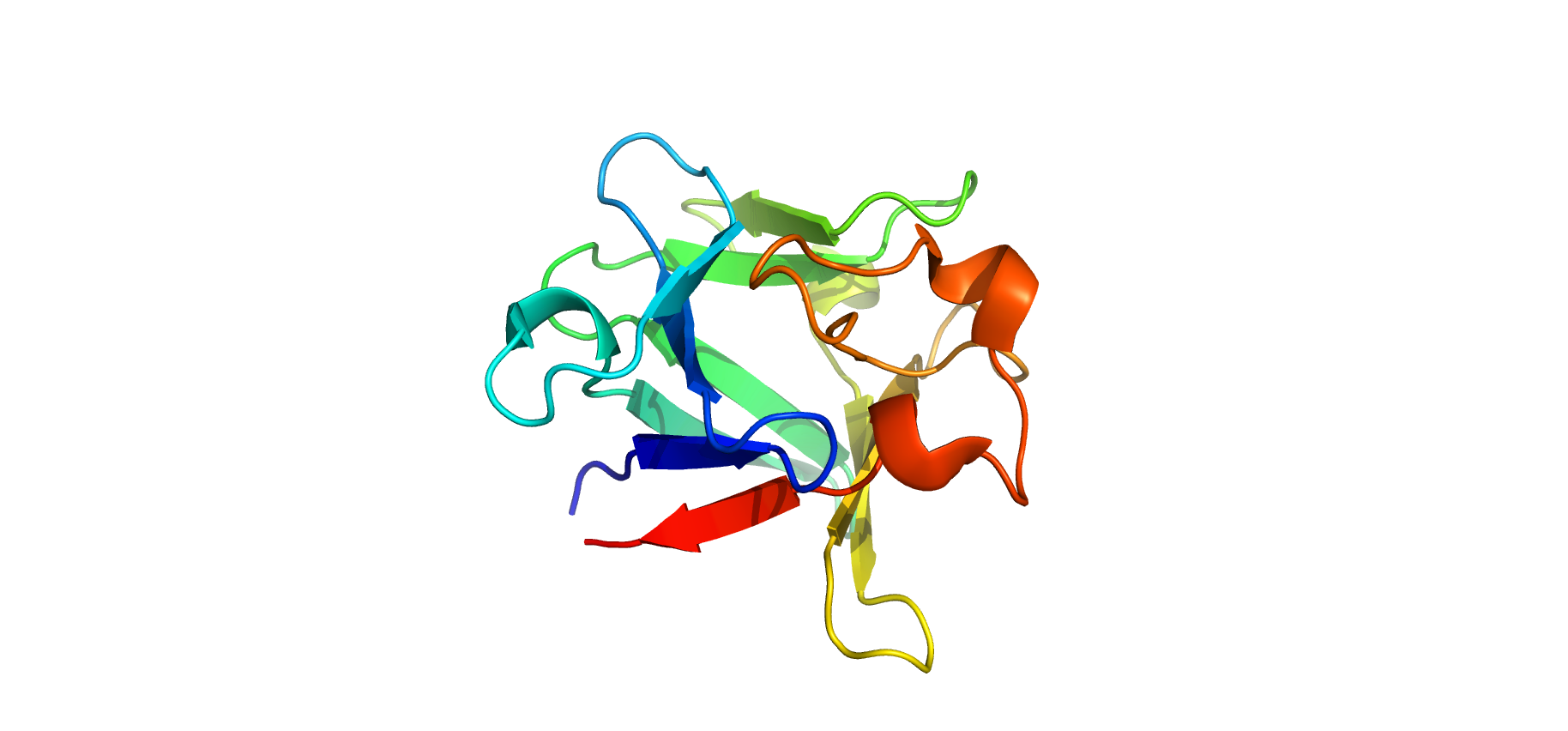}
	\includegraphics[width = 0.45\textwidth]{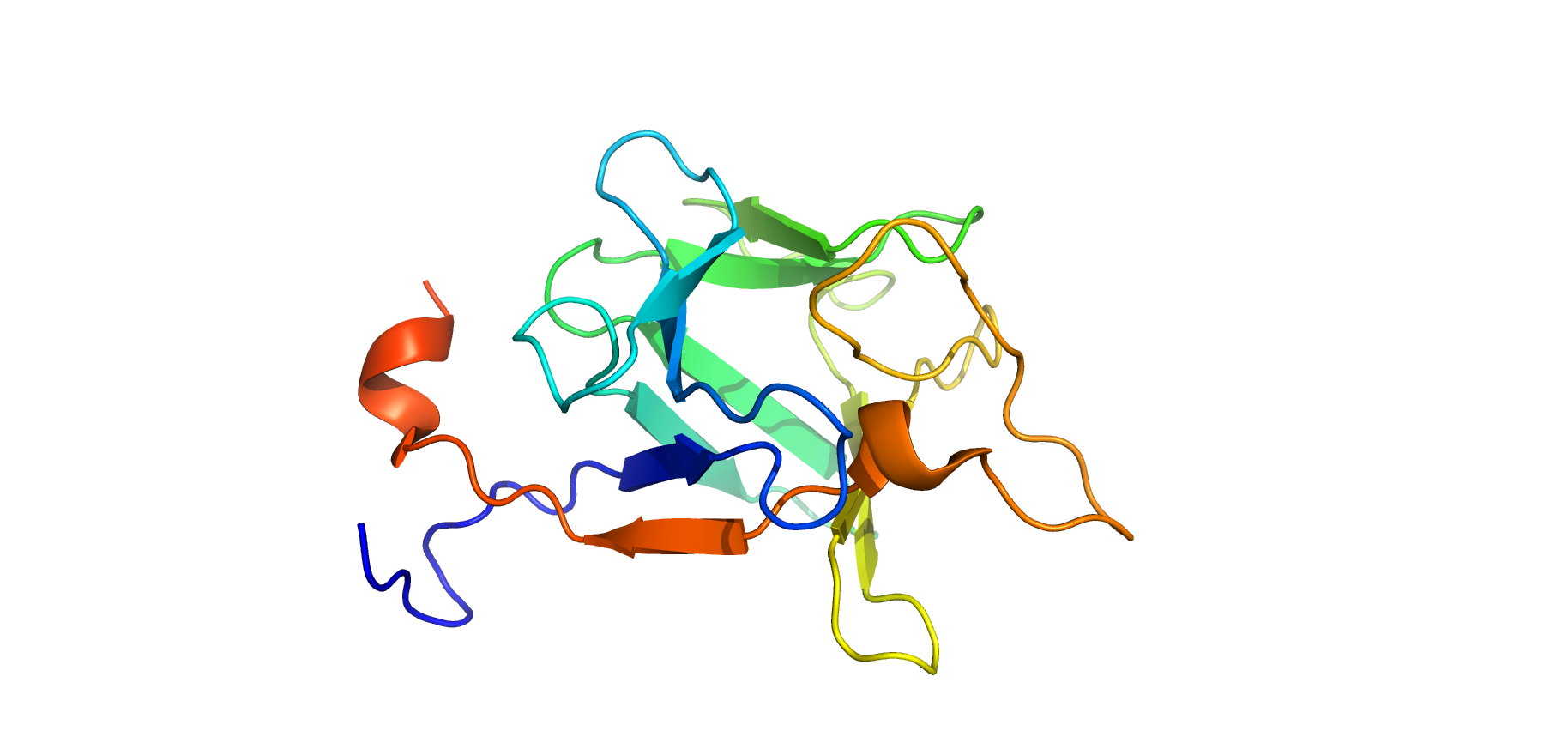}
	
	\includegraphics[width = 0.45\textwidth]{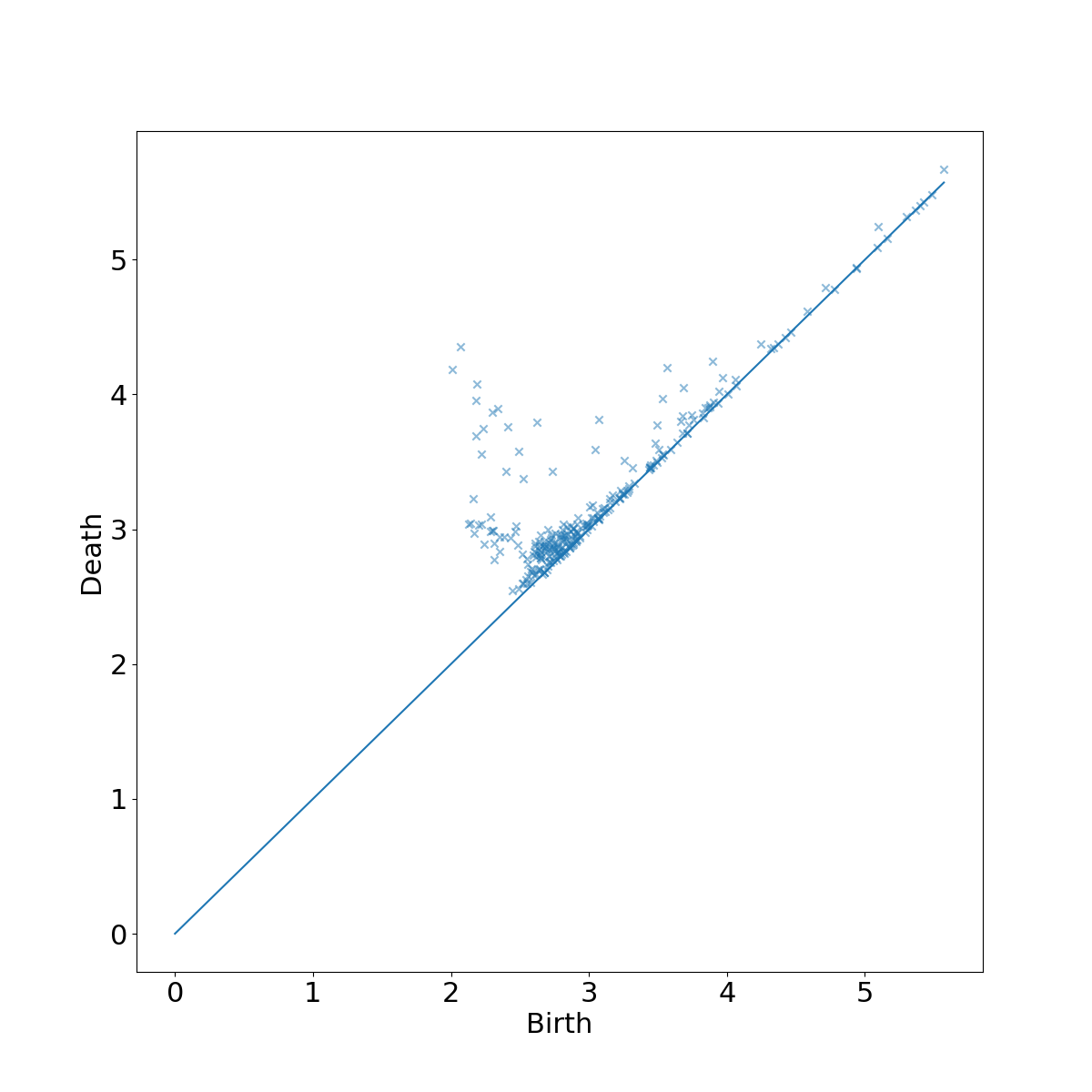}
	\includegraphics[width = 0.45\textwidth]{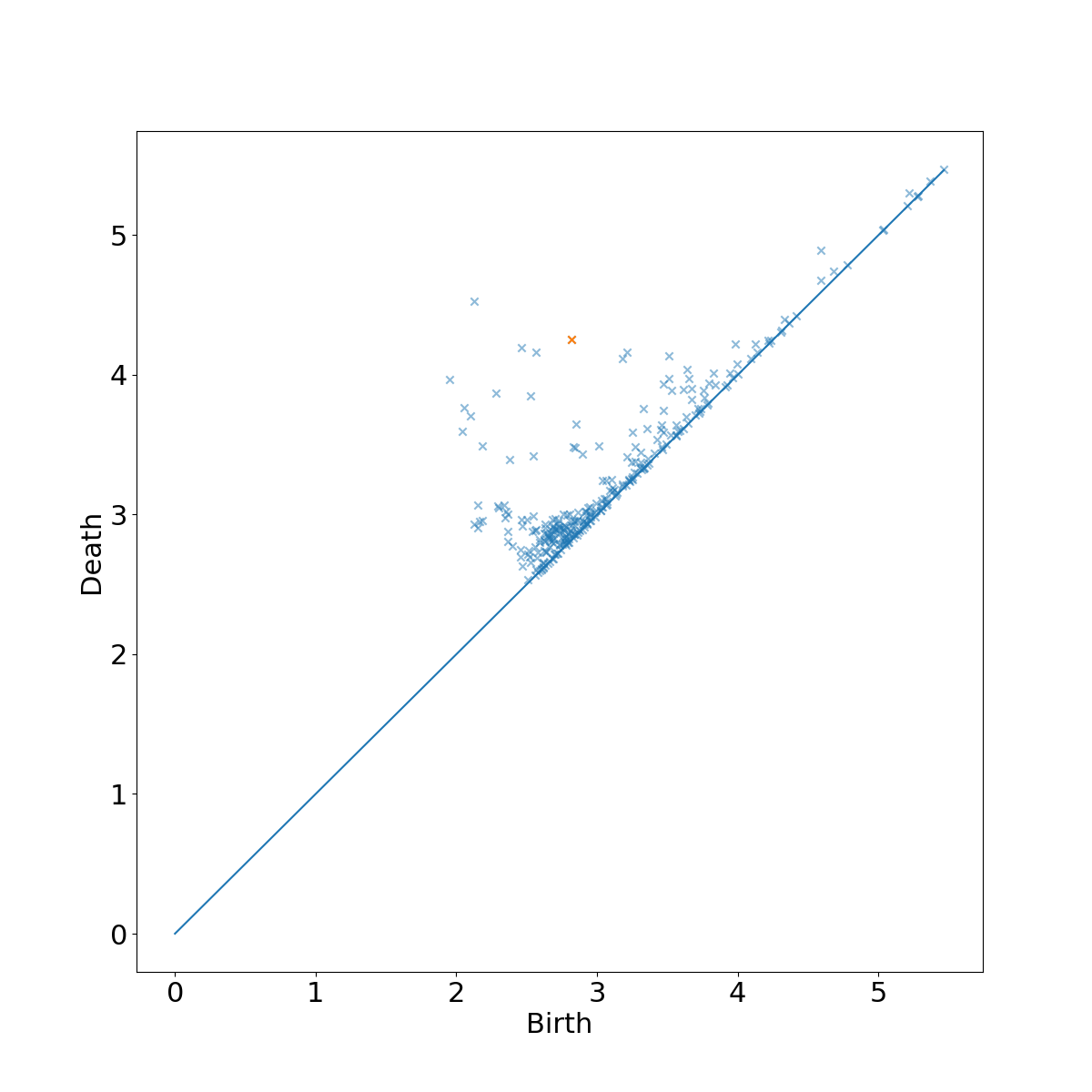}
	
	\includegraphics[width = 0.45\textwidth]{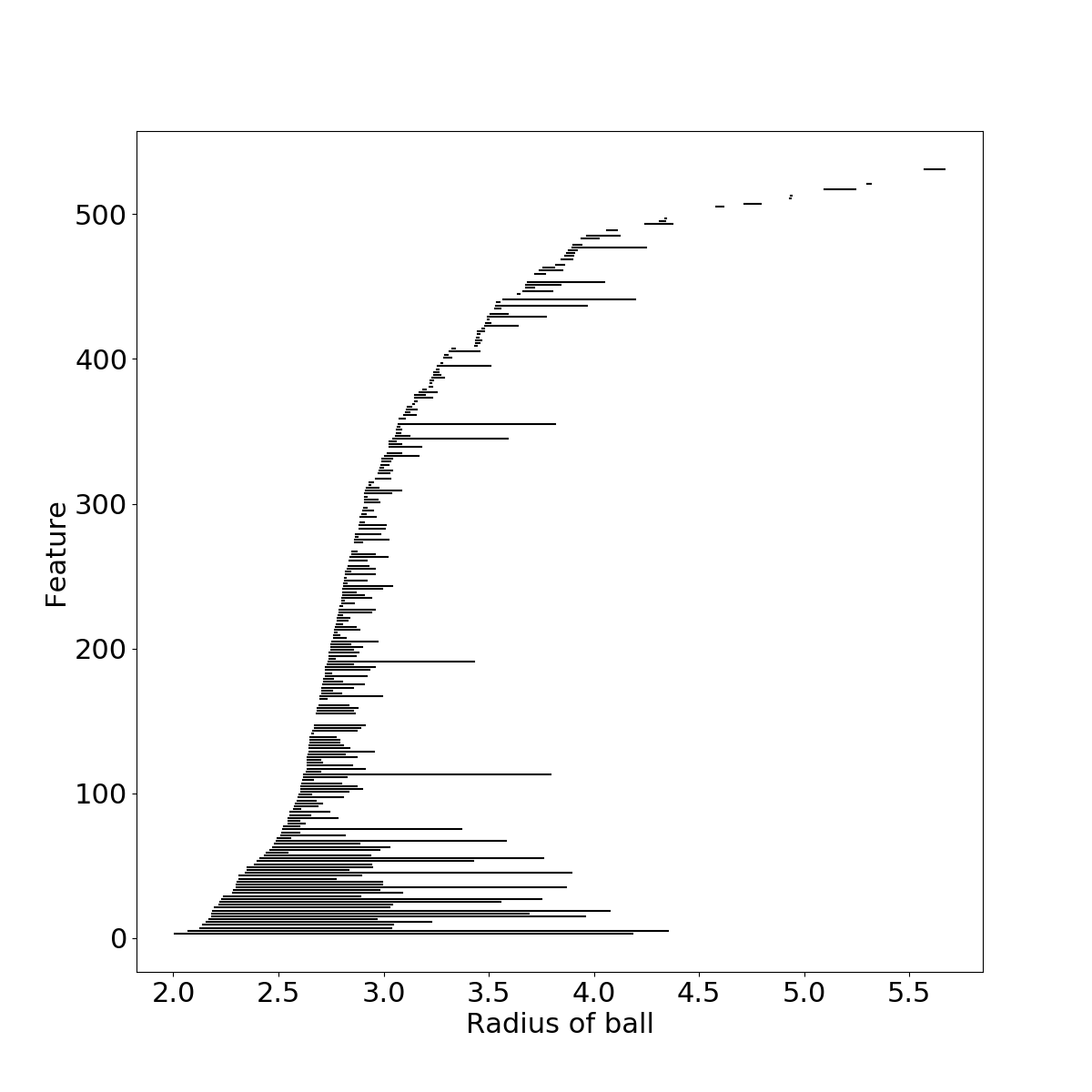}
	\includegraphics[width = 0.45\textwidth]{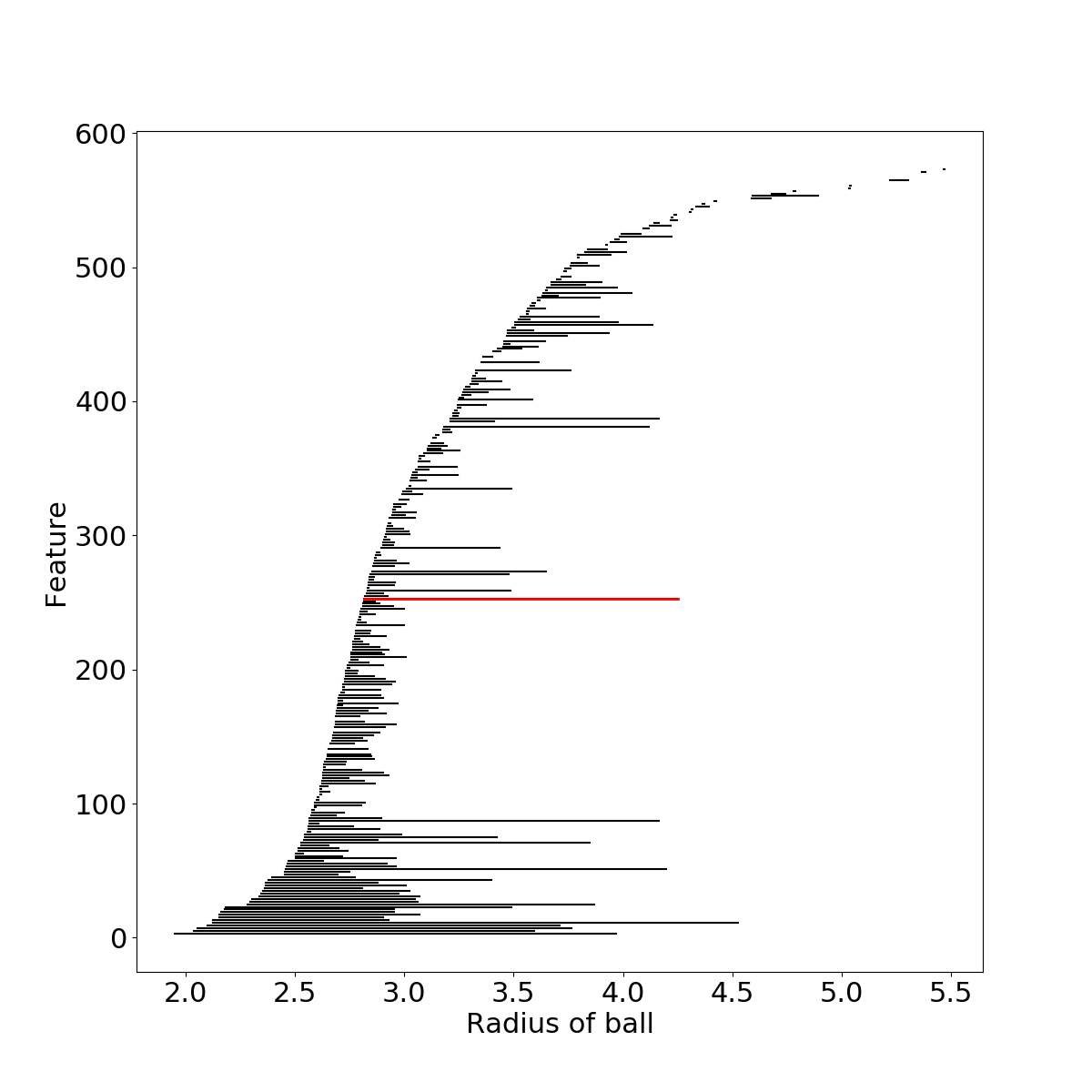}
	
	\caption{Proteins 1BAS (left column) and 2P39 (right column), colored cyan in Figure \ref{fig:CAT_pairs} showing they are close in PH but not GIT. Views in 3-D at the top show these proteins share a common core. Protein 2P39 has a large loop (right side of 2P39) and extra stretches at C-terminus and N-terminus (left side of 2P39, red and blue respectively), which are represented in the barcode by the persistent bar colored red.}
	\label{fig:ex_pair}
\end{figure}

The proteins 2P39 and 1BAS (both of which are monomers, meaning that they exist as individual chains), colored cyan in Figure \ref{fig:CAT_pairs}, demonstrate aspects of protein structure captured better by PH than by GIT and are shown in detail in Figure \ref{fig:ex_pair}. These proteins have a common core, but 2P39, shown on the right panel of Figure \ref{fig:ex_pair}, has a large loop (right side of 2P39) and extra stretches at the C-terminus and the N-terminus (left side of 2P39, red and blue respectively). The GIT representation is very sensitive to these latter features since it focuses on local curvature, making these proteins far apart in that sense. These features do appear in the barcode of 2P39, highlighted as a red bar in the lower right panel of Figure \ref{fig:ex_pair}. While these extra features are present in the barcode, they do not play a big role since PH focuses on packing of the common core, which generates a large number of persistent bars appearing early. The Wasserstein distance between barcodes in Figure \ref{fig:ex_pair} is driven by those early persistent bars, which are weighted more strongly than the red bar. However, such extra features do pose a challenge for the GIT approach that focuses on local curvature. 

\section{Discussion}

Protein fold classification can be done by considering proteins as curves in space, characterizing these curves by their corresponding 30-dimensional GIT vectors and subsequently identifying clusters of GIT vectors. Such clusters correspond to proteins with similar folds. The great advantage of this method is that a computationally expensive pairwise comparison of protein structures is replaced by efficient clustering of vectors. In contrast, the focus of the present paper is on the treatment of protein structures as topological objects characterized by PDs.

Our AJIVE analysis provides a decomposition of the data that allows us to meaningfully compare the PH approach to GIT. In particular, the individual components of the MDS coordinates returned by AJIVE highlights the fact that PH perceives aspects of protein topology different from the GIT vectors. Specifically, this point is made by studying both clusters and pairs of proteins. A direction of interest for future work is to study how additional information from PH improves the GIT protein fold classification. This could be done using the joint components from AJIVE to analyze clustering improvement.

\begin{appendices}
\chapter{Tables}
\label{tables}
\input{appendix}
\end{appendices}

\end{document}

%% file: appendix.tex
\begin{table}[h!]
\centering
\captionsetup{justification=centering}
 \begin{tabular}{||c c||}
 \hline
 Protein &  CATH label \\ [0.5ex] 
 \hline\hline
 2vuj-A00 & 2.60.120.180 \\
 \hline
 2dfb-A00 & 2.60.120.180 \\ 
 \hline
 1pvx-A00 & 2.60.120.180 \\
 \hline
 2z79-B00 & 2.60.120.180 \\
 \hline
 1t6g-D00 & 2.60.120.180 \\
 \hline
 3b5l-B00 & 2.60.120.180 \\ 
 \hline
 1xnk-A00 & 2.60.120.180 \\ [1ex]
 \hline
\end{tabular}
\caption{Proteins contained in the cluster colored red in Figures \ref{fig:ind_cluster_21} and \ref{fig:ind_cluster_21_proteins}.}
\label{table:cluster_table}
\end{table}

\begin{table}[h!]
\centering
 \begin{tabular}{||c c c c||} 
 \hline
 Proteins & CAT label & distance ratio & plot color \\ [0.5ex] 
 \hline\hline
  2xb4-A00, 3nwo-A00 & 3.40.50 & 0.004422826 & red \\ 
 \hline
 2o9s-A00, 2a28-A00 & 2.30.30 & 0.006634789 & orange \\
 \hline
 1f8m-C00, 1s5m-A00 & 3.20.20 & 0.007395661 & green \\
 \hline
 3dac-A00, 3fe7-A00 & 1.10.245 & 0.007683128 & yellow \\
 \hline
 3e6j-A00, 1pgv-A00 & 3.80.10 & 0.012707447 & purple \\ 
 \hline
 2nnr-A00, 2omy-B00 & 2.60.40 & 0.012987783 & blue \\ 
 \hline
 2ozf-A00, 2he4-A00 & 2.30.42 & 0.016355023	& pink \\ 
 \hline
 2x7b-A00, 1n71-C00 & 3.40.630 & 0.036542583 & magenta \\
 \hline
 1urr-A00, 2eky-C00 & 3.30.70 & 0.039149797 & black \\
 \hline
 2p39-A00, 1bas-A00 & 2.80.10 & 0.041134023 & cyan \\ [1ex] 
\hline
\end{tabular}
\caption{CAT pairs closely related by PH but not by GIT.}
\label{table:pairs_table}
\end{table}

%% file: springer_formatted.bbl
\begin{thebibliography}{10}
	\providecommand{\url}[1]{{#1}}
	\providecommand{\urlprefix}{URL }
	\expandafter\ifx\csname urlstyle\endcsname\relax
	\providecommand{\doi}[1]{DOI~\discretionary{}{}{}#1}\else
	\providecommand{\doi}{DOI~\discretionary{}{}{}\begingroup
		\urlstyle{rm}\Url}\fi
	
	\bibitem{burley.2020.NAR}
	Burley, S.K., Bhikadiya, C., Bi, C., Bittrich, S., Chen, L., Crichlow, G.V.,
	Christie, C.H., Dalenberg, K., Di~Costanzo, L., Duarte, J.M., Dutta, S.,
	Feng, Z., Ganesan, S., Goodsell, D.S., Ghosh, S., Green, R.K., Guranović,
	V., Guzenko, D., Hudson, B.P., Lawson, C., Liang, Y., Lowe, R., Namkoong, H.,
	Peisach, E., Persikova, I., Randle, C., Rose, A., Rose, Y., Sali, A., Segura,
	J., Sekharan, M., Shao, C., Tao, Y.P., Voigt, M., Westbrook, J., Young, J.Y.,
	Zardecki, C., Zhuravleva, M.: {RCSB Protein Data Bank: powerful new tools for
		exploring 3D structures of biological macromolecules for basic and applied
		research and education in fundamental biology, biomedicine, biotechnology,
		bioengineering and energy sciences}.
	\newblock Nucleic Acids Research \textbf{49}(D1), D437--D451 (2020).
	\newblock \doi{10.1093/nar/gkaa1038}.
	\newblock \urlprefix\url{https://doi.org/10.1093/nar/gkaa1038}
	
	\bibitem{cang.2015.CMB}
	Cang, Z., Mu, L., Wu, K., Opron, K., Xia, K., Wei, G.W.: A topological approach
	for protein classification.
	\newblock Computational and Mathematical Biophysics \textbf{3}(1) (2015).
	\newblock \doi{doi:10.1515/mlbmb-2015-0009}.
	\newblock \urlprefix\url{https://doi.org/10.1515/mlbmb-2015-0009}
	
	\bibitem{carlsson.2009.BAMS}
	Carlsson, G.: Topology and data.
	\newblock Bull. Amer. Math. Soc. (N.S.) \textbf{46}(2), 255--308 (2009).
	\newblock \doi{10.1090/S0273-0979-09-01249-X}.
	\newblock
	\urlprefix\url{https://doi-org.libproxy.lib.unc.edu/10.1090/S0273-0979-09-01249-X}
	
	\bibitem{chen.2010.AC}
	Chen, V.B., Arendall III, W.B., Headd, J.J., Keedy, D.A., Immormino, R.M.,
	Kapral, G.J., Murray, L.W., Richardson, J.S., Richardson, D.C.: {{\it
			MolProbity}: all-atom structure validation for macromolecular
		crystallography}.
	\newblock Acta Crystallographica Section D \textbf{66}(1), 12--21 (2010).
	\newblock \doi{10.1107/S0907444909042073}.
	\newblock \urlprefix\url{https://doi.org/10.1107/S0907444909042073}
	
	\bibitem{DAVIES1995853}
	Davies, G., Henrissat, B.: Structures and mechanisms of glycosyl hydrolases.
	\newblock Structure \textbf{3}(9), 853--859 (1995).
	\newblock \doi{https://doi.org/10.1016/S0969-2126(01)00220-9}.
	\newblock
	\urlprefix\url{https://www.sciencedirect.com/science/article/pii/S0969212601002209}
	
	\bibitem{edelsbrunner.1995.DCG}
	Edelsbrunner, H.: The union of balls and its dual shape.
	\newblock Discrete Comput. Geom. \textbf{13}(3-4), 415--440 (1995).
	\newblock \doi{10.1007/BF02574053}.
	\newblock
	\urlprefix\url{https://doi-org.libproxy.lib.unc.edu/10.1007/BF02574053}
	
	\bibitem{edelsbrunner.2010.AMS}
	Edelsbrunner, H., Harer, J.L.: Computational topology.
	\newblock American Mathematical Society, Providence, RI (2010).
	\newblock \doi{10.1090/mbk/069}.
	\newblock \urlprefix\url{https://doi-org.libproxy.lib.unc.edu/10.1090/mbk/069}.
	\newblock An introduction
	
	\bibitem{edelsbrunner.1983.TIT}
	Edelsbrunner, H., Kirkpatrick, D.G., Seidel, R.: On the shape of a set of
	points in the plane.
	\newblock IEEE Trans. Inform. Theory \textbf{29}(4), 551--559 (1983).
	\newblock \doi{10.1109/TIT.1983.1056714}.
	\newblock
	\urlprefix\url{https://doi-org.libproxy.lib.unc.edu/10.1109/TIT.1983.1056714}
	
	\bibitem{feng.2018.JMvA}
	Feng, Q., Jiang, M., Hannig, J., Marron, J.S.: Angle-based joint and individual
	variation explained.
	\newblock J. Multivariate Anal. \textbf{166}, 241--265 (2018).
	\newblock \doi{10.1016/j.jmva.2018.03.008}.
	\newblock
	\urlprefix\url{https://doi-org.libproxy.lib.unc.edu/10.1016/j.jmva.2018.03.008}
	
	\bibitem{gamerio.2015.JIAM}
	Gameiro, M., Hiraoka, Y., Izumi, S., Kramar, M., Mischaikow, K., Nanda, V.: A
	topological measurement of protein compressibility.
	\newblock Jpn. J. Ind. Appl. Math. \textbf{32}(1), 1--17 (2015).
	\newblock \doi{10.1007/s13160-014-0153-5}.
	\newblock
	\urlprefix\url{https://doi-org.libproxy.lib.unc.edu/10.1007/s13160-014-0153-5}
	
	\bibitem{ghrist.2008.BAMS}
	Ghrist, R.: Barcodes: the persistent topology of data.
	\newblock Bull. Amer. Math. Soc. (N.S.) \textbf{45}(1), 61--75 (2008).
	\newblock \doi{10.1090/S0273-0979-07-01191-3}.
	\newblock
	\urlprefix\url{https://doi-org.libproxy.lib.unc.edu/10.1090/S0273-0979-07-01191-3}
	
	\bibitem{ghrist.2014.createspace}
	Ghrist, R.: Elementary Applied Topology.
	\newblock Createspace (2014)
	
	\bibitem{gronbaek.2020.PJ}
	Gr{\o}nb{\ae}k, C., Hamelryck, T., R{\o}gen, P.: Gisa: using gauss integrals to
	identify rare conformations in protein structures.
	\newblock PeerJ \textbf{8} (2020).
	\newblock \doi{10.7717/peerj.9159}.
	\newblock \urlprefix\url{https://doi.org/10.7717/peerj.9159}
	
	\bibitem{ichinomiya.2020.BJ}
	Ichinomiya, T., Obayashi, I., Hiraoka, Y.: Protein-folding analysis using
	features obtained by persistent homology.
	\newblock Biophysical Journal \textbf{118}(12), 2926--2937 (2020).
	\newblock \doi{https://doi.org/10.1016/j.bpj.2020.04.032}.
	\newblock
	\urlprefix\url{https://www.sciencedirect.com/science/article/pii/S0006349520303763}
	
	\bibitem{jolliffe.2002.springer}
	Jolliffe, I.T.: Principal component analysis, second edn.
	\newblock Springer Series in Statistics. Springer-Verlag, New York (2002)
	
	\bibitem{kerber.2017.ACM}
	Kerber, M., Morozov, D., Nigmetov, A.: Geometry helps to compare persistence
	diagrams.
	\newblock ACM J. Exp. Algorithmics \textbf{22}, Art. 1.4, 20 (2017).
	\newblock \doi{10.1145/3064175}.
	\newblock \urlprefix\url{https://doi-org.libproxy.lib.unc.edu/10.1145/3064175}
	
	\bibitem{top8000}
	Laboratory, T.R.: The top8000 protein dataset (2010).
	\newblock
	\urlprefix\url{http://kinemage.biochem.duke.edu/databases/top8000.php}
	
	\bibitem{vanderMaaten.2008.JMLR}
	van~der Maaten, L., Hinton, G.: Visualizing data using t-sne.
	\newblock Journal of Machine Learning Research \textbf{9}(86), 2579--2605
	(2008).
	\newblock \urlprefix\url{http://jmlr.org/papers/v9/vandermaaten08a.html}
	
	\bibitem{morozov}
	Morozov, D.: Dionysus.
	\newblock \urlprefix\url{https://mrzv.org/software/dionysus2/}
	
	\bibitem{panaretos.2020.springer}
	Panaretos, V.M., Zemel, Y.: An invitation to statistics in Wasserstein space.
	\newblock Springer Nature (2020)
	
	\bibitem{roegen.2005.JPCM}
	Roegen, P.: Evaluating protein structure descriptors and tuning gaussian
	integral based descriptors.
	\newblock J. Phys.: Condens. Matter  (2005)
	
	\bibitem{roegen.2003.PNAS}
	Roegen, P., Fain, B.: Automatic classification of protein structure by using
	gauss integrals.
	\newblock PNAS  (2003)
	
	\bibitem{CATH.2019}
	Sillitoe, I., Dawson, N., Lewis, T., Das, S., Lees, J., Ashford, P., Tolulope,
	A., Scholes, H.M., Senatorov, I., Bujan, A., Rodriguez-Conde, F.C., Dowling,
	B., Thornton, J., Orengo, C.: Cath: expanding the horizons of structure-based
	functional annotations for genome sequences.
	\newblock Nucleic Acids Res.  (2019)
	
	\bibitem{torgerson.1952.psych}
	Torgerson, W.S.: Multidimensional scaling. {I}. {T}heory and method.
	\newblock Psychometrika \textbf{17}, 401--419 (1952).
	\newblock \doi{10.1007/BF02288916}.
	\newblock
	\urlprefix\url{http://dx.doi.org.libproxy.lib.unc.edu/10.1007/BF02288916}
	
	\bibitem{winter.2009.IEEE}
	Winter, P., Sterner, H., Sterner, P.: Alpha shapes and proteins.
	\newblock IEEE  (2009)
	
	\bibitem{xia.2014.IJNMBE}
	Xia, K., Wei, G.W.: Persistent homology analysis of protein structure,
	flexibility, and folding.
	\newblock International journal for numerical methods in biomedical engineering
	\textbf{30}(8), 814--844 (2014).
	\newblock \doi{10.1002/cnm.2655}.
	\newblock \urlprefix\url{https://pubmed.ncbi.nlm.nih.gov/24902720}.
	\newblock 24902720[pmid]
	
\end{thebibliography}
